\documentclass[aps,showpacs]{revtex4} 
\usepackage{amsmath}
\usepackage{amssymb}
\usepackage[dvips]{graphicx}
\begin{document}

\title{ Exact General Relativistic  Perfect Fluid  Disks with Halos  }

\author{Daniel  Vogt\footnote{e-mail: danielvt@ifi.unicamp.br}
  }

\affiliation{
Instituto de F\'{\i}sica Gleb Wataghin, Universidade Estadual 
de Campinas 13083-970 Campinas, S.P., Brazil
}

\author{ Patricio. S. Letelier\footnote{e-mail: letelier@ime.unicamp.br}
 } 
 
\affiliation{
 Departamento de Matem\'atica Aplicada-IMECC,
Universidade Estadual de Campinas,
13081-970 Campinas,  S.P., Brazil}

\begin{abstract}
Using the well-known ``displace, cut and reflect'' method used to generate 
disks from given solutions of Einstein field equations, we construct static 
disks made of perfect fluid based on vacuum Schwarzschild's solution in isotropic coordinates. The same method is applied to different exact
 solutions to the Einstein's
equations that represent   
static spheres of perfect fluids. We construct several models of disks 
with axially symmetric perfect fluid halos. 
 
All disks  have some common features: surface energy density 
and pressures decrease monotonically and rapidly with radius. As the 
``cut'' parameter $a$ decreases, 
the disks become more relativistic, with surface energy density and 
pressure more concentrated 
near the center. Also regions of unstable circular orbits are more likely to 
appear for high relativistic disks. Parameters can be chosen so 
that the sound velocity in the fluid and the tangential velocity of test particles 
in circular motion are less then the velocity of light. 
This tangential velocity first increases with radius and reaches a maximum.
\end{abstract}


\maketitle

\setlength{\parindent}{3em}

\section{Introduction}

Axially symmetric solutions of Einstein's field equations corresponding to 
disklike configurations of matter are of great astrophysical interest, since 
they can be used as models of galaxies or accretion disks. These solutions can 
be static or stationary and with or without radial pressure. Solutions for 
static disks without radial pressure were first studied by Bonnor and Sackfield \cite{Bonnor}, and Morgan and Morgan \cite{Morgan1}, and with
 radial pressure by 
Morgan and Morgan \cite{Morgan2}. Disks with radial 
tension have been considered 
in \cite{Gonzalez}, and models of disks with electric fields \cite{Bicak}, 
magnetic fields \cite{Letelier1}, and both magnetic and electric fields have 
been introduced recently \cite{Katz}. 
Solutions for self-similar static disks were
analyzed by Lynden-Bell and Pineault \cite{LP}, and Lemos \cite{LEM}. The
superposition of static disks with black holes were considered by
Lemos and Letelier \cite{LL1,LL2,LL3}, and Klein \cite{KLE}. Also
Bi\u{c}\'{a}k, Lynden-Bell and Katz \cite{BLK} studied static disks as
sources of known vacuum spacetimes and Bi\u{c}\'{a}k, Lynden-Bell and
Pichon \cite{BLP} found an infinity number of new static solutions.
For a recent survey on relativistic 
gravitating disks, see \cite{Semerak}.

The principal method to generate the above mentioned solution is 
the ``displace, cut and reflect'' method.
One of the main  problem of 
the solutions generated by using   this simple
 method  is that usually the matter content of the disk is
 anisotropic i.e.,  the
 radial pressure is different from the azimuthal  pressure. In most of the 
solutions  the radial pressure is null. This made these  solutions 
 rather unphysical. Even though, one can argue that when no radial pressure is present  stability can be achieved if we have two circular 
streams of particles moving in opposite directions (counter rotating 
hypothesis, see for instance \cite{BLK}). 

In this article we apply the ``displace, cut and reflect'' method
 on spherically symmetric  solutions
of Einstein's field equations in isotropic coordinates to generate static
 disks made of a
{\em perfect fluid}, i.e., with  radial   pressure equal to tangential
 pressure 
 and also disks of perfect fluid surrounded by an halo made of 
  perfect fluid matter.

The article is divided as follows. Section II gives an overview of
 the   ``displace, cut and reflect''  method. Also we 
present   the basic equations 
used to calculate the main physical variables of the disks. 
In section III we apply the formalism to obtain 
 the simplest model of disk, that is based on 
Schwarzschild's vacuum  solution in isotropic coordinate. The  
  generated class of  disks is made of 
a perfect fluid with   well behaved density and pressure. 
Section IV presents some models of disks with halos
obtained from different known exact solutions of Einstein's field 
equations for static spheres of perfect fluid in isotropic coordinates. 
 In section 
V we give some examples of disks with halo generated from spheres
 composed of fluid layers. Section 
VI is devoted to discussion of the results.

\section{Einstein equations and disks}

For a static spherically symmetric spacetime
the general line element in isotropic spherical coordinates
can be cast as, 
\begin{equation} \label{eq_line1}
\mathrm{d}s^2=e^{\nu(r)}\mathrm{d}t^2-e^{\lambda(r)}\left[ 
\mathrm{d}r^2+r^2(\mathrm{d}\theta^2+\sin^2 \theta \mathrm{d}\varphi^2) 
\right] \mbox{.}
\end{equation}
In cylindrical coordinates $(t\text{,}R\text{,}z\text{,}\varphi)$ the line 
element (\ref{eq_line1}) takes the form,
\begin{equation} \label{eq_line2}
\mathrm{d}s^2=e^{\nu(R,z)}\mathrm{d}t^2-e^{\lambda(R,z)}\left( 
\mathrm{d}R^2+\mathrm{d}z^2+R^2 \mathrm{d}\varphi^2 \right) \mbox{.}
\end{equation}

\begin{figure} 
\centering
\includegraphics[scale=0.8]{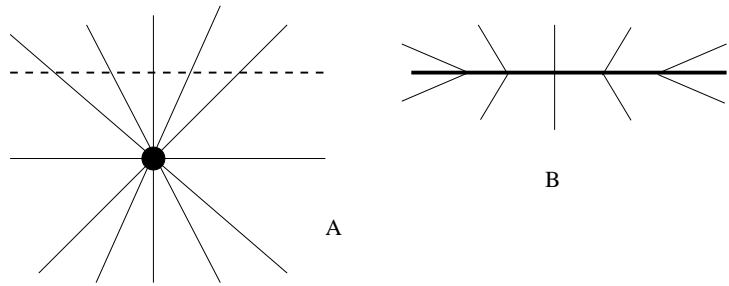}
\caption{An illustration of the ``displace, cut and reflect'' method for the generation of disks. 
In A the spacetime with a singularity is displaced and cut by 
 a plane (dotted line), in B the part with singularities is disregarded and the
upper part is reflected on the plane.} \label{fig_schem1}
\end{figure}

The metric of the disk will be constructed using the well known ``displace, cut and 
reflect'' 
method that was used by Kuzmin \cite{Kuzmin} in Newtonian gravity and later in general 
relativity by many authors \cite{Gonzalez}-\cite{Semerak}. The
 material content of the disk will be described by functions that are 
distributions with support on the disk. The method can be divided in the following steps that are illustrated in 
Fig.\ \ref{fig_schem1}: 
First, in a space wherein we have a compact source of gravitational field,
 we choose a surface (in our case, the plane $z=0$) that divides the  space in 
two pieces:  one with no singularities or sources and the other with the
 sources. Then we 
disregard the part of the space with singularities and use the surface to make an inversion of the 
nonsingular part of the space. This results in a space with a 
singularity that is a delta 
function with support on $z=0$.
This procedure is mathematically equivalent to make the transformation
$z \rightarrow |z|+a$, with $a$ constant. In the Einstein tensor we
have first and second derivatives of $z$.  Since $\partial_z |z|=2
\vartheta(z)-1$ and $\partial_{zz} |z|=2\delta(z)$, where
$\vartheta(z)$ and $\delta(z)$ are, respectively, the Heaviside
function and the Dirac distribution. Therefore the Einstein field
equations will separate in two different pieces \cite{Taub}: one valid
for $z\not =0$ (the usual Einstein's equations), and other involving
distributions with an associated energy-momentum tensor,
$T_{ab}=Q_{ab}\delta(z)$, with support on $z=0$. For the metric
(\ref{eq_line2}), the non-zero components of $Q_{ab}$ are
\begin{align}
Q^t_t &=\frac{1}{16 \pi} \left[
-b^{zz}+g^{zz}(b^R_R+b^z_z+b^{\varphi}_{\varphi}) \right] \mbox{,}
\label{eq_qtt}\\ Q^R_R &= Q^{\varphi}_{\varphi} =\frac{1}{16 \pi}
\left[ -b^{zz}+g^{zz}(b^t_t+b^R_R+b^z_z) \right] \mbox{,}
\label{eq_qrr}
\end{align}
where $b_{ab}$ denote the jump of the first derivatives of the metric
tensor on the plane $z=0$,
\begin{equation}
b_{ab}=g_{ab,z}|_{z=0^+}-g_{ab,z}|_{z=0^-} \mbox{,}
\end{equation}
and the other quantities are evaluated at $z=0^+$. The ``true''
surface energy-momentum tensor of the disk can be written as
$S_{ab}=\sqrt{-g_{zz}}Q_{ab}$, thus the surface energy density
$\sigma$ and the radial and azimuthal pressures or tensions $(P)$
read:
\begin{equation} \label{eq_disk_surf}
\sigma=\sqrt{-g_{zz}}Q^t_t \text{,} \quad
P=-\sqrt{-g_{zz}}Q^R_R=-\sqrt{-g_{zz}}Q^\varphi_\varphi \mbox{.}
\end{equation}
Note that when the same procedure is applied to an axially symmetric
spacetime in Weyl coordinates we have $Q^R_R=0$, i.e., we have no
radial pressure or tension.

\begin{figure} 
\centering 
\includegraphics[scale=0.8]{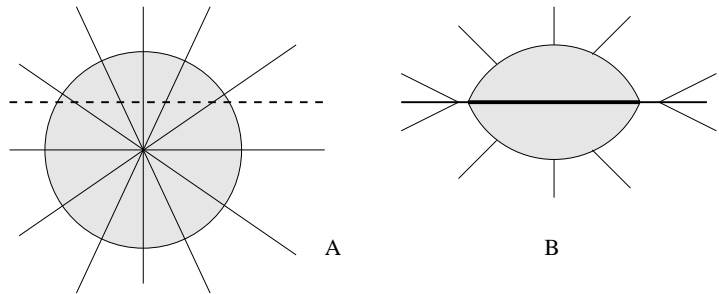}
\caption{An illustration of the ``displace, cut and reflect'' method
for the generation of disks with halos. In A the sphere of perfect
fluid is displaced and cut by a plane (dotted line), in B the lower
part is disregarded and the upper part is reflected on the plane.}
\label{fig_schem2}
\end{figure}

 This procedure in principle can be applied to any spacetime solution
of the Einstein equations with or without source (stress tensor).  The
application to a static sphere of perfect fluid is schematized in
Fig.\ \ref{fig_schem2}. The sphere is displaced and cut by a distance
$a$ less then its radius $r_b$. The part of the space that contains
the center of the sphere is disregarded. After the inversion of the
remaining space, we end up with a disk surrounded by a cap of perfect
fluid. The properties of the inner part of the disk will depend on the
internal fluid solution, but if the internal spherical fluid solution
is joined to the standard external Schwarzschild solution, the
physical properties of the outer part of the disk will be those
originated from Schwarzschild's vacuum solution.

In isotropic coordinates the matching at the boundary of the fluid
sphere leads to four continuity conditions 
metric functions $e^{\lambda}$ and $e^{\nu}$ together with their first
derivatives with respect to the radial coordinate should be continuous
across the boundary. In addition, to have a compact body the pressure
at the surface of the material sphere has to drop to zero.  Also to have
a meaningfully solution the velocity of sound,
$\mathrm{V}^2=\frac{dp}{d\rho}$ should be restricted to the interval
$0 \leq \mathrm{V}<1$.

The Einstein equations for a static spherically symmetric space time
in isotropic coordinates for a perfect fluid source give us that
density $\rho$ and pressure $p$ are related to the metric functions by
\begin{align}
\rho &= -\frac{e^{-\lambda}}{8 \pi} \left[
\lambda^{''}+\frac{1}{4}(\lambda')^2+\frac{2\lambda'}{r} \right]
\mbox{,} \\ p &= \frac{e^{-\lambda}}{8 \pi}\left[
\frac{1}{4}(\lambda')^2+\frac{1}{2}\lambda'\nu'+\frac{1}{r}(\lambda'+\nu')
\right] \mbox{,}
\end{align}
where primes indicate differentiation with respect to $r$.

\begin{figure} 
\centering 
\includegraphics[scale=0.8]{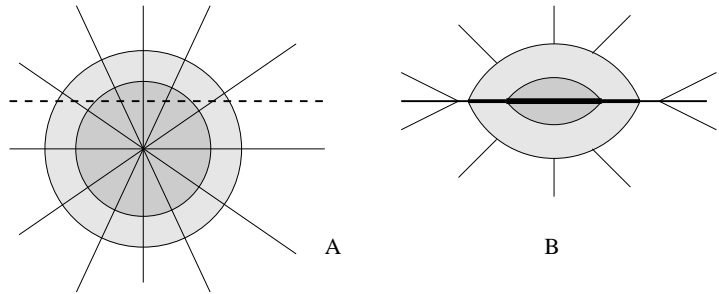}
\caption{An illustration of the ``displace, cut and reflect'' method
for the generation of disks with various layers of halos. In A the
sphere with different layers of fluid is displaced and cut with a
plane (dotted line), in B the field is reflected on the
plane.}\label{fig_schem3}
\end{figure}

Also static spheres composed of various layers of fluid can be used to
generate disks with halos of fluid layers (see figure
\ref{fig_schem3}). The disk will then be composed of different axial
symmetric ``pieces'' glued together.
The matching conditions at the boundary of adjacent spherical fluid
 layers in isotropic coordinates involves four continuity conditions:
 the two metric functions $e^{\lambda}$ and $e^{\nu}$, the first
 derivative of $\lambda$ with respect to the radial coordinate, and
 the pressure should be continuous across the boundary. At the most
 external boundary, the metric functions $e^{\lambda}$ and $e^{\nu}$,
 and their first derivatives with respect to the radial coordinate
 should be continuous across the boundary; also the pressure there
 should go to zero.

\section{The simplest disk }

We first apply the ``displace, cut and reflect'' method to generate
disks discussed in the previous section and depicted in Fig.\
\ref{fig_schem1} to the Schwarzschild metric in isotropic coordinates
$(t,r,\theta,\varphi)$,
\begin{equation} 
\mathrm{d}s^2=\frac{\left(1-\frac{m}{2r}\right)^2}{\left(1+\frac{m}{2r}\right)^2}\mathrm{
d}t^2- \left( 1+\frac{m}{2r} \right)^4
\left[\mathrm{d}r^2+r^2(\mathrm{d}\theta^2+\sin^2 \theta
\mathrm{d}\varphi^2) \right] \mbox{.} \label{eq_metrica_sch}
\end{equation}
Expressing solution (\ref{eq_metrica_sch}) in cylindrical coordinates,
and using Eq.\ (\ref{eq_qtt}) -- (\ref{eq_disk_surf}), we obtain a
disk with surface energy density $\sigma$ and radial and azimuthal
pressures (or tensions) $P$ given by
\begin{align}
\sigma &= \frac{4ma}{\pi(m+2\sqrt{R^2+a^2})^3}\mbox{,}
\label{eq_en_sch} \\ P &=
-\frac{2m^2a}{\pi(m+2\sqrt{R^2+a^2})^3(m-2\sqrt{R^2+a^2})}
\label{eq_P_sch}\mbox{.}
\end{align}
The total mass of the disk can be calculated with the help of Eq.\
(\ref{eq_en_sch}):
\begin{equation}
\mathcal{M}=\int_0^{\infty}\! \int_0^{2\pi} \sigma
\sqrt{g_{RR}g_{\varphi \varphi}} \, \mathrm{d}R \,\mathrm{d}\varphi =
\frac{m}{4a} (m+4a) \mbox{.}
\end{equation}

Eq.\ (\ref{eq_en_sch}) shows that the disk's surface energy density is
always positive (weak energy condition). Positive values (pressure) for
the stresses in azimuthal and radial directions are obtained if
$m<2\sqrt{R^2+a^2}$. The strong energy condition, $\sigma+ P_{\varphi
\varphi}+P_{RR}= \sigma+2P>0$ is then satisfied. These properties
characterize a fluid made of matter with the usual gravitational
attractive property. This is not a trivial property of these disks
since it is known that the
``displace, cut and reflect'' method sometimes gives disks made of exotic matter like cosmic strings, see for instance \cite{LLstrings}.

 Another useful parameter is the velocity of sound propagation 
$V$, defined as $V^2=\frac{dP}{d\sigma}$, which 
can be calculated using Eq.\ (\ref{eq_en_sch}) and Eq.\ (\ref{eq_P_sch}):
\begin{equation}
V^2=\frac{m(4\sqrt{R^2+a^2}-m)}{3(m-2\sqrt{R^2+a^2})^2} \mbox{.} \label{eq_sound_sch}
\end{equation}
The condition $V^2<1$ (no tachyonic matter) imposes the inequalities $m<\sqrt{R^2+a^2}$ or $m>3\sqrt{R^2+a^2}$. 
If the pressure condition and the speed of sound less then the speed of light condition 
are to be simultaneously satisfied, then $m<\sqrt{R^2+a^2}$. This inequality will be valid
in all the disk if $m<a$.
\begin{figure} 
\centering
\includegraphics[scale=0.6]{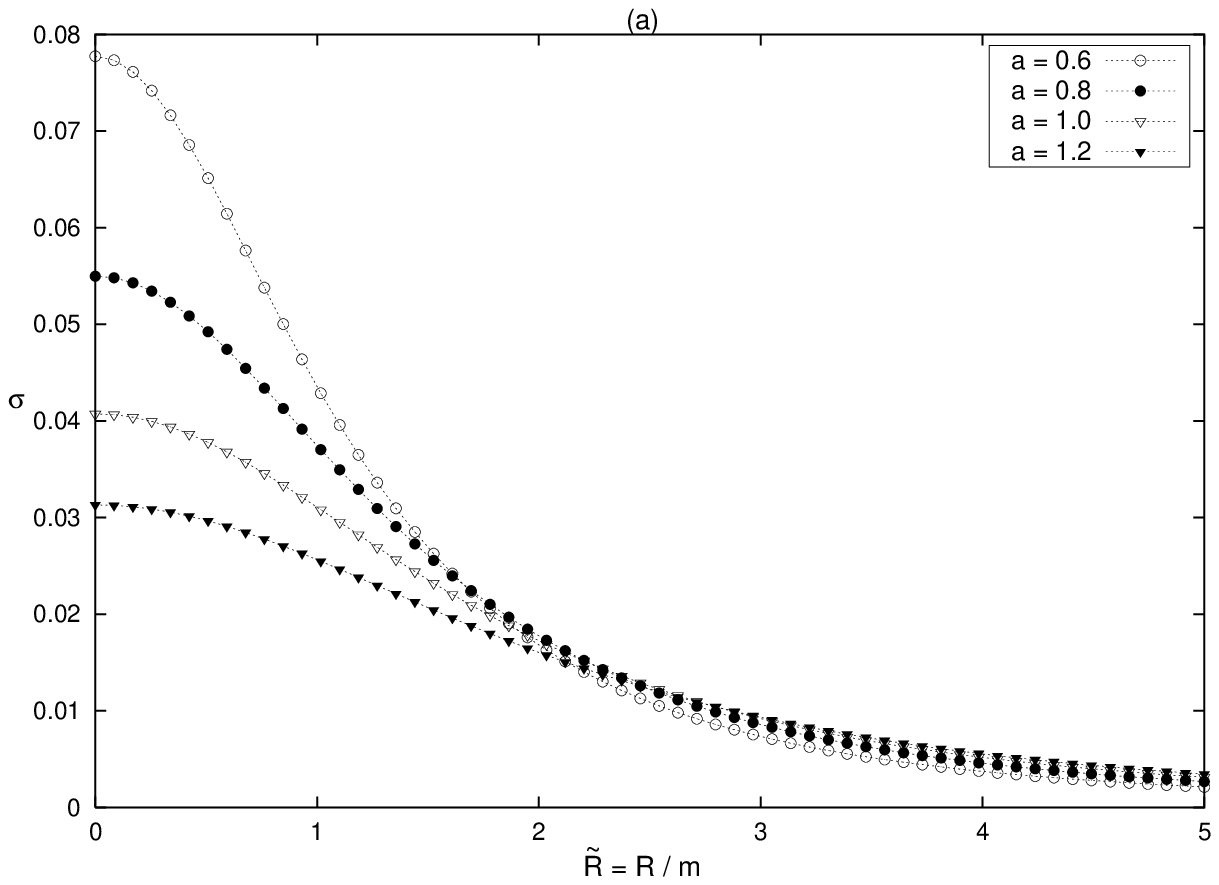}%
\includegraphics[scale=0.6]{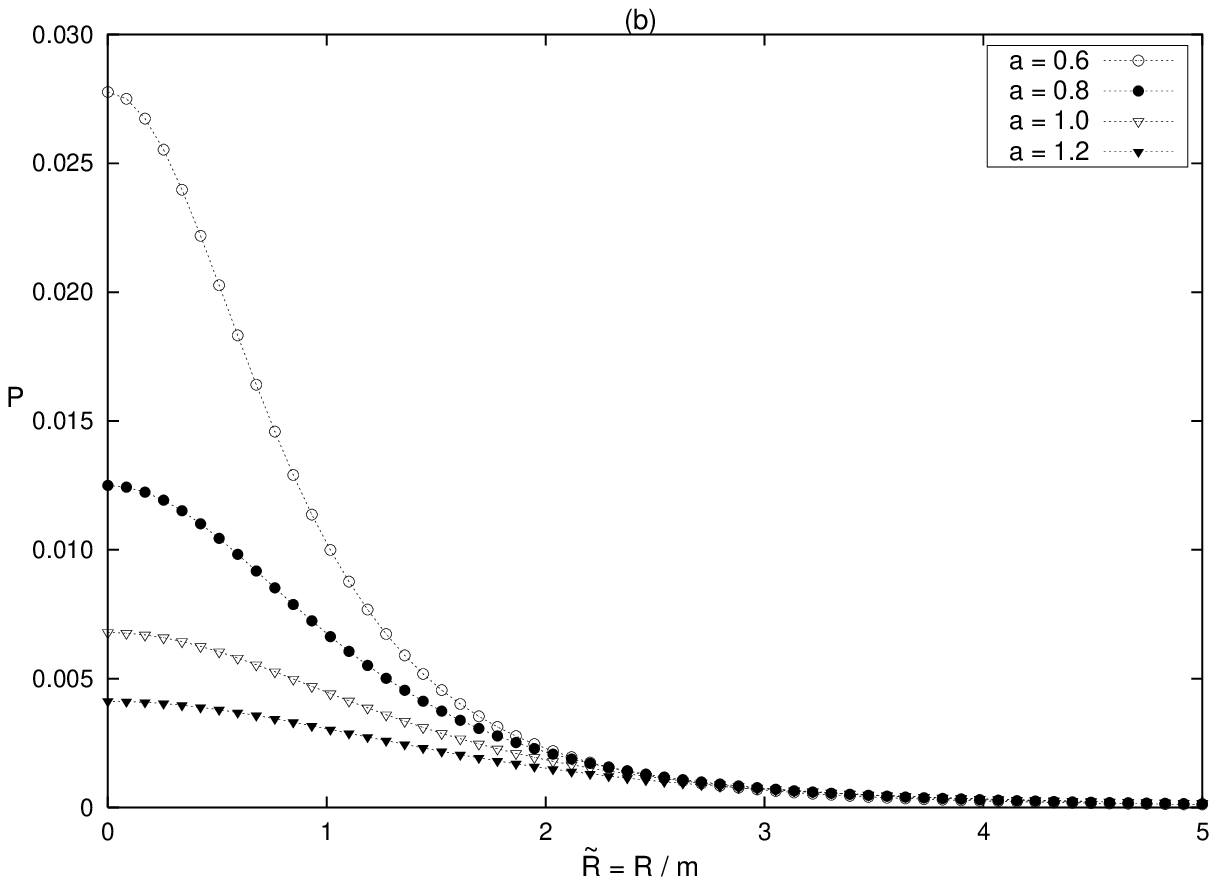} \\
\vspace{0.3cm}
\includegraphics[scale=0.6]{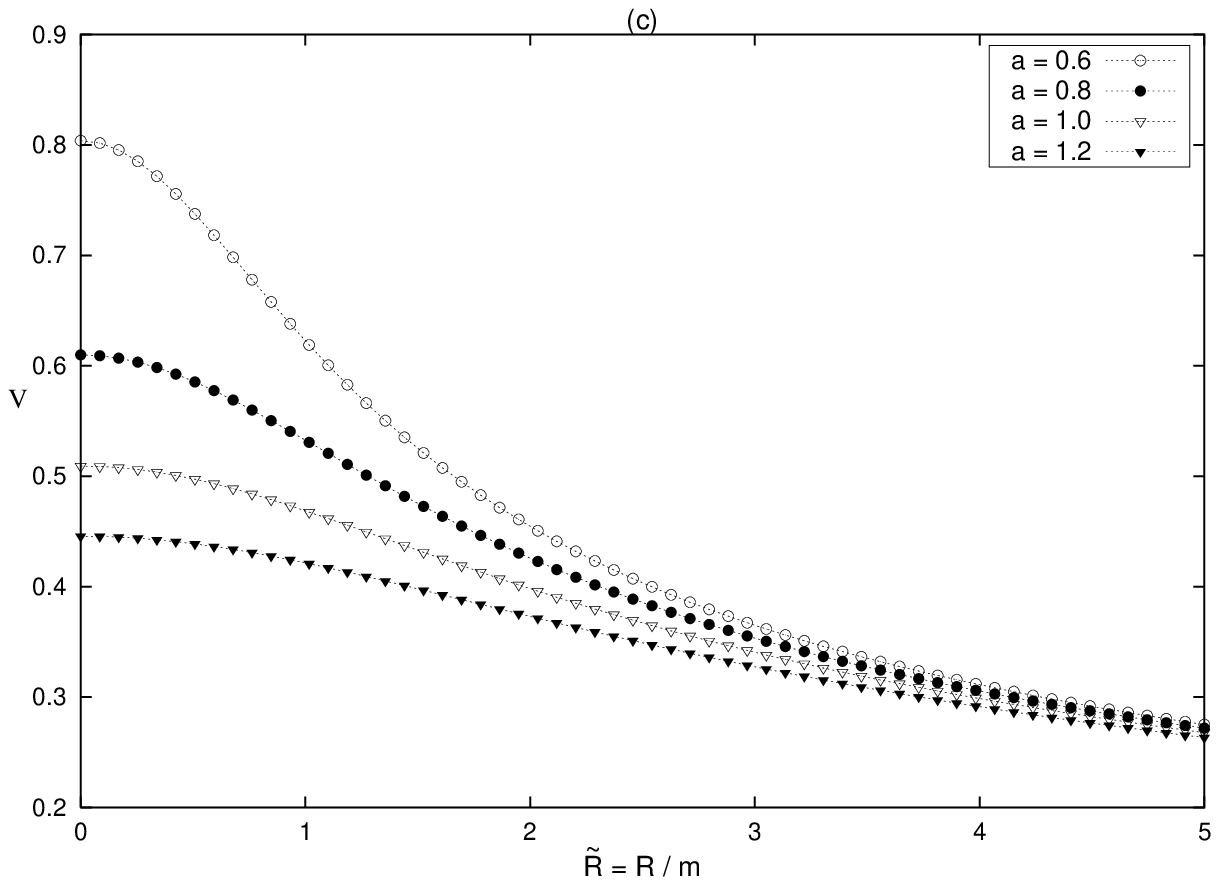}%
\includegraphics[scale=0.6]{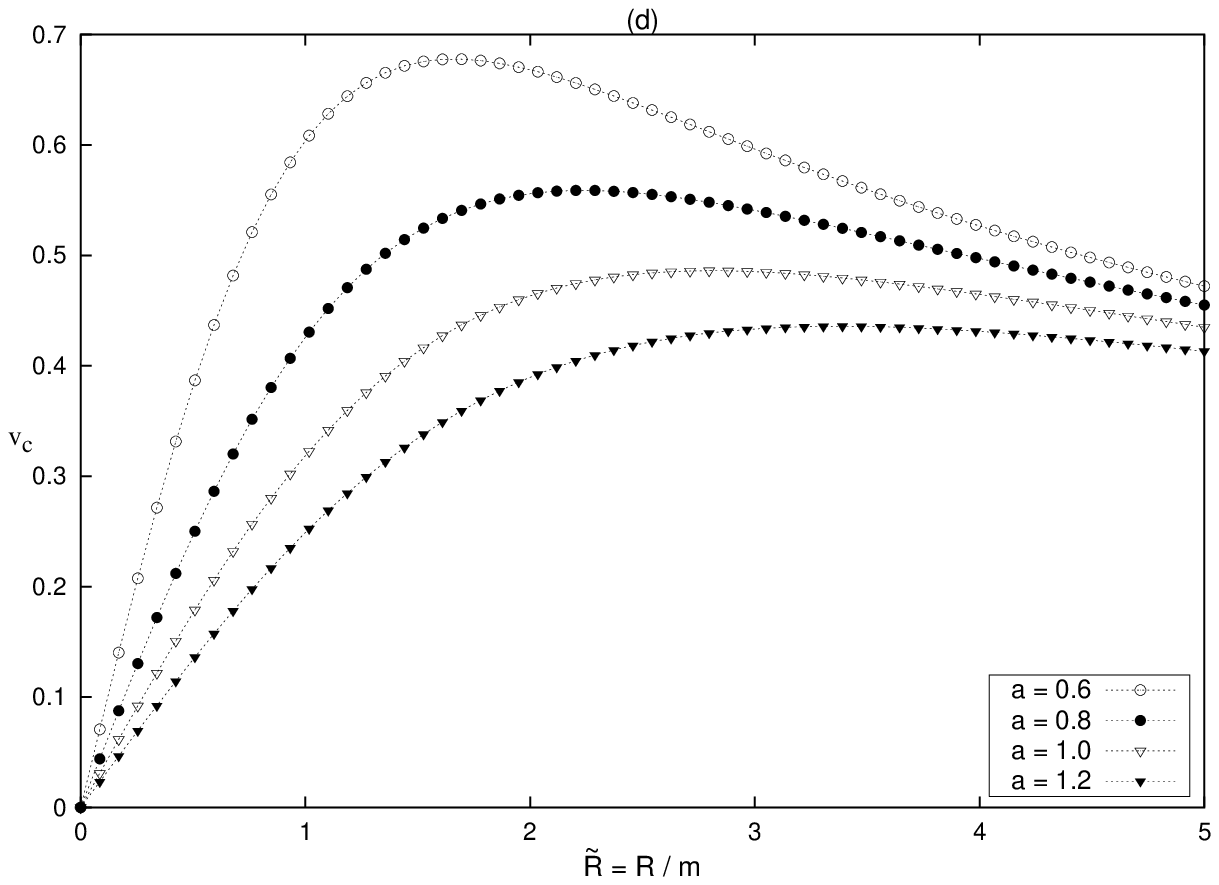}
\caption{(a) The surface energy density $\sigma$, (b) pressures $P$, (c) sound velocity $V$ and (d) tangential velocity $\mathrm{v}_c$ (rotation curve or rotation profile) with $m=0.5$ and $a=0.6 \text{, }0.8\text{, 
}1.0$, 
and $1.2$ as function of $\tilde{R}=R/m$. We use geometric units $G=c=1$.} \label{fig_1}
\end{figure}

\begin{figure} 
\centering
\includegraphics[scale=0.6]{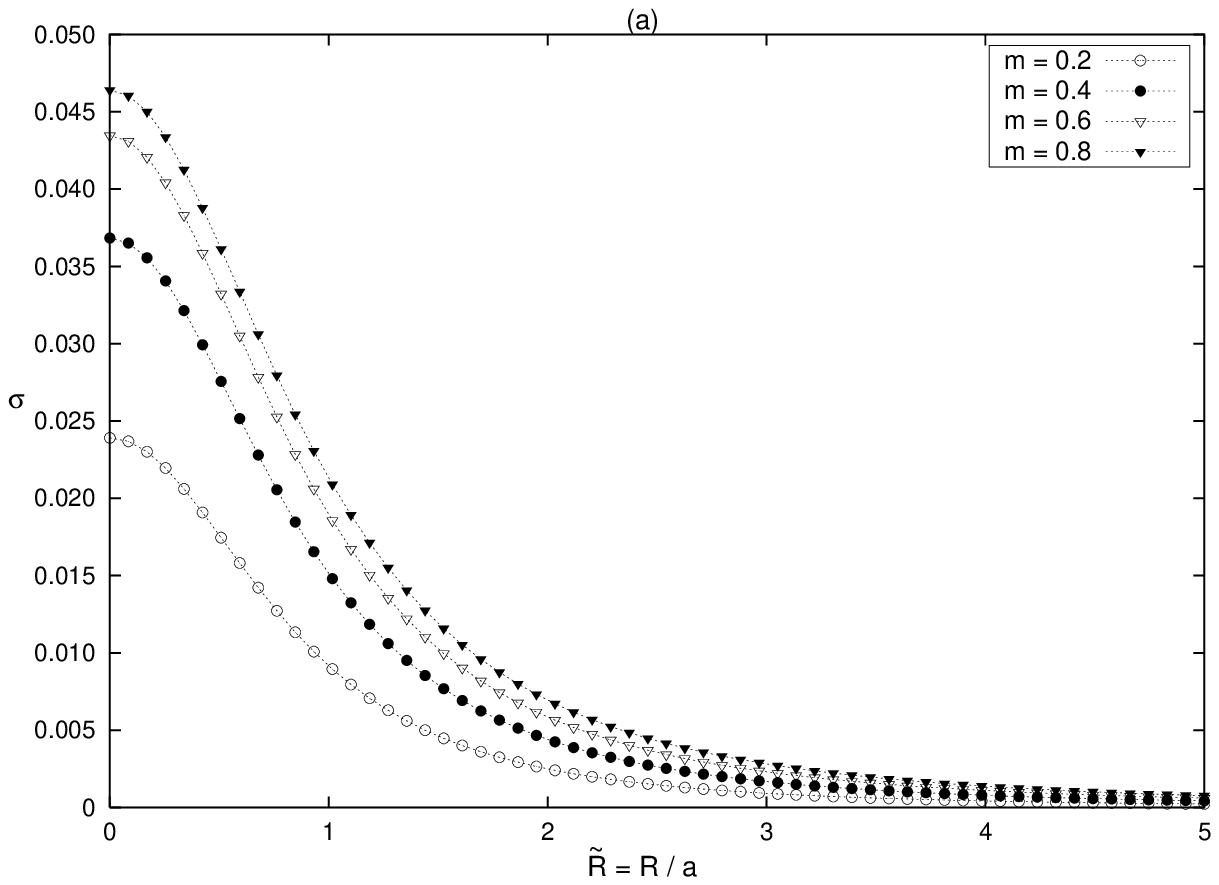}%
\includegraphics[scale=0.6]{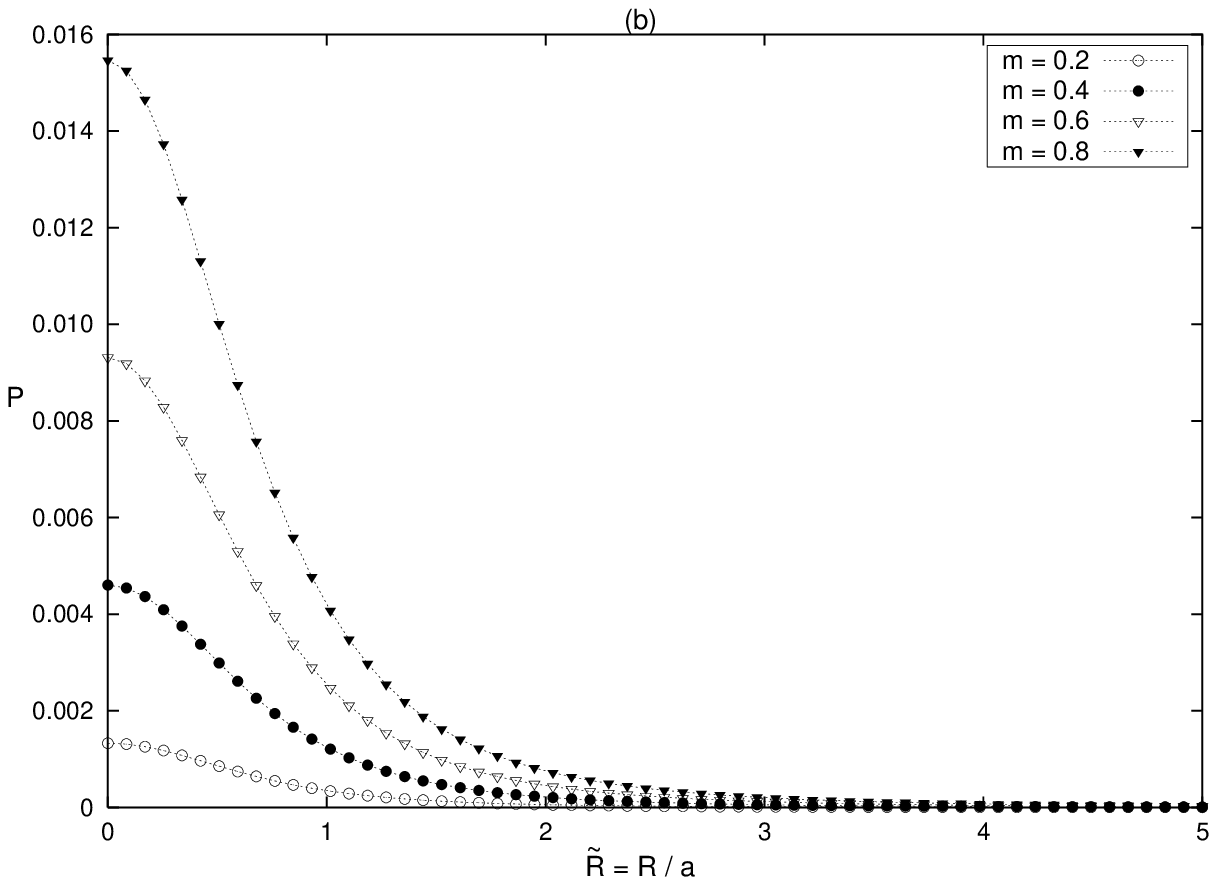} \\
\vspace{0.3cm}
\includegraphics[scale=0.6]{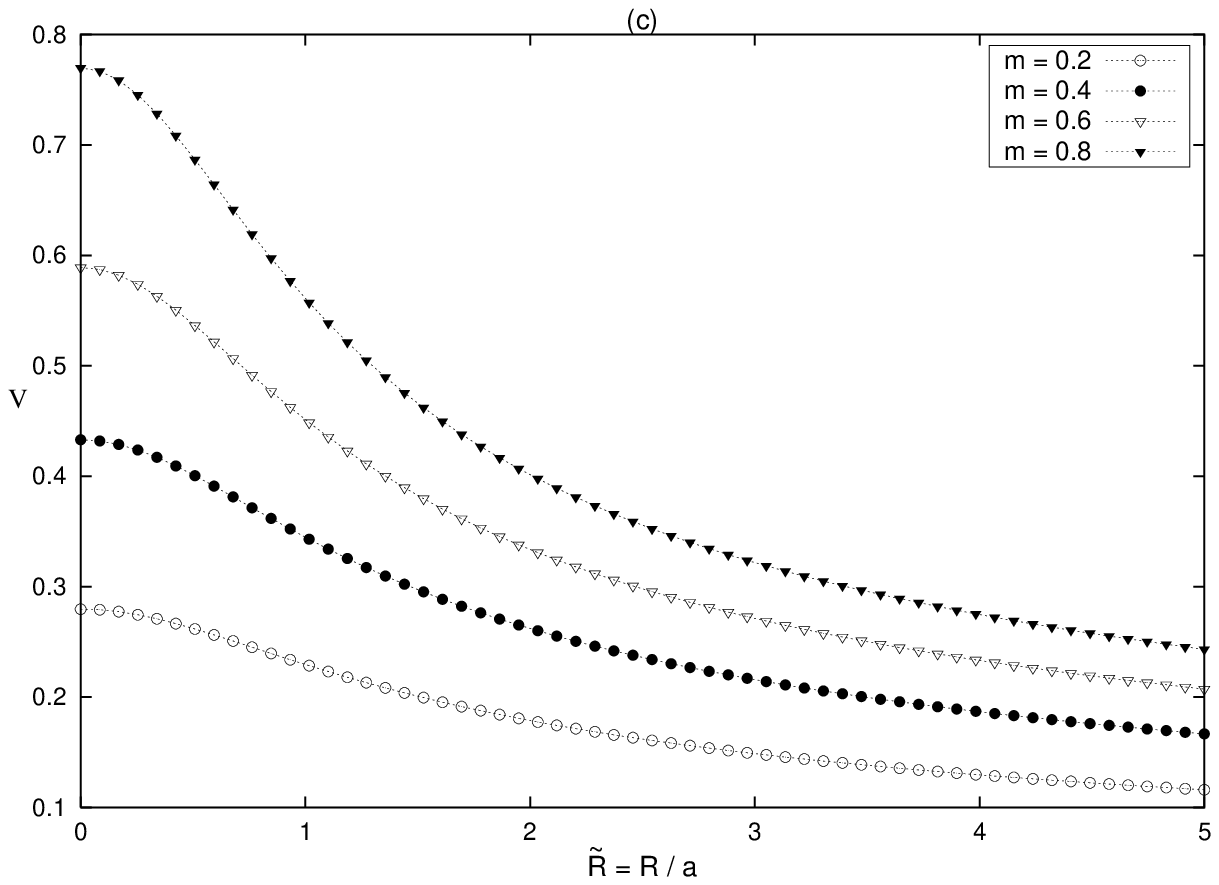}
\caption{(a) The surface energy density $\sigma$, (b) pressures $P$ and (c) sound velocity $V$ with $a=1.0$ and $m=0.2 \text{, }0.4\text{, }0.6$, and $0.8$ as function of $\tilde{R}=R/a$.} \label{fig_2}
\end{figure}

With the presence of radial pressure one does not need the assumption of streams of rotating and
counter rotating matter usually used to explain the stability of static disk models. However, a
tangential velocity (rotation profile)
 can be calculated by assuming a test particle
 moves in a circular geodesic on
the disk. We tacitly assume that this particle only interact
 gravitationally with the fluid.  This assumption is valid for the case of
 a particle moving
 in a very  diluted  gas like the gas made of stars that models a galaxy disk.

 The geodesic equation for the $R$ coordinate obtained from metric (\ref{eq_line2}) is
\begin{equation} \label{eq_geod}
e^{\lambda}\ddot{R}+\frac{1}{2}(e^{\nu})_{,R}\dot{t}^2-\frac{1}{2}(e^{\lambda})_{,R}
(\dot{R}^2+\dot{z}^2)-\frac{1}{2}\left( e^{\lambda}R^2 \right)_{,R}\dot{\varphi}^2=0 \mbox{.}
\end{equation}
For circular motion on the $z=0$ plane, $\dot{R}=\ddot{R}=0$ and $\dot{z}=0$, then Eq.\ (\ref{eq_geod})
reduces to
\begin{equation} \label{eq_phi_t}
\frac{\dot{\varphi}^2}{\dot{t}^2}=\frac{(e^{\nu})_{,R}}{(e^{\lambda}R^2)_{,R}} \mbox{.}
\end{equation}
The tangential velocity measured by an observer at infinity is then
\begin{equation} \label{eq_veloc_tang}
\mathrm{v}_{c}^2=-\frac{g_{\varphi \varphi}}{g_{tt}}\left( \frac{d\varphi}{dt} \right)^2=
R^2\frac{e^{\lambda}(e^{\nu})_{,R}}{e^{\nu}(R^2e^{\lambda})_{,R}} \mbox{.}
\end{equation}
From the metric on the disk,
\begin{equation} \label{eq_sch_coef}
e^{\nu}=\frac{\left( 1-\frac{m}{2\sqrt{R^2+a^2}}\right)^2}{\left( 1+\frac{m}{2\sqrt{R^2+a^2}}\right)^2}
\quad \text{and} \quad e^{\lambda}=\left( 1+\frac{m}{2\sqrt{R^2+a^2}} \right)^4 \mbox{,}
\end{equation}
we find that Eq.\ (\ref{eq_veloc_tang}) can be cast as,
\begin{equation} \label{eq_v_fi_sch}
\mathrm{v}_{c}^2=\frac{mR^2}{\left( 1-\frac{m}{2\sqrt{R^2+a^2}}\right)\left[ (R^2+a^2)^{3/2}+\frac{m}{2}
(a^2-R^2)\right]} \mbox{.}
\end{equation}
For $R>>a$, Eq.\ (\ref{eq_v_fi_sch}) goes as $\mathrm{v}_{c}=(m/R)^{1/2}$, the Newtonian circular velocity. 

To determine the stability of circular orbits on the disk's plane, we use an extension 
of Rayleigh \cite{lord,Landau}  criteria of stability of a fluid at rest in a gravitational field 
\begin{equation} \label{eq_stab}
h\frac{dh}{dR}>0 \mbox{,}
\end{equation}
where $h$ is the specific angular momentum of a particle on the disk's plane:
\begin{equation}
h=-g_{\varphi \varphi}\frac{d\varphi}{ds}=-g_{\varphi \varphi}\frac{d\varphi}{dt}\frac{dt}{ds} \mbox{.}
\end{equation}
Using Eq.\ (\ref{eq_phi_t}) and the relation
\begin{equation}
1=e^{\nu}\left( \frac{dt}{ds} \right)^2-R^2e^{\lambda}\left( \frac{d\varphi}{ds} \right)^2 \mbox{,}
\end{equation}
one obtains the following expression for $h$:
\begin{equation} \label{eq_h}
h=R^2e^{\lambda}\sqrt{\frac{(e^{\nu})_{,R}}{e^{\nu}(R^2e^{\lambda})_{,R}-R^2e^{\lambda}(e^{\nu})_{,R}}}
\mbox{.}
\end{equation}

For the functions (\ref{eq_sch_coef}), Eq.\ (\ref{eq_h}) reads
\begin{equation} \label{eq_h_sch}
h=\frac{2\sqrt{m}R^2\left( 1+\frac{m}{2\sqrt{R^2+a^2}}\right)^2(R^2+a^2)^{1/4}}
{\sqrt{4(R^2+a^2)^2-8mR^2\sqrt{R^2+a^2}+m^2(R^2-a^2)}} \mbox{.}
\end{equation}
The stability criterion is always satisfied for $\frac{a}{m}\gtrapprox 1.016$. 

In figure \ref{fig_1} (a)--(d) we show, respectively, the surface energy density, pressures, the sound velocity and curves of the tangential velocity (rotation curves) Eq.\ (\ref{eq_v_fi_sch})  with $m=0.5$ and $a=0.6 \text{, }0.8\text{, 
}1.0$, and $1.2$ as functions of $\tilde{R}=R/m$. Figure \ref{fig_2} (a)--(c) display, respectively, the surface energy density, pressures and sound velocity with parameters $a=1.0$ and $m=0.2 \text{, }0.4 \text{, }0.6 \text{, }0.8$ as as functions of $\tilde{R}=R/m$. We see that the first three
quantities decrease monotonically with the radius of the disk, as can be checked from 
Eq.\ (\ref{eq_en_sch}), (\ref{eq_P_sch}) and (\ref{eq_sound_sch}). Energy density decreases rapidly enough to, in principle, define a cut off radius and consider the disk as finite.

\section{Disks with halos}
Now we study  some disks with halos constructed  from several
exact solutions of the Einstein equations for static spheres of perfect fluid.
A survey of these class of solutions is presented  in \cite{Kuchowicz}.

\subsection{Buchdahl's Solution} 
The first situation that we shall study is similar to the one depicted in 
Fig.\ \ref{fig_schem2} wherein we start with a sphere of perfect 
 fluid. This case will not be exactly the same as the one presented in the 
mentioned figure 
 because the  sphere  has  no  boundary. Hence  the 
generated disk will be completely immersed in the fluid. An example
of exact solution  of  the Einstein equations  that represent
 a fluid sphere with no boundary is the
 the  Buchdahl solution  that may be regarded as a reasonably close 
analogue to the classical Lane-Emden index 5 polytrope  \cite{Buchdahl}. The metric functions for this solution
are: 
\begin{equation} \label{eq_sol_buch}
e^{\nu}=\left( 
\frac{1-\frac{A}{\sqrt{1+kr^2}}}{1+\frac{A}{\sqrt{1+kr^2}}}\right)^2 \text{,} 
\quad e^{\lambda}=\left( 1+\frac{A}{\sqrt{1+kr^2}}\right)^4 \mbox{,}
\end{equation}
where $A$ and $k$ are constants. Far from the origin the solution goes over into the
external Schwarzschild metric, when $m=\frac{2A}{\sqrt{k}}$. The density, pressure and 
sound velocity are given by:
\begin{gather}
\rho =\frac{3Ak}{2\pi(A+\sqrt{1+kr^2})^5} \text{,} \quad p = 
\frac{kA^2}{2\pi(-A+\sqrt{1+kr^2})(A+\sqrt{1+kr^2})^5} \text{,} \label{eq_rhop_buch}\\
\mathrm{V}^2 =\frac{2A(-2A+3\sqrt{1+kr^2})}{15(A-\sqrt{1+kr^2})^2} \mbox{.} \label{eq_v_buch}
\end{gather}
The condition $\mathrm{V}<1$ is satisfied for  $A<\frac{(18-\sqrt{39})}{19}\sqrt{1+kr^2}$.

\begin{figure} 
\centering
\includegraphics[scale=0.6]{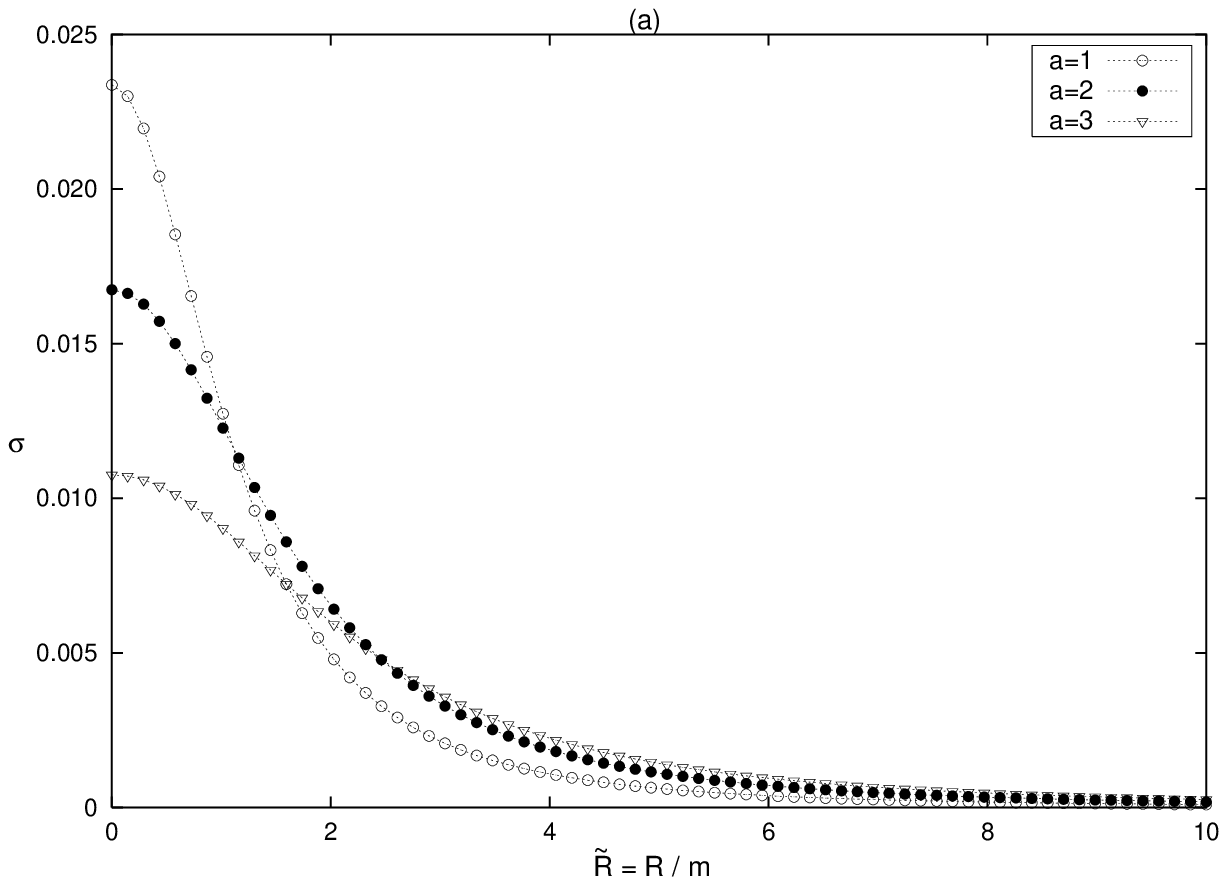}%
\includegraphics[scale=0.6]{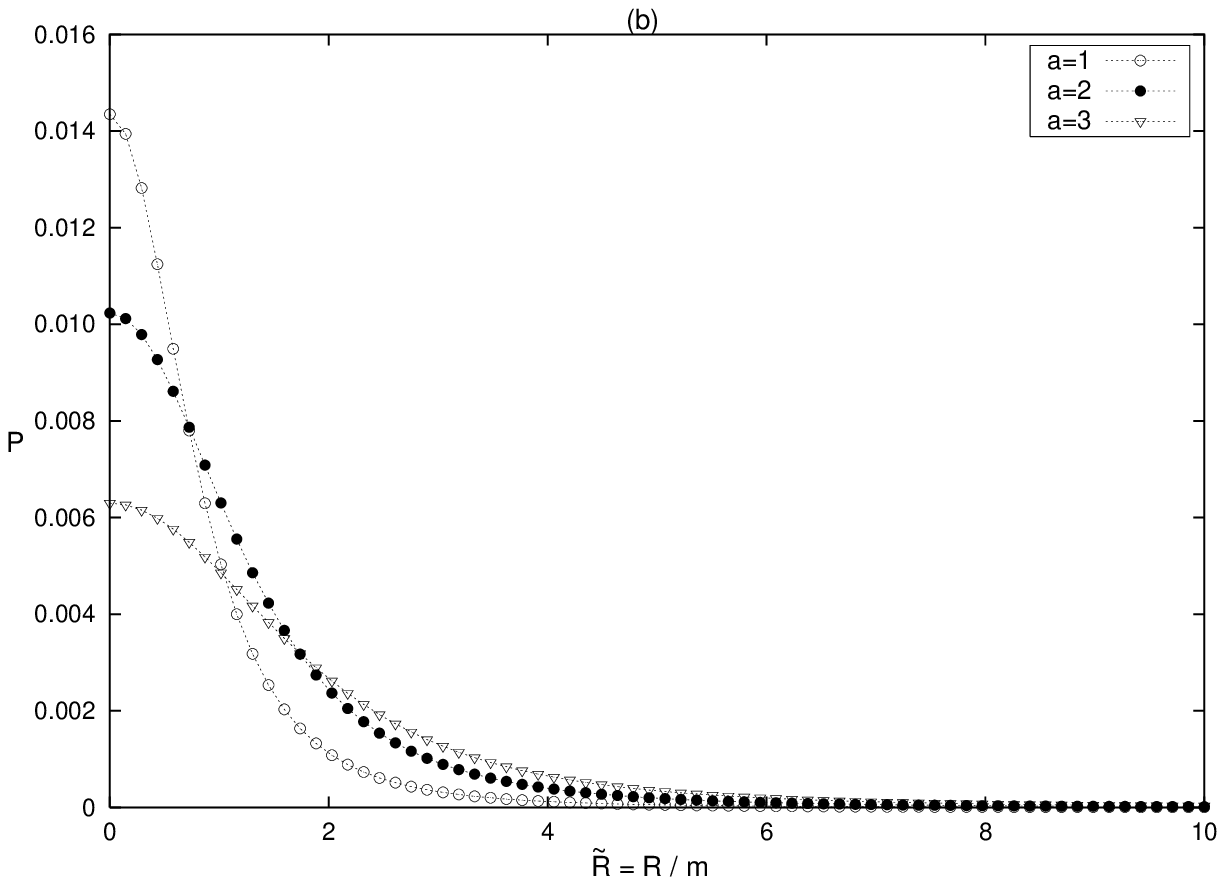} \\
\vspace{0.3cm}
\includegraphics[scale=0.6]{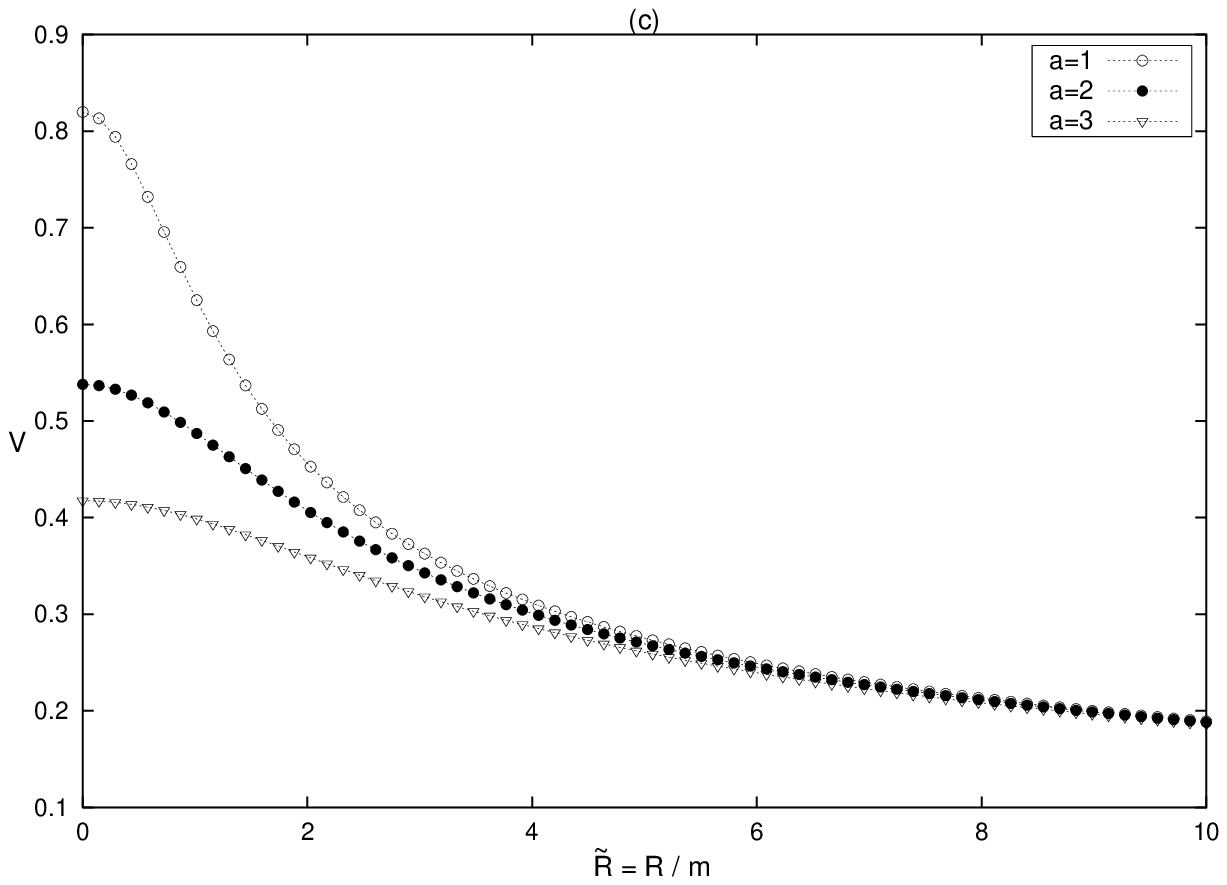}%
\includegraphics[scale=0.6]{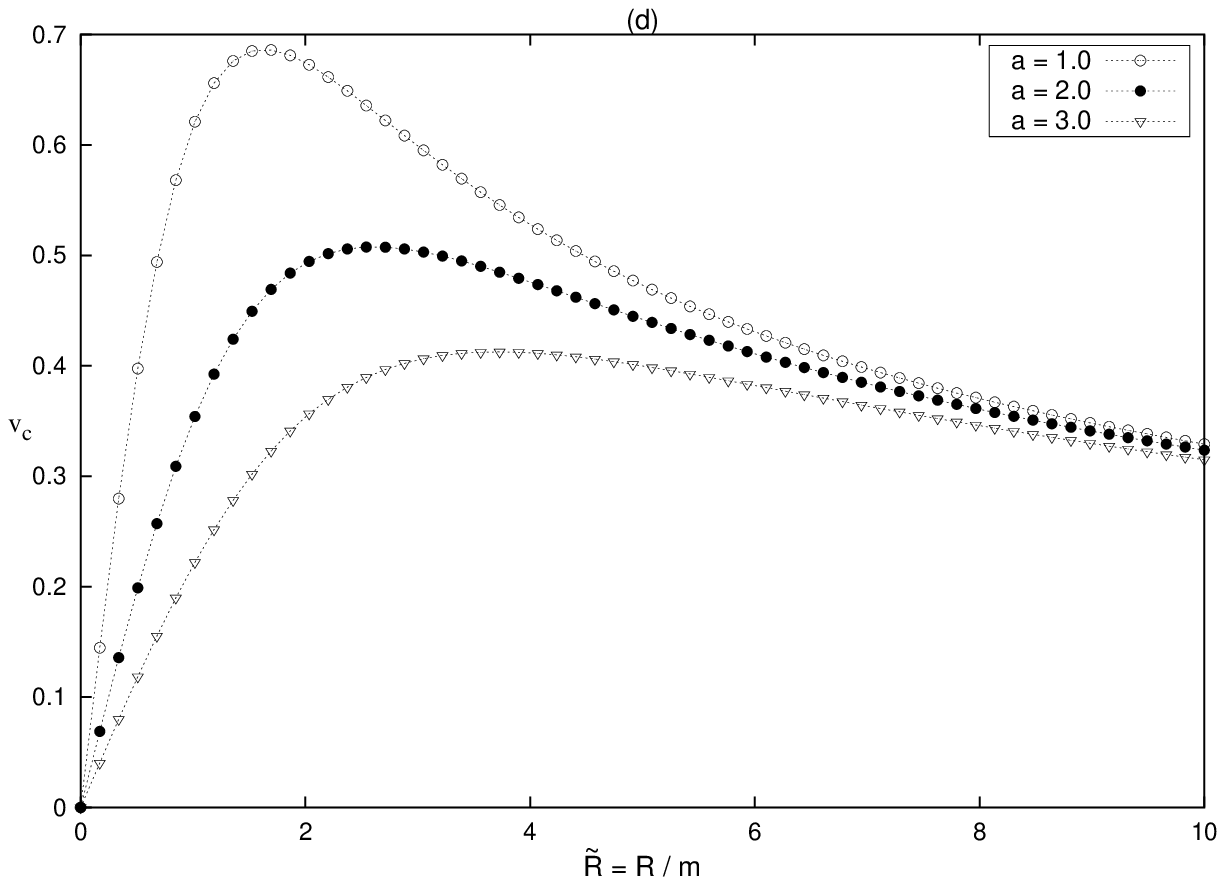}
\caption{(a) The surface energy density $\sigma$, Eq.\ (\ref{eq_en_buch}), (b) the pressure $P$ Eq.\ (\ref{eq_P_buch}), (c) the velocity of sound $V$ Eq.\ (\ref{eq_V_buch}), (d) the tangential velocity $\mathrm{v}_c$ Eq.\ (\ref{eq_v_fi_buch}) for the disk with 
$A=0.6\text{; }k=1\text{; for }a=1\text{, }2\text{ and }3$ as function of $\tilde{R}=R/m$.} \label{fig_3}
\end{figure}

 Using Eq.\ (\ref{eq_sol_buch}) and  Eqs.\ (\ref{eq_qtt})--(\ref{eq_disk_surf}), we get the following 
expressions for the energy density, pressure and sound velocity of the disk:
\begin{align}
\sigma &=\frac{akA}{\pi \left[A+\sqrt{1+k(R^2+a^2)} \right]^3} \mbox{,} \label{eq_en_buch}\\
P &=\frac{akA^2}{2 \pi \left[-A+\sqrt{1+k(R^2+a^2)} \right]
\left[A+\sqrt{1+k(R^2+a^2)} \right]^3} \mbox{,} \label{eq_P_buch}\\
V^2 &=\frac{A\left[-A+2\sqrt{1+k(R^2+a^2)} \right]}{3\left[A-\sqrt{1+k(R^2+a^2)} \right]^2}
\mbox{.} \label{eq_V_buch}
\end{align}
The conditions $V<1$ and $P>0$ are both satisfied if $A<\frac{1}{2}\sqrt{1+ka^2}$. Figure
\ref{fig_3} (a)--(d) shows, respectively, $\sigma$, 
$P$, $V$ and rotation curves,  Eq.\ (\ref{eq_v_fi_buch}),
as functions of $\tilde{R}=R/m$ for the disk calculated from Buchdahl's
solution. 

\begin{figure} 
\centering
\includegraphics[scale=0.6]{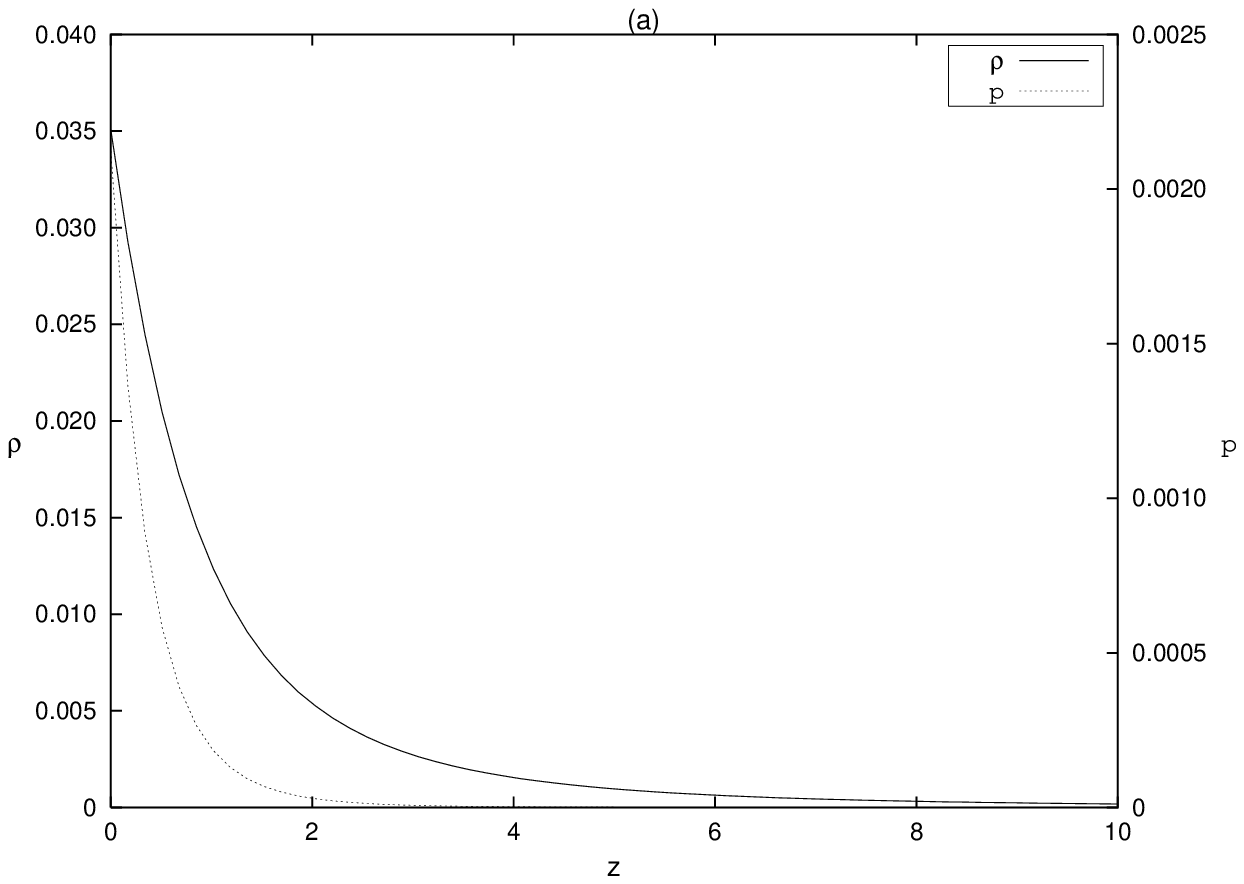}%
\hspace{0.2cm}
\includegraphics[scale=0.6]{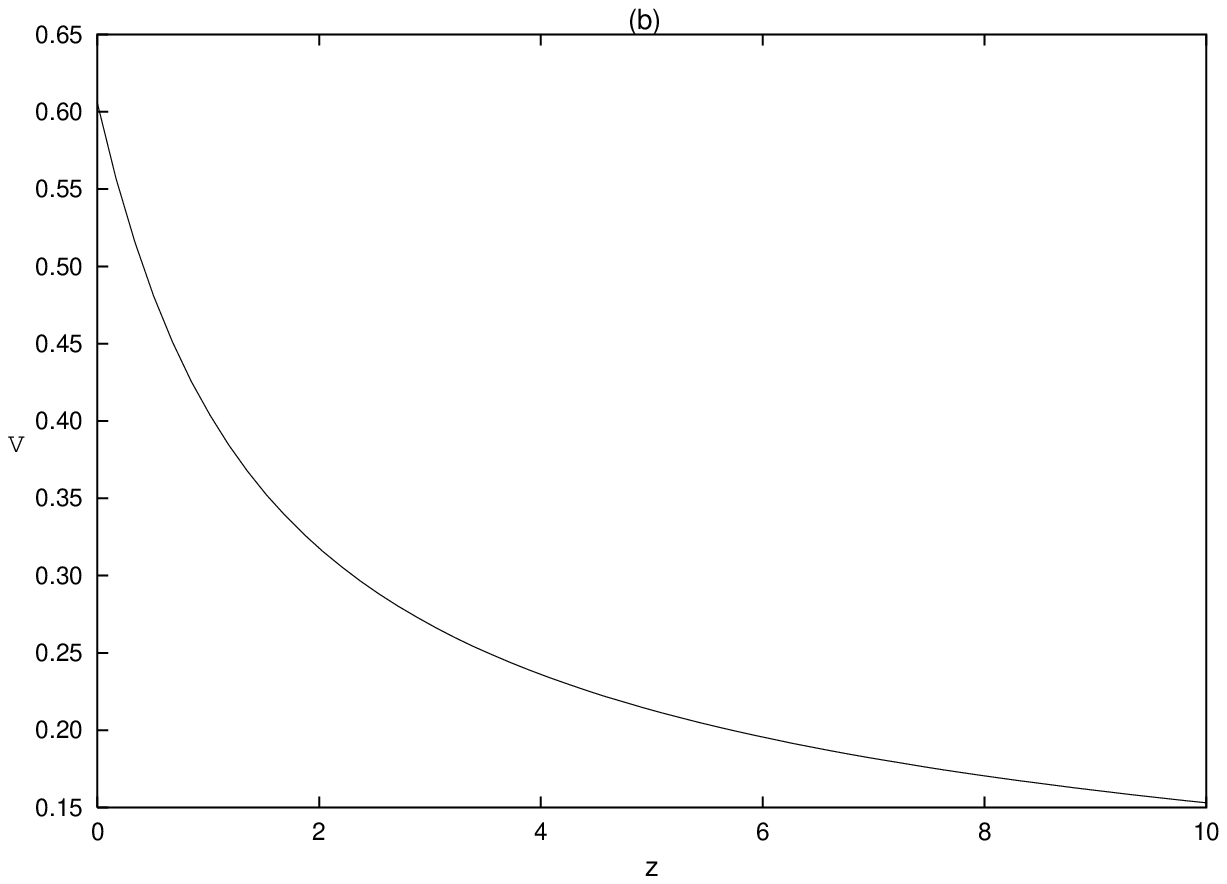}
\caption{(a) The density $\rho$ and pressure $p$ Eq.\ (\ref{eq_rhop_buch}) and (b) the velocity of sound V Eq.\ (\ref{eq_v_buch}) for the halo with 
$A=0.6\text{; }k=1\text{ and }a=1$ along the axis $z$.} \label{fig_4}
\end{figure}

In figure \ref{fig_4} (a)--(b) we show, 
respectively, the density $\rho$ together with pressure $p$, and sound velocity $\mathrm{V}$
of the halo along the axis $z$ for $A=0.6\text{; }k=1\text{ and }a=1$. Note that in this solution
there is no boundary of the fluid sphere: the disk is completely immersed in the fluid.

The tangential velocity $\mathrm{v}_c$ calculated from metric coefficients (\ref{eq_sol_buch}) is
\begin{equation} \label{eq_v_fi_buch}
\mathrm{v}_c^2=\frac{2AkR^2}{ \left( 1- A/{\sqrt{1+k(R^2+a^2)} }\right)\left\{ [1+k(R^2+a^2)]^{3/2}+A[1+k(a^2-R^2)]\right\}} \mbox{.}
\end{equation}
For $R>>a$, Eq.\ (\ref{eq_v_fi_buch}) goes as $\mathrm{v}_c=(2A)^{1/2}/(R^{1/2}k^{1/4})$.
The specific angular momentum follows from Eq.\ (\ref{eq_h}) and Eq.\ (\ref{eq_sol_buch}):
\begin{equation} \label{eq_h_buch}
h=\frac{\sqrt{2Ak}R^2\left( 1+\frac{A}{\sqrt{1+k(R^2+a^2)}}\right)^2[1+k(R^2+a^2)]^{1/4}}
{\sqrt{[1+k(R^2+a^2)]^2-4AkR^2\sqrt{1+k(R^2+a^2)}-A^2[1+k(a^2-R^2)]}} \mbox{.}
\end{equation} 

\subsection{Narlikar-Patwardhan-Vaidya  Solutions 1a and 1b}

Now we shall study the generation of a disk  solution with an halo exactly as the  one depicted in Fig.\ \ref{fig_schem2}. We start with a solution of the Einstein equations in isotropic coordinates 
that represents a  sphere with radius $r_b$ of perfect fluid that on $r=r_b$  will be continuously matched  to the vacuum Schwarzschild  solution.
Narlikar, Patwardhan and Vaidya  \cite{Narlikar} gave the following two exact 
solutions of the Einstein equations  for a static sphere of
perfect fluid characterized by the metric functions $(\lambda,  \nu_{1a})$ and
$(\lambda,  \nu_{1b})$,

\begin{eqnarray} 
& e^{\lambda} =Cr^k \mbox{,} & \label{eq_nar_sol1a}\\
& e^{\nu_{1a}} =\left( A_{1a}r^{1-n+k/2}+B_{1a}r^{1+n+k/2} \right)^2 &\text{ for }  -2+\sqrt{2} < k \leq 0 \text{,} \label{eq_nar_sol1b}\\
&e^{\nu_{1b}} =r^{\sqrt{2}}\left[ A_{1b}+B_{1b} \ln (r) \right]^2 &\text{ for } k=-2+\sqrt{2} \text{,} \label{eq_nar_sol1c}
\end{eqnarray}
where $A_{1a}\text{, }A_{1b}\text{, }B_{1a}\text{, }B_{1b}\text{, }C$ are constants, and $n=\sqrt{1+2k+k^2/2}$.  We shall refer to these solutions as 
NPV 1a and NPV 1b, respectively.

The density, pressure and sound velocity  for for the solutions
  $(\lambda, \nu_{1a})$ and
$(\lambda,  \nu_{1b})$,   will be denoted by $(\rho, p_{1a},V_{1a})$ and
 $(\rho, p_{1b},V_{1b})$, respectively. We find,
\begin{align}
\rho &=\frac{-k(k+4)r^{-2-k}}{32\pi C} \mbox{,} \label{eq_rho_nar1}\\
p_{1a} &=\frac{1}{32 \pi C r^{2+k}(A_{1a}+B_{1a}r^{2n})} \left\{ A_{1a}\left[ 3k^2+8(1-n)-4k(n-3) \right] \right. \notag \\
 &\left.+B_{1a} \left[3k^2+8(1+n)+4k(n+3) \right] r^{2n} \right\} \mbox{,} \label{eq_p1_nar1}\\
p_{1b} &= \frac{A_{1b}+B_{1b} \ln (r)+2\sqrt{2}B_{1b}}{16 \pi C [A_{1b}+B_{1b} \ln (r)]}\mbox{,} \label{eq_p2_nar1} \\
\mathrm{V}_{1a}^2 &=\frac{1}{k(k+4)(A_{1a}+B_{1a}r^{2n})^2} \left\{ A_{1a}^2 \left[ -3k^2+8(n-1)+4k(n-3) \right] \right. \notag \\
 &\left. -B_{1a}^2r^{4n} \left[ 3k^2+8(1+n)+4k(n+3) \right] -2A_{1a}B_{1a}r^{2n}\left[ 3k(k+4)+8(1-n^2) \right] \right\} \mbox{,}\label{eq_v1_nar1}\\
\mathrm{V}_{1b}^2 &=\frac{2B_{1b}^2r^{\sqrt{2}}}{[A_{1b}+B_{1b} \ln (r)]^2} \mbox{.} \label{eq_v2_nar1}
\end{align}

The condition of  continuity of the metric functions  $(\lambda, \nu)$ 
 given by   (\ref{eq_nar_sol1a})--(\ref{eq_nar_sol1c})  and 
the corresponding functions in Eq.\ (\ref{eq_metrica_sch})
 at the boundary $r=r_b$    leads  to following expressions:
\begin{gather}
\frac{m}{2r_b} =-\frac{k}{k+4}\text{,} \quad C =r_b^{-k}\left( \frac{4}{k+4}\right)^4 \mbox{,} \label{eq_cond_sol1a}\\
A_{1a} =-\frac{3k^2+8(1+n)+4k(n+3)}{16nr_b^{1-n+k/2}}\text{,} \quad B_{1a} =\frac{3k^2+8(1-n)-4k(n-3)}{16nr_b^{1+n+k/2}}\mbox{,} \label{eq_cond_sol1b}\\
A_{1b} =-\frac{2\sqrt{2}+\ln (r_b)}{4r_b^{1/ \sqrt{2}}}\text{,} \quad B_{1b} =\frac{1}{4r_b^{1/ \sqrt{2}}}
\mbox{.} \label{eq_cond_sol1c}
\end{gather}
$\mathrm{V}_{1a}$ has its maximum at $r=0$, and $\mathrm{V}_{1b}$ at $r=r_b$. Condition 
$\mathrm{V}_{1b}(r_b)<1$ is satisfied if $r_b<4^{1/\sqrt{2}}$.

\begin{figure} 
\centering
\includegraphics[scale=0.6]{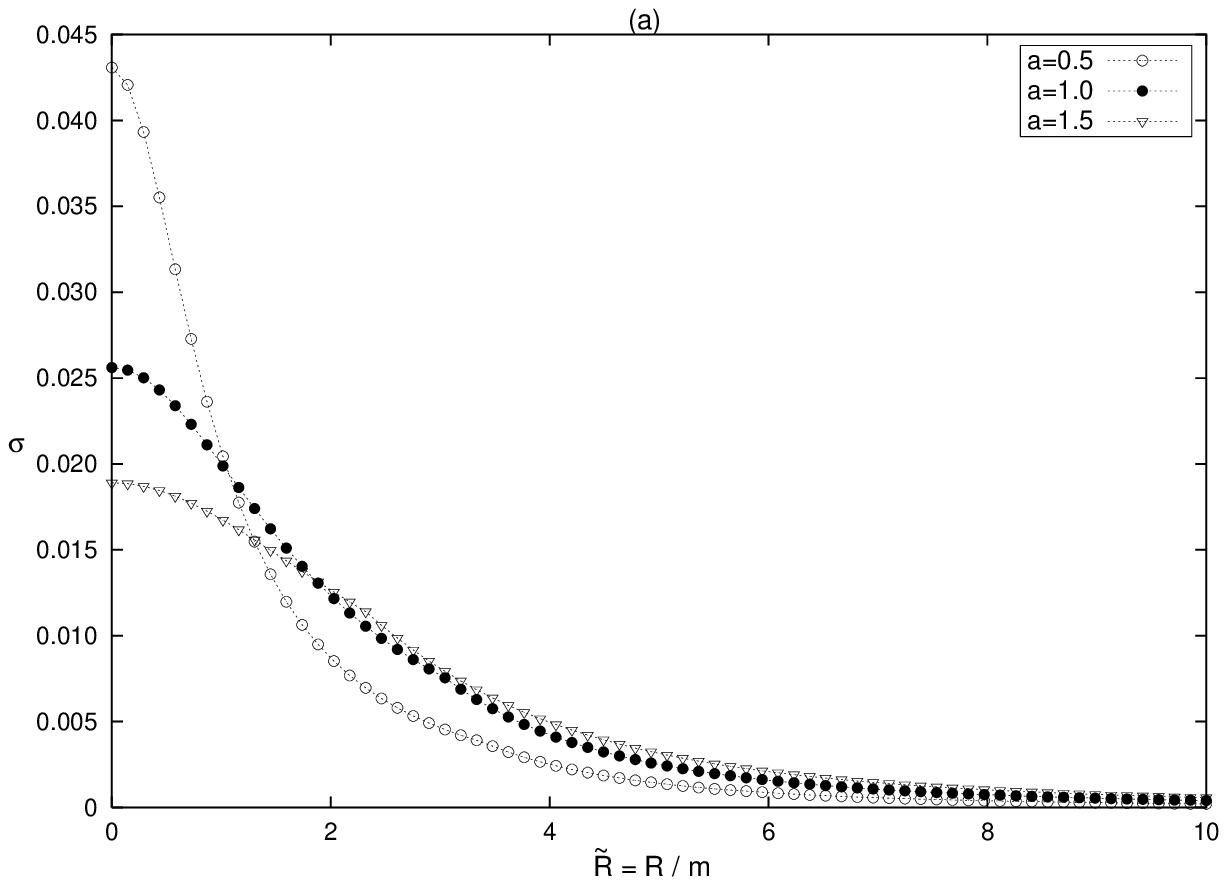}%
\includegraphics[scale=0.6]{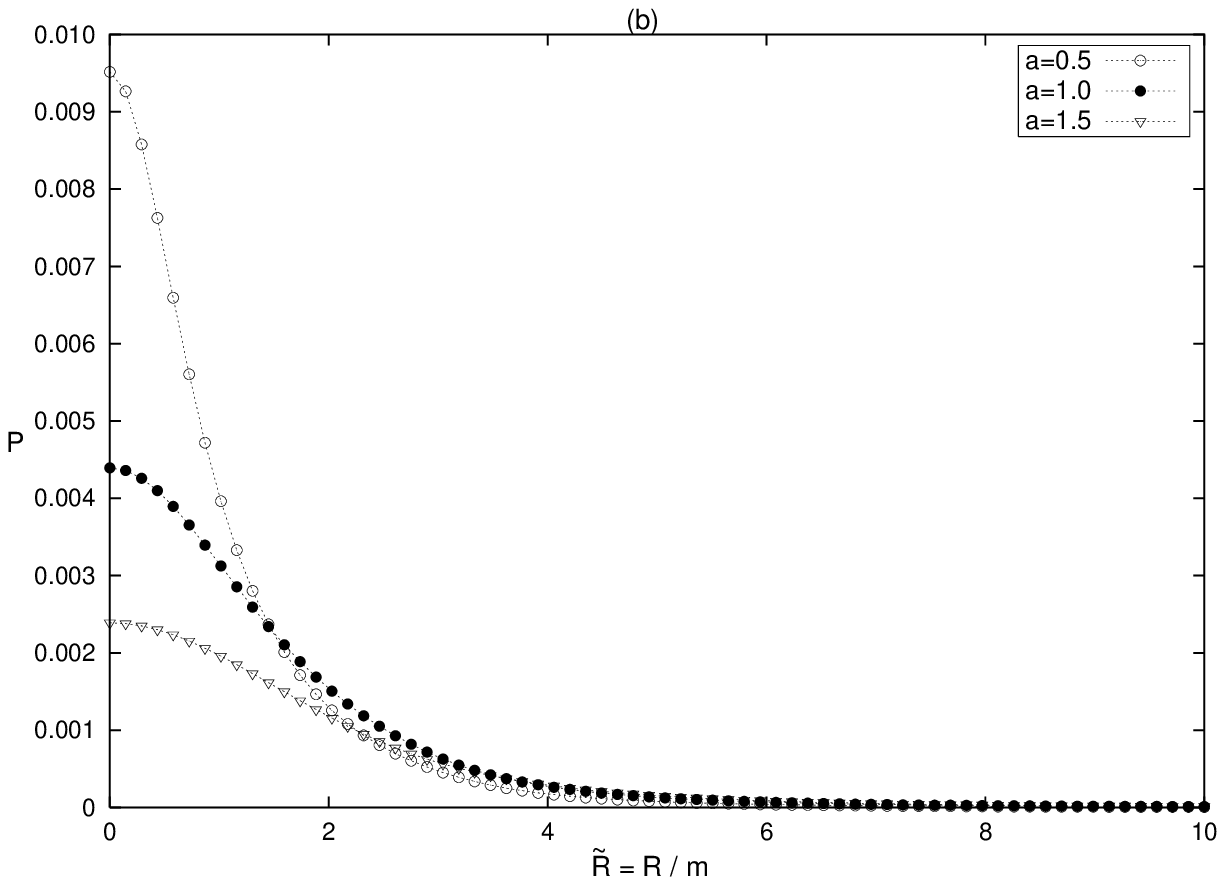} \\
\vspace{0.3cm}
\includegraphics[scale=0.6]{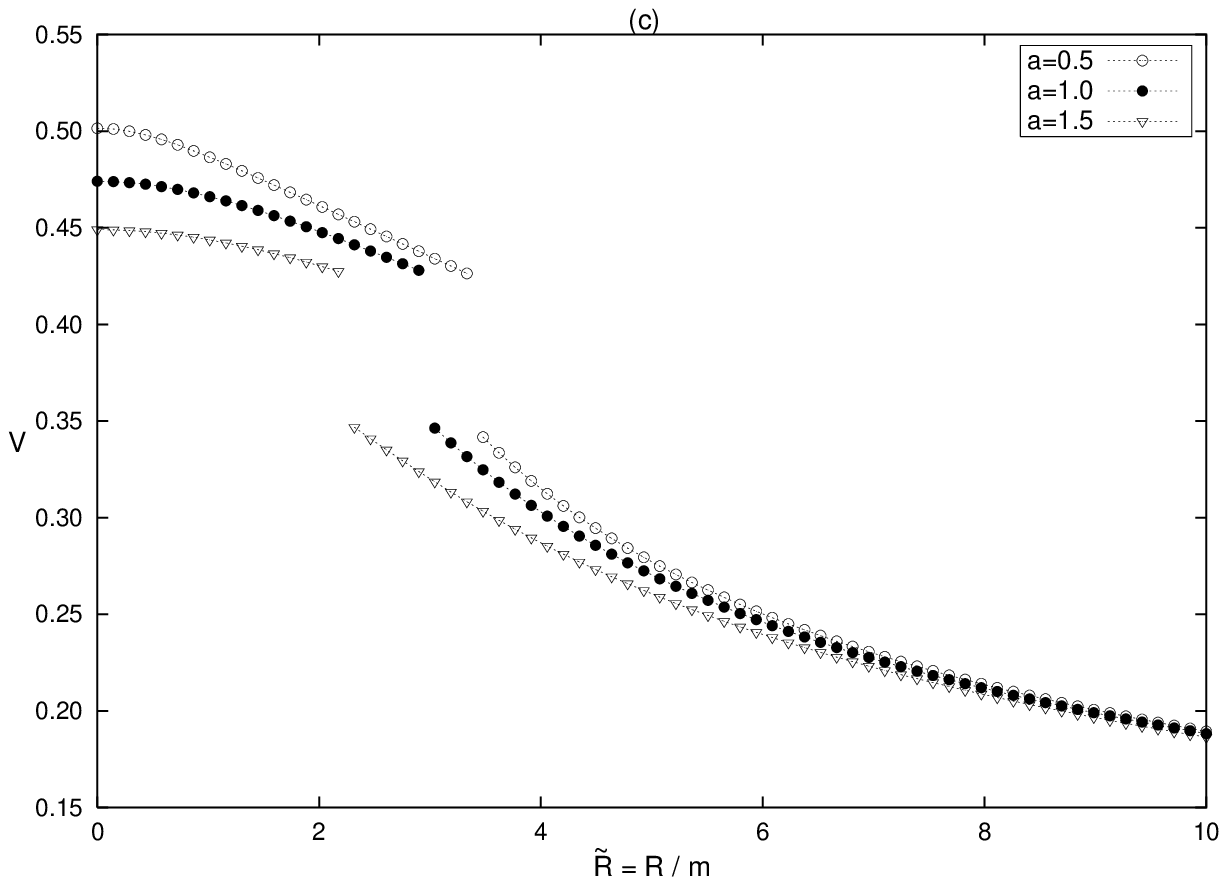}
\caption{(a) The surface energy density $\sigma$ Eq.\ (\ref{eq_en_nar1}), (b) the pressure $P$ Eq.\ (\ref{eq_P1_nar1}) and (c) the velocity of sound $V$ Eq.\ (\ref{eq_V1_nar1}) for the disk with 
$\mathbf{k=-1/2}\text{; }r_b=2\text{; for }a=0.5\text{, }1.0\text{ and }1.5$ as function 
of $\tilde{R}=R/m$.} \label{fig_5}
\end{figure}

\begin{figure} 
\centering
\includegraphics[scale=0.6]{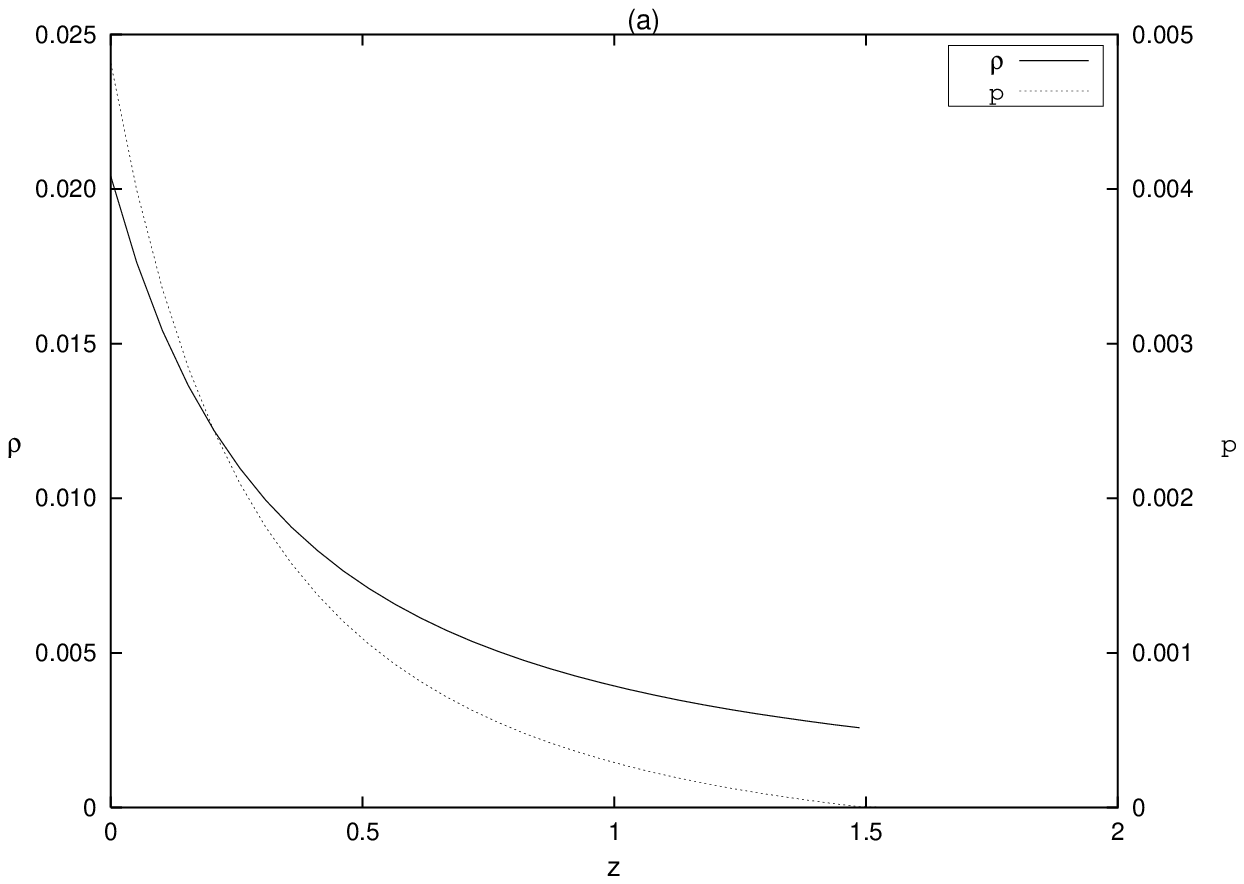}%
\hspace{0.2cm}
\includegraphics[scale=0.6]{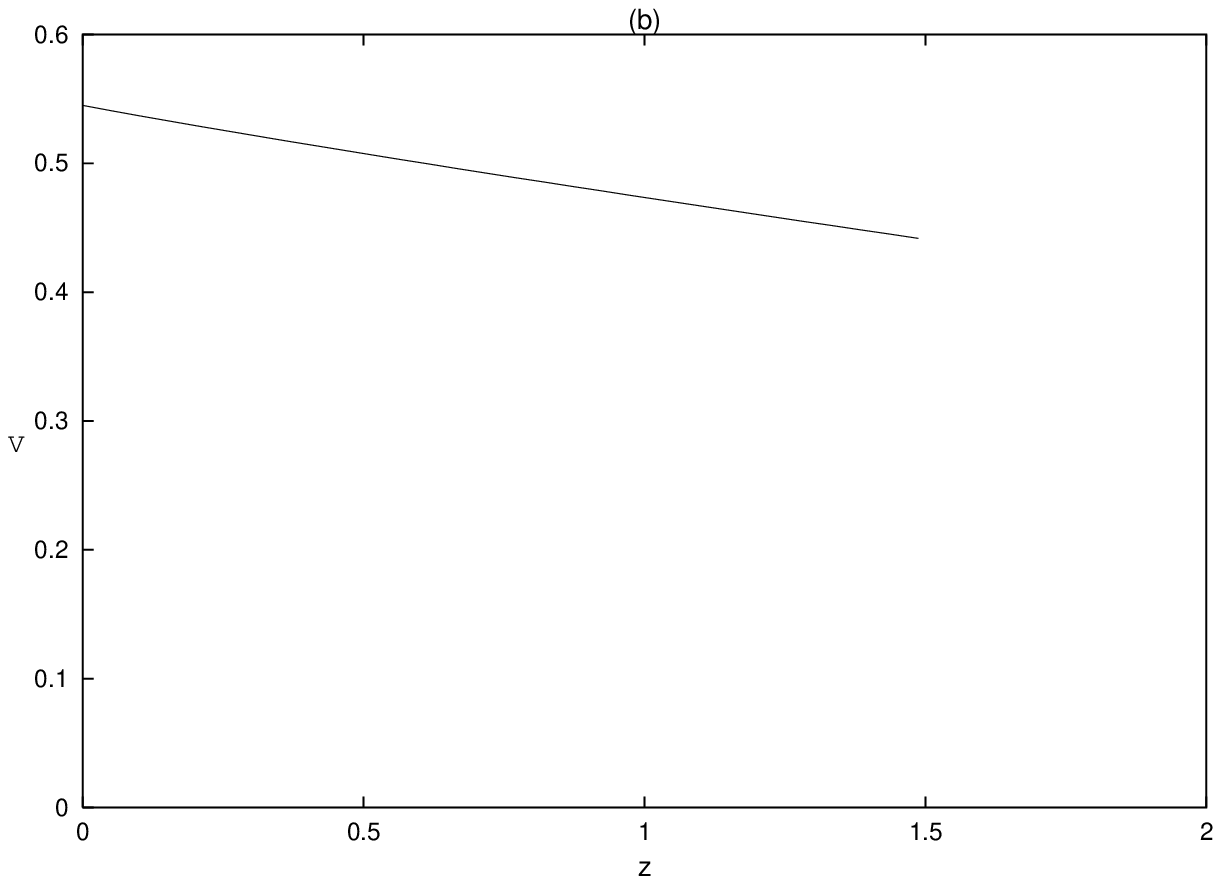}
\caption{(a) The density $\rho$ Eq.\ (\ref{eq_rho_nar1}) and pressure $p$ Eq.\ (\ref{eq_p1_nar1}), (b) the velocity of sound V Eq.\ (\ref{eq_v1_nar1}) 
for the halo with $\mathbf{k=-1/2}\text{; }r_b=2\text{; for }a=0.5$ along the axis $z$.} \label{fig_6}
\end{figure}

\begin{figure} 
\centering
\includegraphics[scale=0.6]{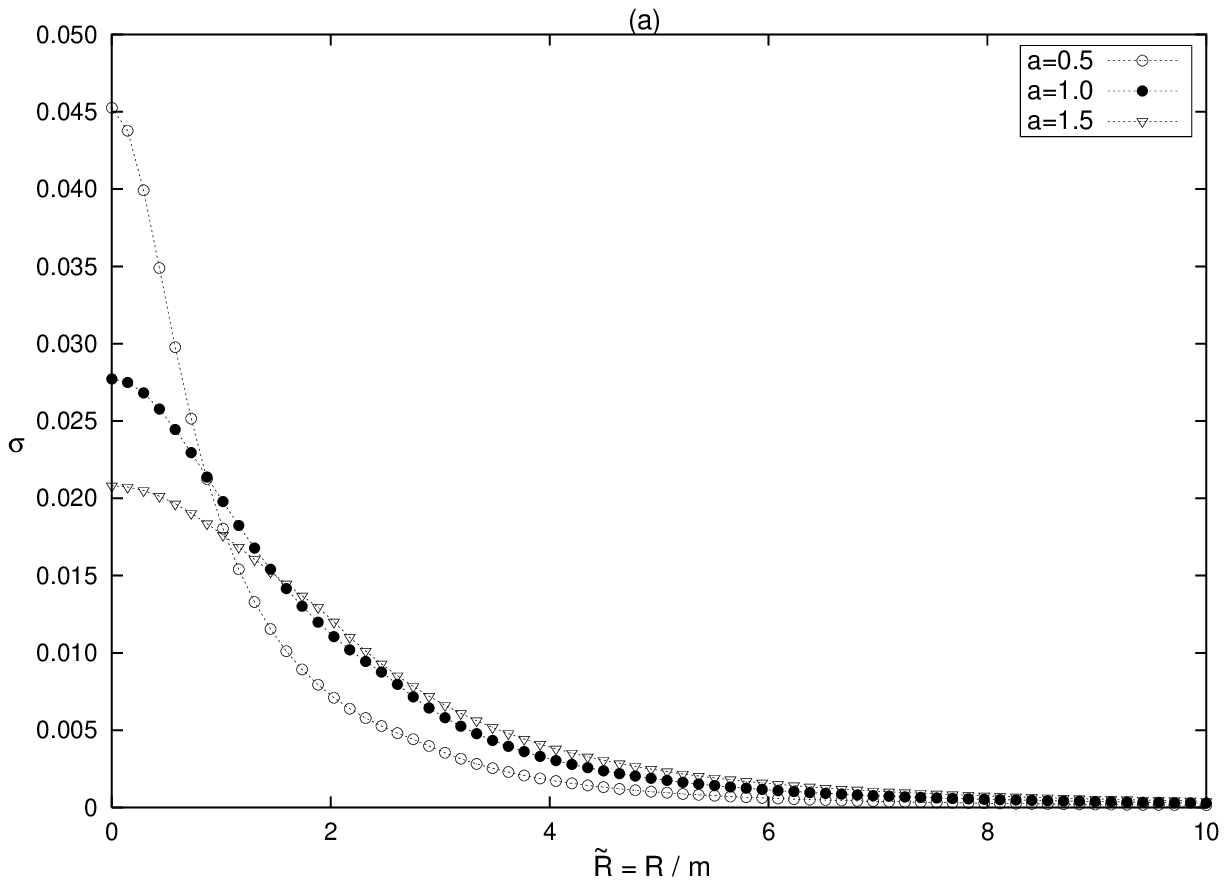}%
\includegraphics[scale=0.6]{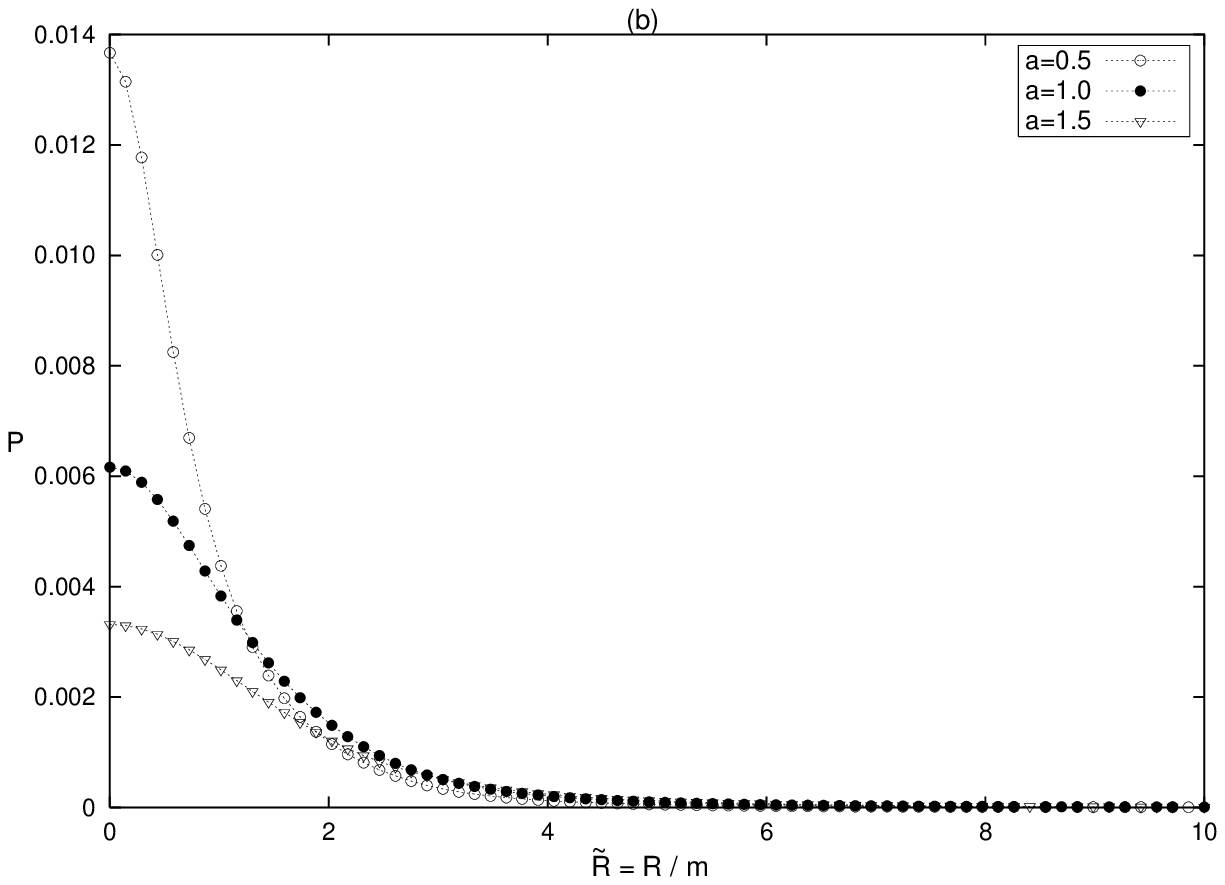} \\
\vspace{0.3cm}
\includegraphics[scale=0.6]{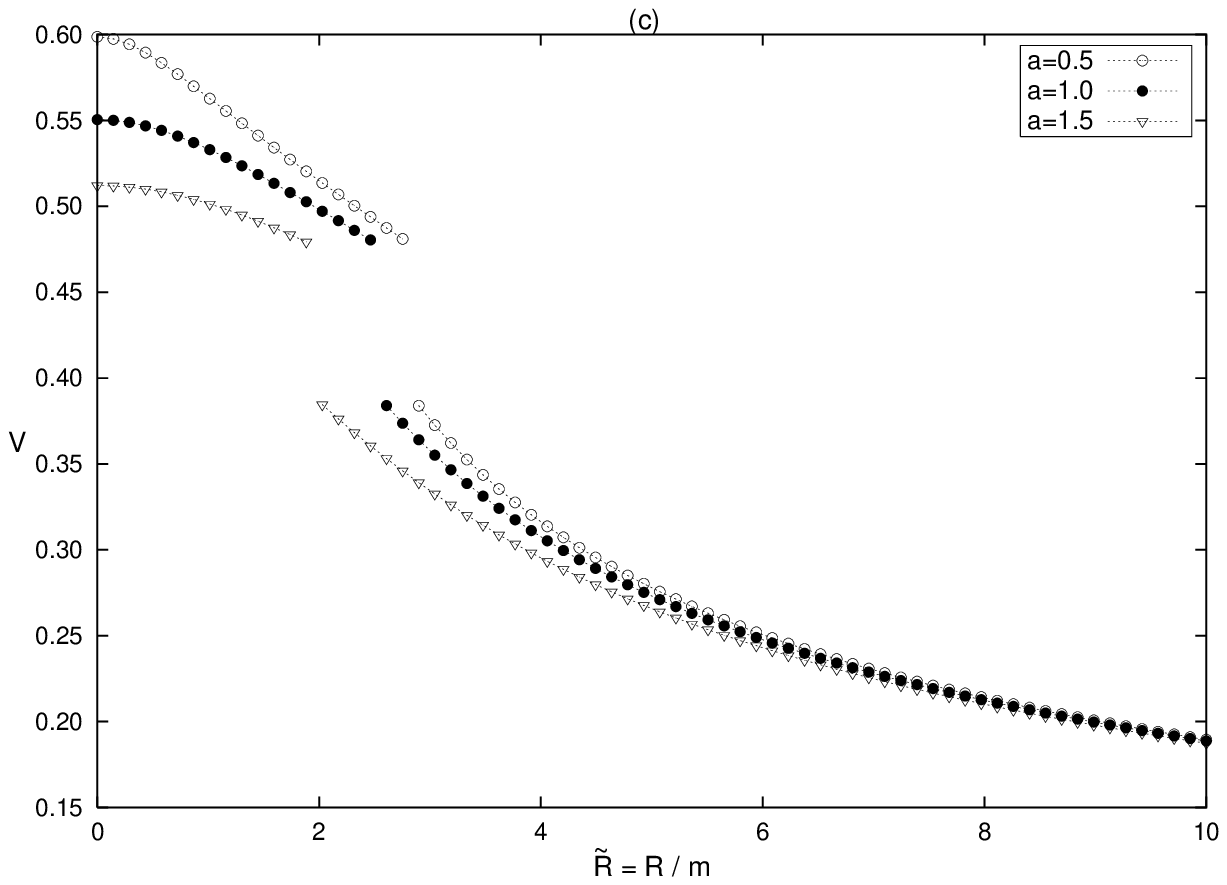}
\caption{(a) The surface energy density $\sigma$ Eq.\ (\ref{eq_en_nar1}), (b) the pressure $P$ Eq.\ (\ref{eq_P2_nar1}), (c) the velocity of sound $V$ Eq.\ (\ref{eq_V2_nar1}) for the disk 
with $k=-2+\sqrt{2}\text{; }r_b=2\text{; for }a=0.5\text{, }1.0\text{ and }1.5$ as 
function of $\tilde{R}=R/m$.} \label{fig_7}
\end{figure}

\begin{figure} 
\centering
\includegraphics[scale=0.6]{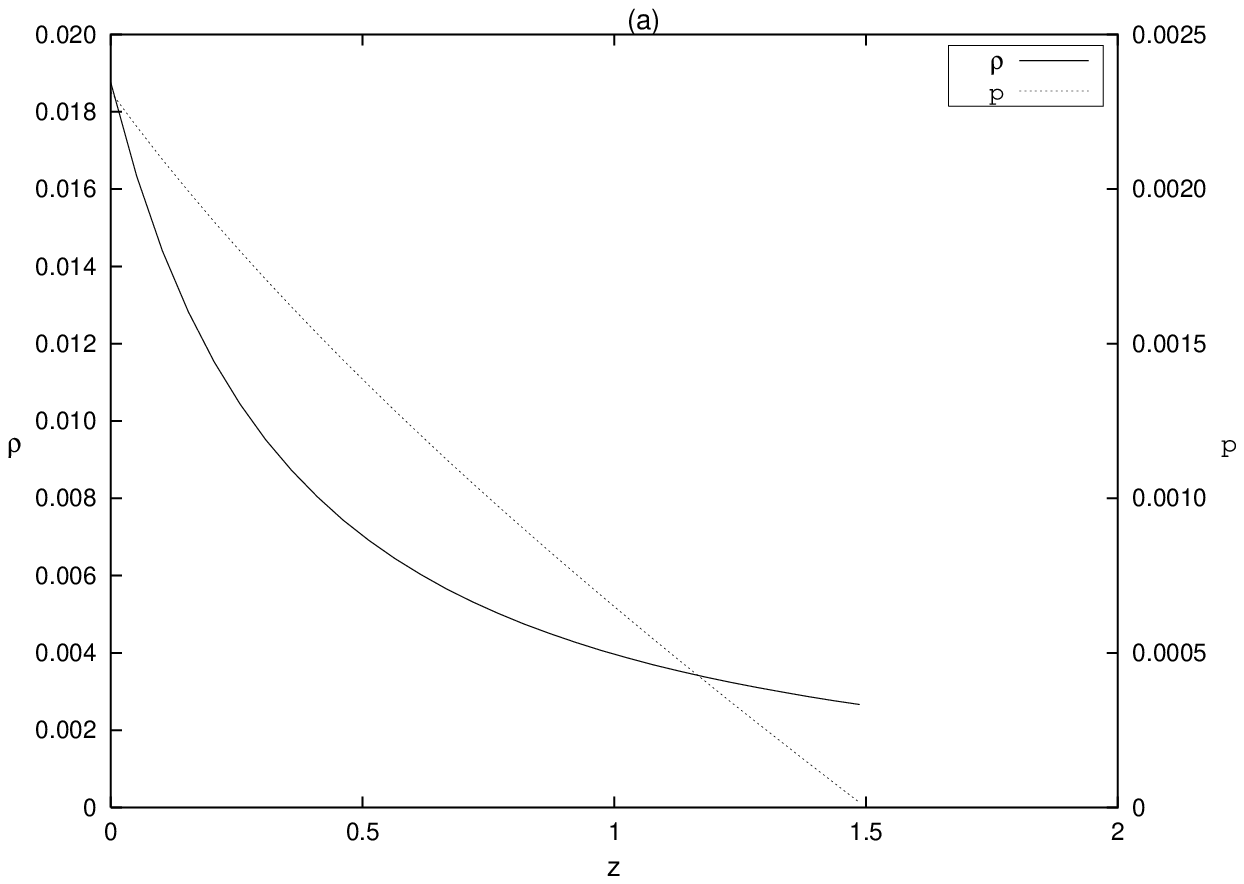}%
\hspace{0.2cm}
\includegraphics[scale=0.6]{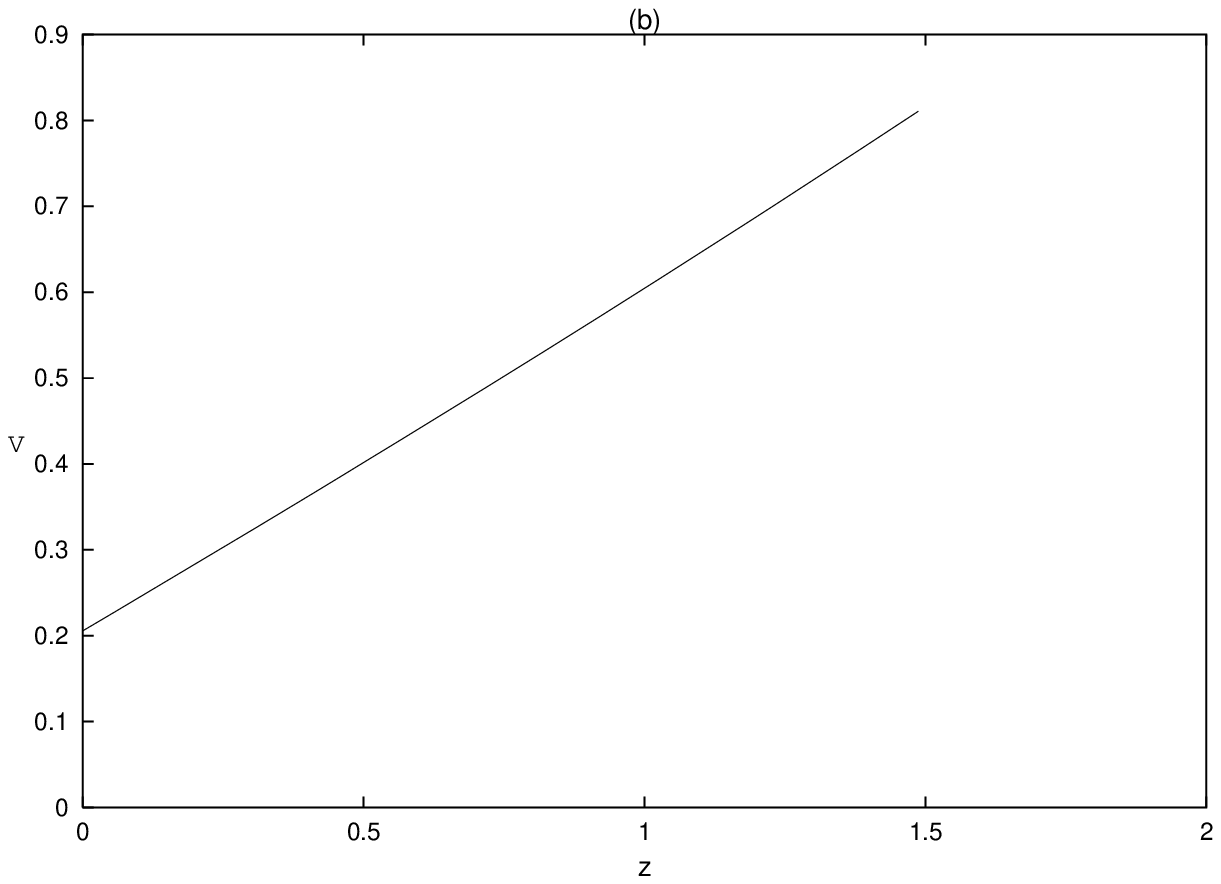}
\caption{(a) The density $\rho$ Eq.\ (\ref{eq_rho_nar1}) and pressure $p$ Eq.\ (\ref{eq_p2_nar1}), (b) the velocity of sound V Eq.\ (\ref{eq_v2_nar1})  for 
the halo with $k=-2+\sqrt{2}\text{; }r_b=2\text{; for }a=0.5$ along the axis $z$.} \label{fig_8}
\end{figure}

Using Eq.\ (\ref{eq_nar_sol1a})--(\ref{eq_nar_sol1c}) in Eq.\ (\ref{eq_qtt})--(\ref{eq_disk_surf}), we get the following expressions for 
the energy density, pressure and sound velocity of the disk:
\begin{align}
\sigma &=-\frac{ka}{4\pi \sqrt{C}\mathcal{R}^{1+k/4}} \text{,}\label{eq_en_nar1}\\
P_{1a} &=\frac{a}{4 \pi \sqrt{C}\mathcal{R}^{1+k/4}} \frac{\left[ A_{1a}(k-n+1) +
B_{1a}(k+n+1)\mathcal{R}^{n}\right]}{\left[ A_{1a}+B_{1a}\mathcal{R}^{n}\right]} \mbox{,} \label{eq_P1_nar1}\\
P_{1b} &=\frac{a}{4 \pi \sqrt{C}\mathcal{R}^{1/2+\sqrt{2}/4}[2A_{1b}+B_{1b} \ln (\mathcal{R})]}
\left[ 2A_{1b}(\sqrt{2}-1)+2B_{1b} \right. \notag \\
& \left. +B_{1b}(\sqrt{2}-1)\ln (\mathcal{R})\right] \mbox{,} \label{eq_P2_nar1}\\
V_{1a}^2 &=\frac{1}{k(k+4)\left[ A_{1a}\mathcal{R}^{-n/2}+B_{1a}\mathcal{R}^{n/2}\right]^2} \left[
-B_{1a}^2\mathcal{R}^n(k^2+5k+nk+4n+4) \right. \notag \\
& \left. +A_{1a}^2\mathcal{R}^{-n} (-k^2-5k+nk+4n-4)
+2A_{1a}B_{1a}(-k^2-5k+4n^2-4) \right] \mbox{,}\label{eq_V1_nar1}\\
V_{1b}^2 &=\frac{1}{2(2A_{1b}+B_{1b} \ln (\mathcal{R}))^2} \left[ B_{1b}^2 \sqrt{2} \ln^2 (\mathcal{R}) +2B_{1b}
 (2B_{1b}+B_{1b}\sqrt{2}+2A_{1b}\sqrt{2})\right. \notag \\
& \left. \times \ln (\mathcal{R}) +4A_{1b}B_{1b}(2+\sqrt{2})+4\sqrt{2}A_{1b}^2+8B_{1b}^2 \right] \mbox{,} \label{eq_V2_nar1}
\end{align}
where $\mathcal{R}=R^2+a^2$. $V_{1a}\text{ and }V_{1b}$ have their maximum values at $R=0$. Because the expressions are rather involved, the
restrictions on the constants to ensure the velocities are positive and less then one is best done 
graphically. The curves of $\sigma$, $P$ and $V$ as function of $\tilde{R}=R/m$ with parameters 
$k=-1/2\text{; }r_b=2\text{ and }a=0.5\text{, }1.0\text{,}1.5$ are displayed in figure 
\ref{fig_5} (a)--(c), respectively. Figure \ref{fig_6} (a)--(b)
show the density $\rho$, pressure $p$ and velocity of sound V for the halo with 
parameters $k=-1/2\text{; }r_b=2\text{, for }a=0.5$ along the axis $z$. The same physical quantities are 
shown in figures \ref{fig_7} (a)--(c) and \ref{fig_8} (a)--(b) with $k=-2+\sqrt{2}$. We note that $\sigma$ and $P$
are continuous at the boundary between the internal and external parts of the disk, but the velocity of
sound has a discontinuity. 

\begin{figure} 
\centering
\includegraphics[scale=0.6]{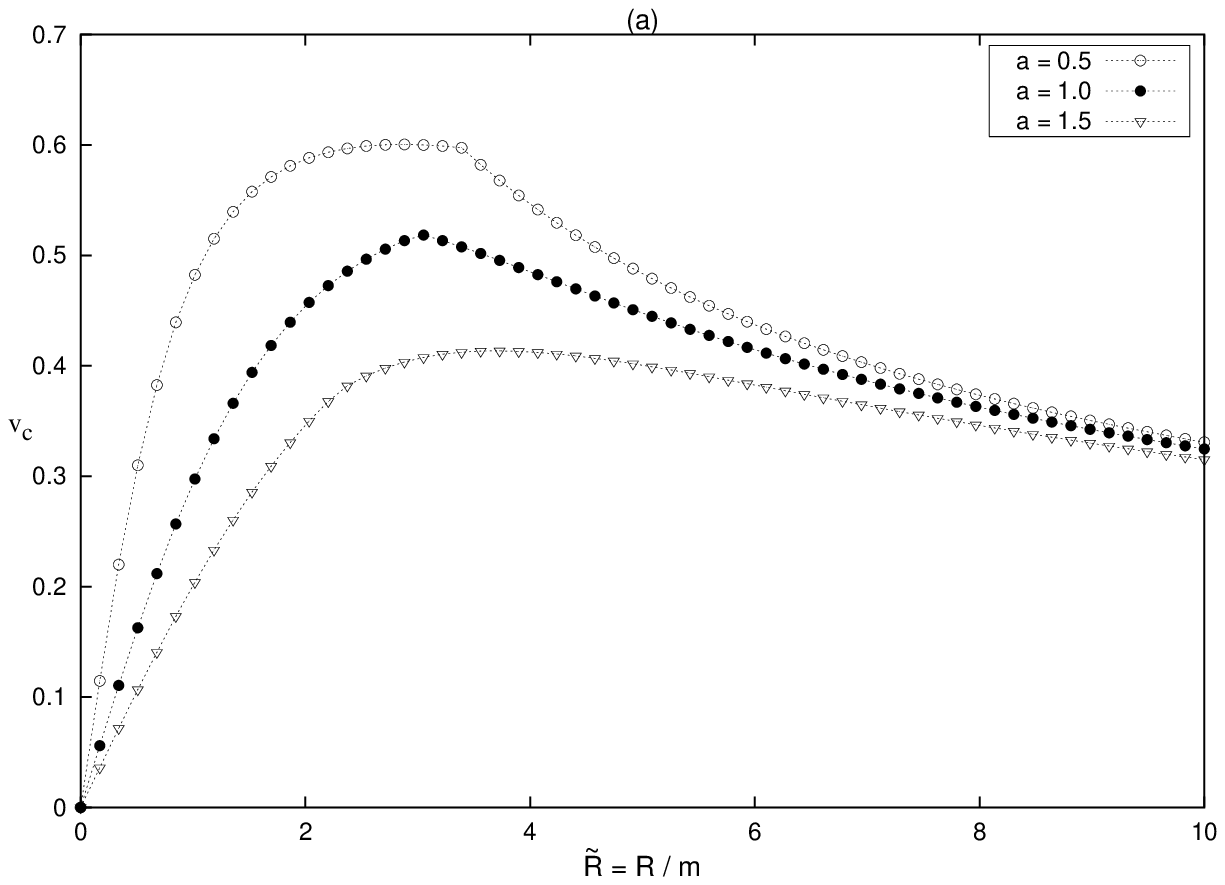}%
\includegraphics[scale=0.6]{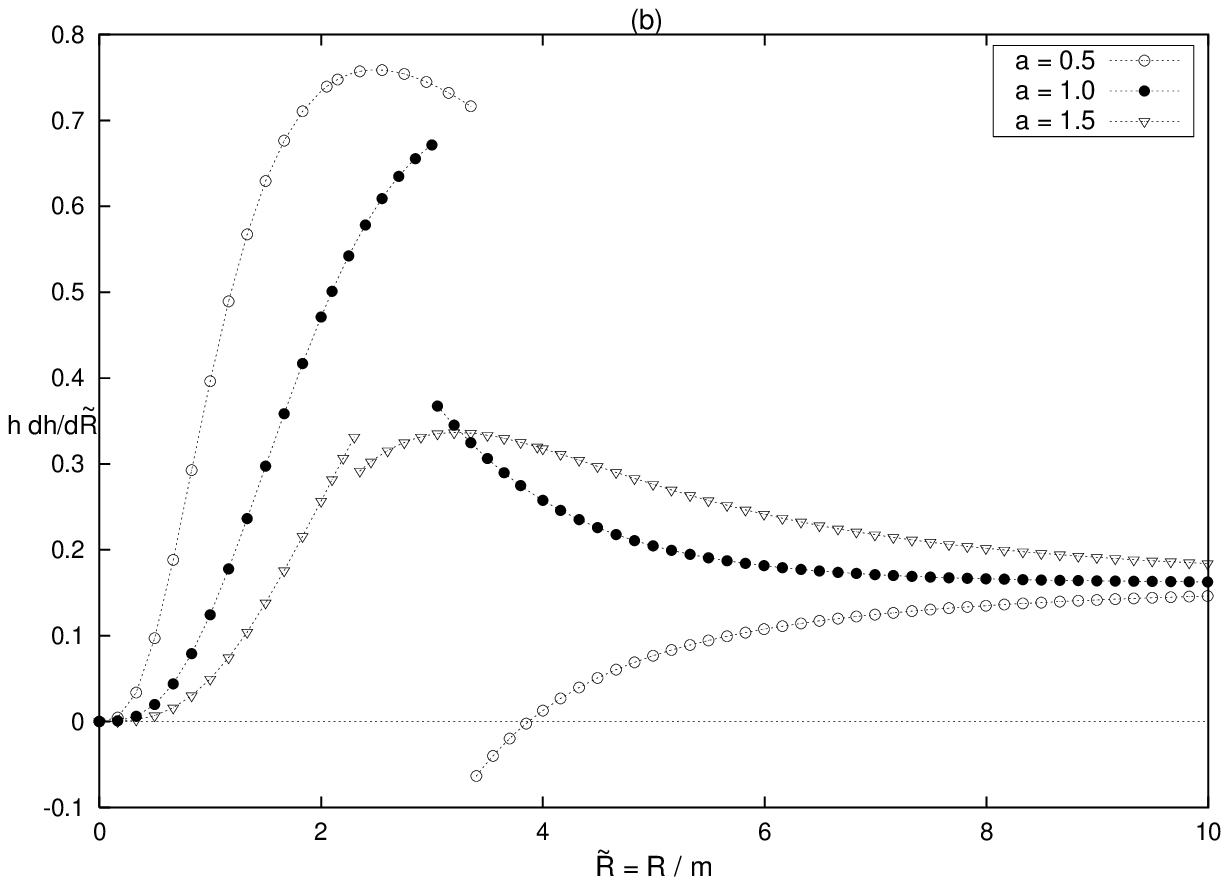}
\caption{(a) The tangential velocity $\mathrm{v}_{c1}$ Eq.\ (\ref{eq_v_fi_nar1}) and (b) the curves of $h\frac{dh}{d\tilde{R}}$ Eq.\ (\ref{eq_h_nar1}) with $\mathbf{k=-1/2}\text{; }r_b=2\text{; for }a=0.5 \text{, }1.0\text{, and }1.5$
as function of $\tilde{R}=R/m$. A region of 
instability appears on the disk generated with parameter $a=0.5$.} \label{fig_9}
\end{figure}

\begin{figure} 
\centering
\includegraphics[scale=0.6]{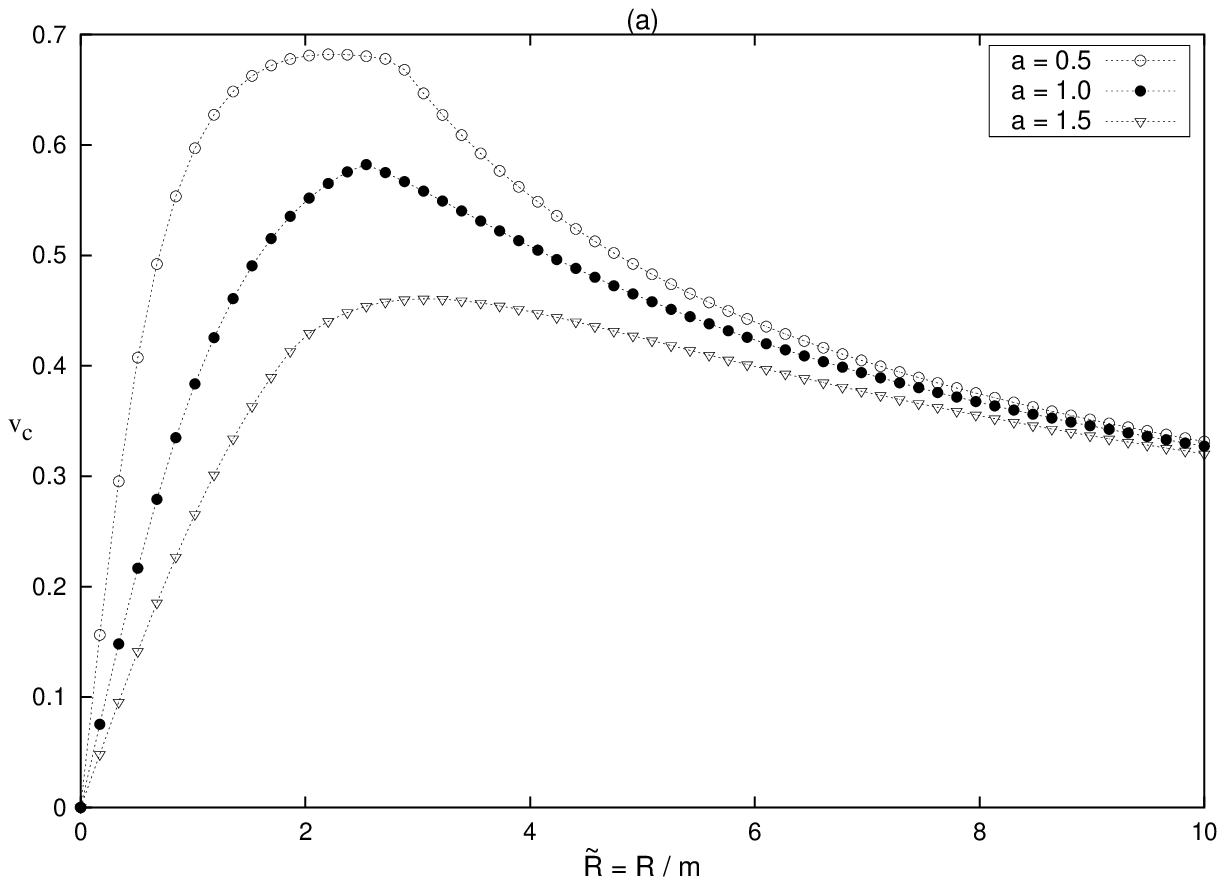}
\includegraphics[scale=0.6]{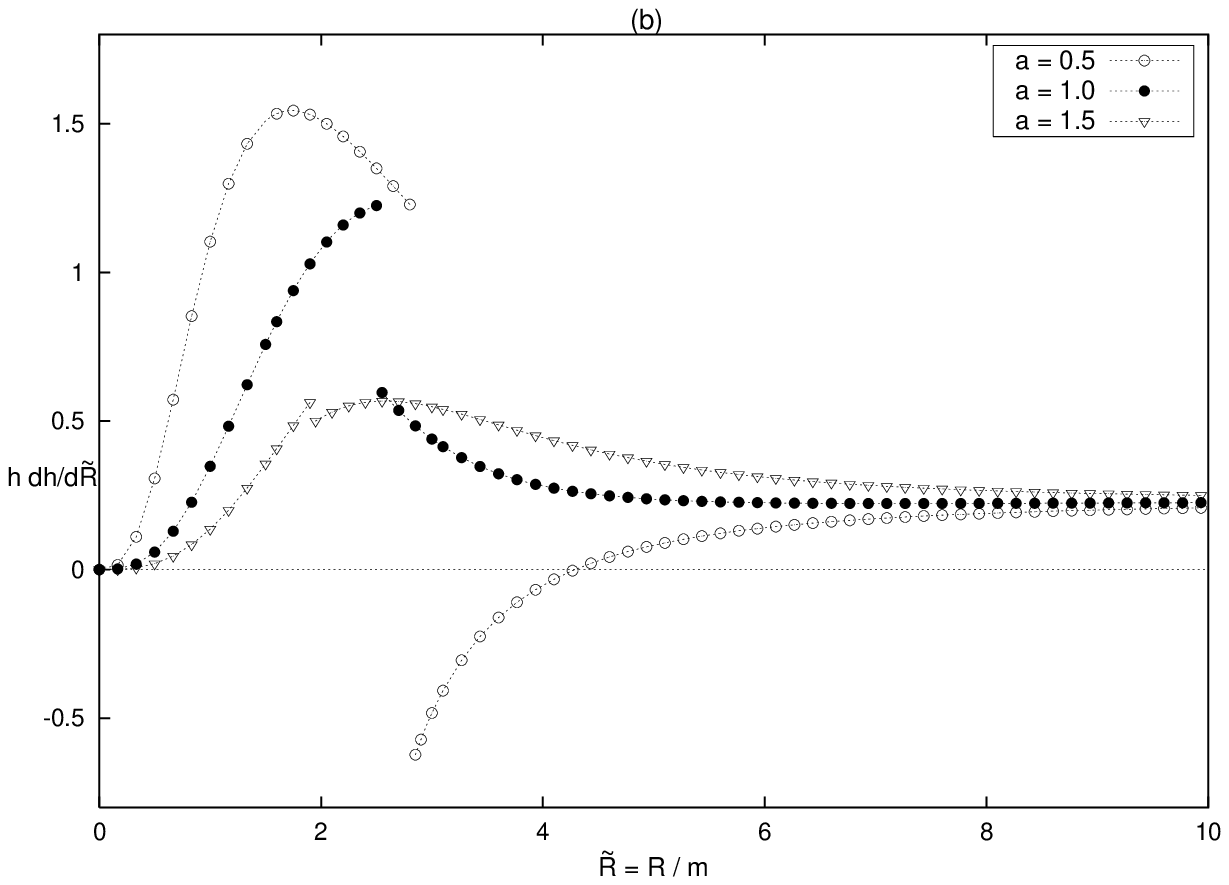}
\caption{(a) The tangential velocity $\mathrm{v}_{c2}$ Eq.\ (\ref{eq_v_fi_nar2}), (b) the curves of $h\frac{dh}{d\tilde{R}}$ Eq.\ (\ref{eq_h_nar2}) with $k=-2+\sqrt{2}\text{; }r_b=2\text{; for }a=0.5 \text{, }1.0\text{, and }1.5$
as function of $\tilde{R}=R/m$. As in the previous case, the same region of 
instability occurs.}  \label{fig_10}
\end{figure}

The tangential velocity $\mathrm{v}_c$ is given by
\begin{align}
\mathrm{v}_{c1a}^2 &=2R^2 \frac{A_{1a}(1-n+k/2)+B_{1a}(1+n+k/2)(R^2+a^2)^n}{[A_{1a}+B_{1a}(R^2+a^2)^n]
[R^2(k+2)+2a^2]} \mbox{,} \label{eq_v_fi_nar1} \\
\mathrm{v}_{c1b}^2 &=R^2\frac{2\sqrt{2}A_{1b}+B_{1b}[4+\sqrt{2}\ln (R^2+a^2)]}{[2A_{1b}+B_{1b}\ln (R^2+a^2)]
[2a^2+\sqrt{2}R^2]} \mbox{,} \label{eq_v_fi_nar2}
\end{align}
and the specific angular momentum $h$,
\begin{align}
h_{1a} &=\sqrt{C}R^2(R^2+a^2)^{k/4}\sqrt{\frac{A_{1a}(1-n+k/2)+B_{1a}(1+n+k/2)(R^2+a^2)^n}
{A_{1a}(a^2+nR^2)+B_{1a}(a^2-nR^2)(R^2+a^2)^n}}\mbox{,} \label{eq_h_nar1} \\
h_{1b} &=\sqrt{C}R^2(R^2+a^2)^{-1/2+\sqrt{2}/4}\sqrt{\frac{2B_{1b}+\sqrt{2}[A_{1b}+B_{1b}\ln (\sqrt{R^2+a^2})]}
{2a^2[A_{1b}+B_{1b}\ln(\sqrt{R^2+a^2})]-2B_{1b}R^2}}\mbox{.} \label{eq_h_nar2} 
\end{align}
In figure \ref{fig_9} (a)--(b), the curves of tangential velocity Eq.\ (\ref{eq_v_fi_nar1}) and $h\frac{dh}{d\tilde{R}}$ Eq.\ (\ref{eq_h_nar1}), respectively, are displayed as functions of $\tilde{R}=R/m$ with $ k=-1/2\text{; }r_b=2\text{; }a=0.5\text{, }1.0\text{,}1.5$. The same quantities are shown in figure \ref{fig_10}(a)--(b) with $k=-2+\sqrt{2}$. For $a=0.5$ the disks have a small region of unstable orbits immediately after the ``boundary radius''. 
  
\subsection{Narlikar-Patwardhan-Vaidya Solutions 2a and 2b}

Like in the previous subsections 
 we  study the generation of a disk  solution with an halo exactly 
as the  one depicted  in Fig.\ \ref{fig_schem2}.
 We also start with a solution of the Einstein equations in isotropic coordinates 
that represents a sphere of radius $r_b$ of perfect fluid that on $r=r_b$  will the continuously matched  to the vacuum Schwarzschild solution. We will
 use another two solutions found by
Narlikar, Patwardhan and Vaidya \cite{Narlikar} that we shall refer as NPV 2a and NPV 2b, respectivelly.
that are  characterized by the metric functions $(\lambda,  \nu_{2a})$ and 
$(\lambda,  \nu_{2b})$, 

\begin{eqnarray} 
&e^{\lambda} =\frac{1}{\left( A_1r^{1+n/2}+A_2r^{1-n/2}\right)^2} & \label{eq_nar_sol2a}\\
&e^{\nu_{2a}} =\frac{\left( B_{1a}r^{1+x/2}+B_{2a}r^{1-x/2}\right)^2}{\left( A_1r^{1+n/2}+A_2r^{1-n/2}\right)^2} &\text{ for } \sqrt{2}<n\leq 2 \mbox{,} \label{eq_nar_sol2b}\\
&e^{\nu_{2b}} =\frac{\left[ B_{1b}+B_{2b} \ln (r)\right]^2}{\left( A_1r^{1/\sqrt{2}}+A_2r^{-1/\sqrt{2}}\right)^2} &\text{ for } n=\sqrt{2}. \label{eq_nar_sol2c}
\end{eqnarray}
where the $A$'s and $B$'s are constants and $x=\sqrt{2n^2-4}$. 
The solution  $(\lambda,\nu_{2a})$ with
 $n=2$ corresponds to Schwarzschild's internal solution in isotropic coordinates (see for instance Ref.\ \cite{Kuchowicz}). This solution has constant density and is conformally flat  when $B_{1a}=0$.

The density, pressure and sound velocity  for the solutions
  $(\lambda, \nu_{2a})$ and
$(\lambda,  \nu_{2b})$,   will be denoted by $(\rho, p_{2a},V_{2a})$ and
 $(\rho, p_{2b},V_{2b})$, respectively. We find,
 
\begin{align}
\rho &=\frac{1}{32 \pi} \left[ (4-n^2)\left( A_1r^{n/2}+A_2r^{-n/2}\right)^2 +12n^2A_1A_2 \right] \mbox{,} \label{eq_rho_nar2}\\
p_{2a} &=\frac{1}{32 \pi} \left[ -12n^2A_1A_2+(3n^2-4)\left( A_1r^{n/2}+A_2r^{-n/2}\right)^2\right. \notag \\
& \left. +\frac{2nx(B_{2a}-B_{1a}r^x)}{B_{2a}+B_{1a}r^x}\left( A_1^2r^{n}-A_2^2r^{-n}\right) \right] \mbox{,} \label{eq_pa_nar2}\\
p_{2b} &=\frac{1}{16 \pi (B_{1b}+B_{2b} \ln (r))} \left[ (B_{1b}+B_{2b} \ln(r)) \left( A_1^2r^{\sqrt{2}}+A_2^2r^{-\sqrt{2}}-10A_1A_2 \right) \right. \notag \\
& \left. +2\sqrt{2}B_{2b} \left( A_2^2r^{-\sqrt{2}}-A_1^2r^{\sqrt{2}} \right) \right] \mbox{,} \label{eq_pb_nar2}\\
\mathrm{V}_{2a}^2 &= \frac{1}{(4-n^2)\left( A_1^2r^{n}-A_2^2r^{-n}\right)(B_{2a}+B_{1a}r^x)^2} \left\{ B_{2a}^2 \left[A_1^2r^n(3n^2+2nx-4)\right. \right. \notag \\
& \left. \left. +A_2^2r^{-n}(-3n^2+2nx+4) \right]+B_{1a}^2r^{2x} \left[ A_1^2r^n(3n^2-2nx-4) \right. \right. \notag \\
& \left. \left. +A_2^2r^{-n}(-3n^2-2nx+4) \right]+2B_{1a}B_{2a}r^x \left[A_1^2r^n(3n^2-2x^2-4)\right. \right. \notag \\
& \left. \left. +A_2^2r^{-n}(-3n^2+2x^2+4) \right]\right\} \mbox{,} \label{eq_va_nar2}\\
\mathrm{V}_{2b}^2 &=\frac{1}{[B_{1b}+B_{2b} \ln (r)]^2\left( A_1^2r^{\sqrt{2}}-A_2^2r^{-\sqrt{2}}\right)} \left\{ A_1^2r^{\sqrt{2}} \left[ (B_{1b}+B_{2b} \ln (r)) \right. \right. \notag \\
& \left. \left. \times (B_{1b}+B_{2b} \ln (r)-2\sqrt{2}B_{2b})+2B_{2b}^2 \right]-
A_2^2r^{-\sqrt{2}} \right. \notag \\
& \left. \times \left[ (B_{1b}+B_{2b} \ln (r))(B_{1b}+B_{2b} \ln (r)+2\sqrt{2}B_{2b})+2B_{2b}^2 \right]\right\} \mbox{.} \label{eq_vb_nar2}
\end{align}

The condition of  continuity of the metric functions  $(\lambda, \nu)$ 
 given by   (\ref{eq_nar_sol2a})--(\ref{eq_nar_sol2c})  and 
the corresponding functions in Eq.\ (\ref{eq_metrica_sch})
 at the boundary $r=r_b$    leads  to following expressions:

\begin{align}
A_1 &=\frac{1}{nr_b^{2+n/2}\left( 1+\frac{m}{2r_b} \right)^3} \left[ \frac{m}{2}-r_b \left( 1-\frac{n}{2}-\frac{mn}{4r_b} \right) \right] \text{,}
\label{eq_cond_sol2a}\\
A_2 &=\frac{1}{r_b^{2-n/2}\left( 1+\frac{m}{2r_b} \right)^3} \left[ -\frac{m}{2n}+r_b \left( \frac{1}{2}+\frac{1}{n}+\frac{m}{4r_b} \right) \right] \text{,}
\label{eq_cond_sol2b}\\
B_{1a} &=\frac{-4r_b^2\left(1-\frac{2m}{r_b}\right)-m^2+2xr_b^2
\left(1-\frac{m^2}{4r_b^2}\right)}{4xr_b^{3+x/2}\left(1+\frac{m}{2r_b}\right)^4} \mbox{,} \label{eq_cond_sol2c}\\
B_{2a} &=\frac{4r_b^2\left(1-\frac{2m}{r_b}\right)+m^2+2xr_b^2
\left(1-\frac{m^2}{4r_b^2}\right)}{4xr_b^{3-x/2}\left(1+\frac{m}{2r_b}\right)^4} \mbox{,} \label{eq_cond_sol2d}\\
B_{1b} &=\frac{1}{4r_b^3\left( 1+\frac{m}{2r_b} \right)^4} \left[4r_b^2-m^2+(m^2-8mr_b+4r_b^2) \ln(r_b) \right] \mbox{,} \label{eq_cond_sol2e}\\
B_{2b} &=-\frac{m^2-8mr_b+4r_b^2}{4r_b^3\left( 1+\frac{m}{2r_b} \right)^4} \mbox{.} \label{eq_cond_sol2f}
\end{align}
$\mathrm{V}_{2a}$ has its maximum at $r=r_b$, and $\mathrm{V}_{2b}$ at $r=0$.

\begin{figure} 
\centering
\includegraphics[scale=0.6]{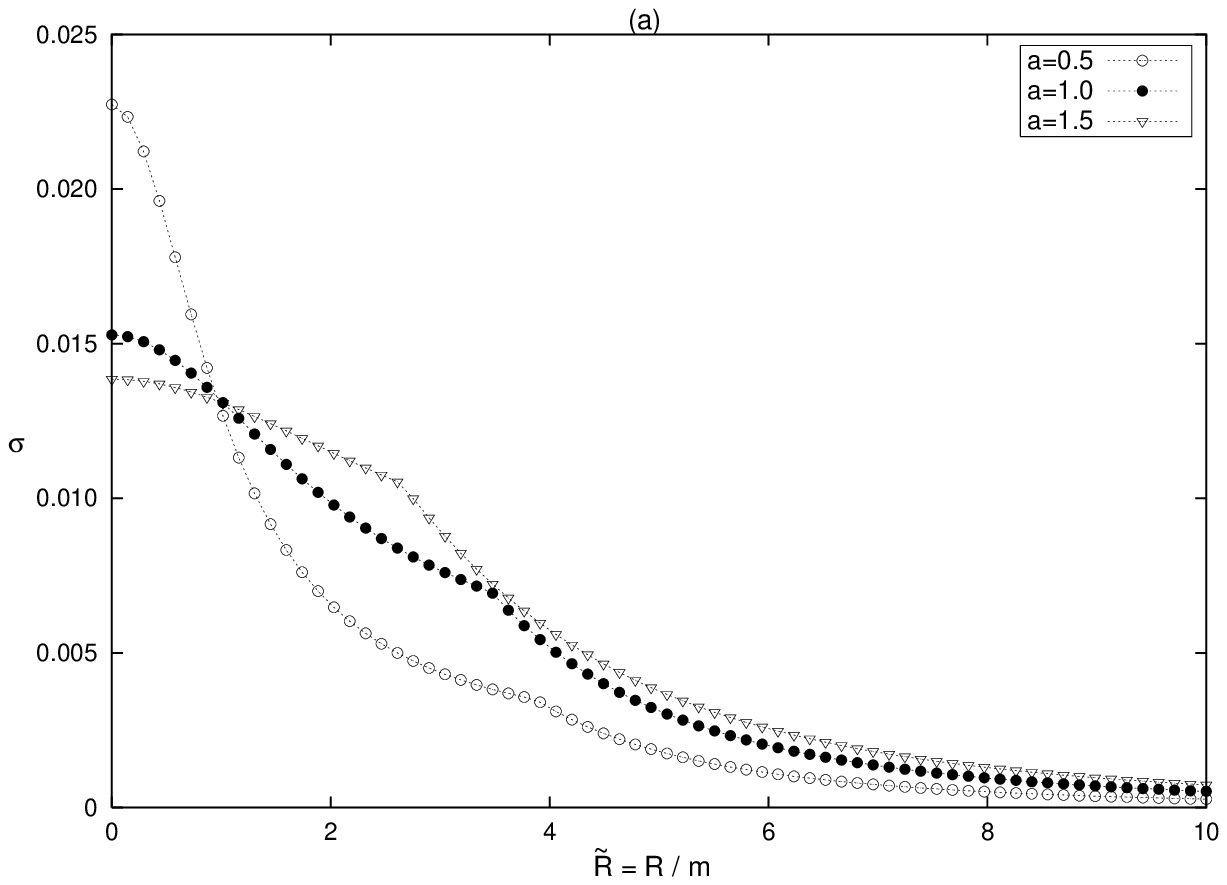}%
\includegraphics[scale=0.6]{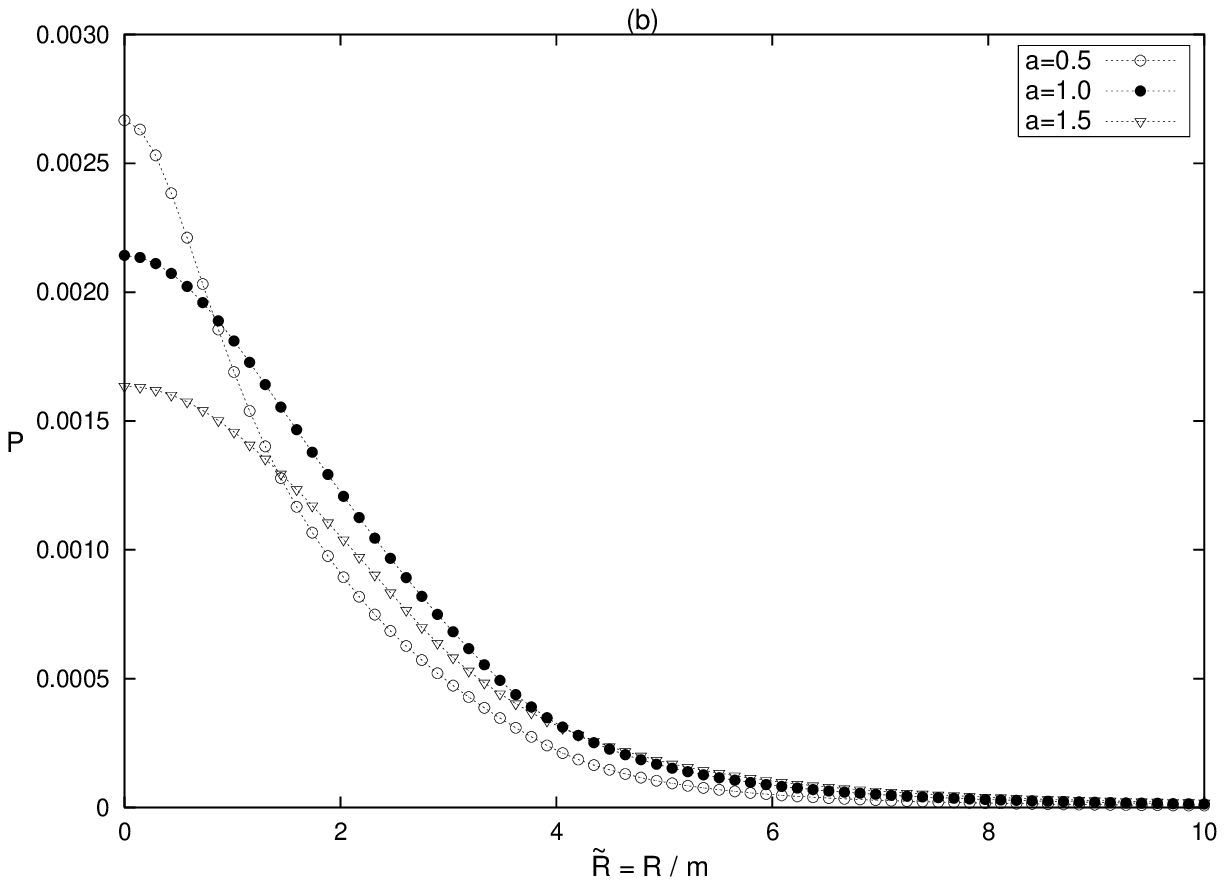} \\
\includegraphics[scale=0.6]{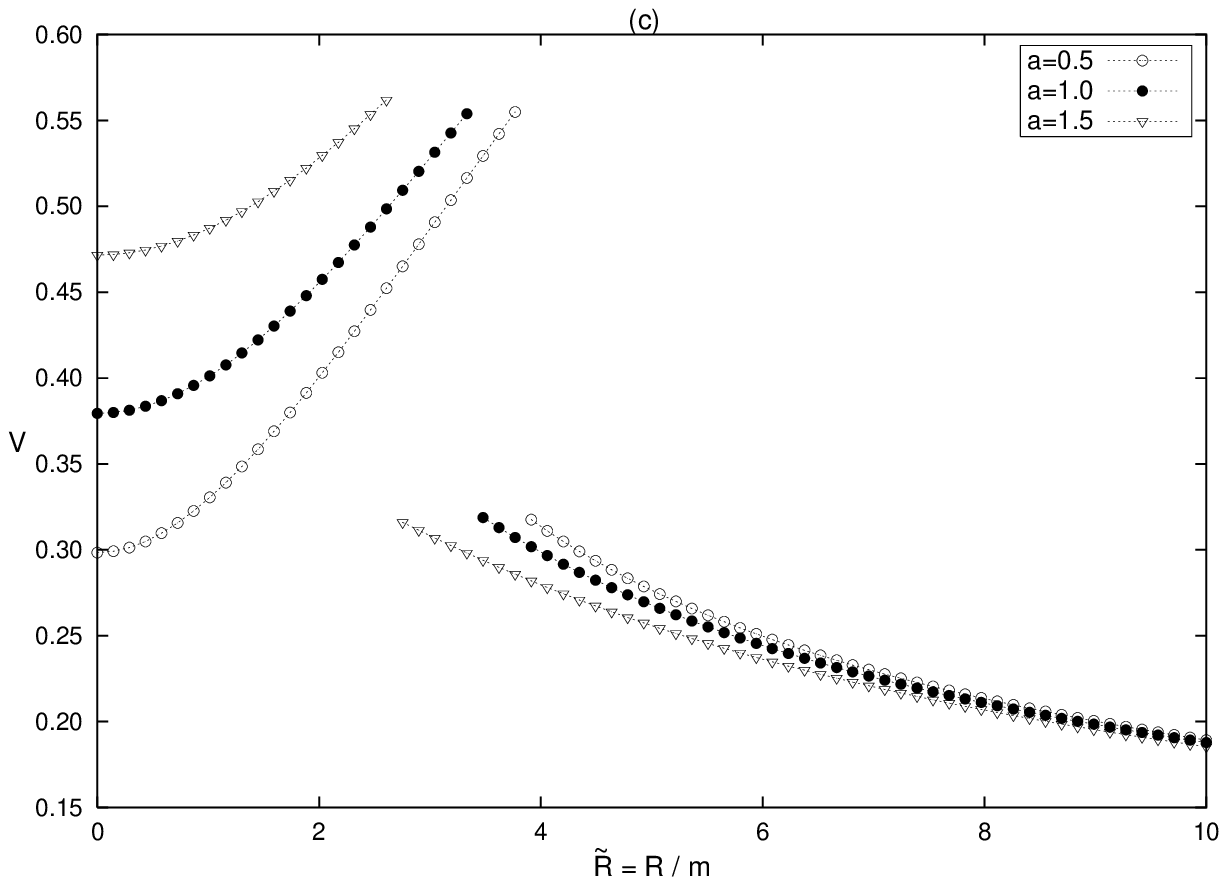}
\caption{(a) The surface energy density $\sigma$ Eq.\ (\ref{eq_en_nar2}), (b) the pressure $P$ Eq.\ (\ref{eq_Pa_nar2}), (c) the velocity of sound $V$ Eq.\ (\ref{eq_Va_nar2})  for the disk with 
$\mathbf{n=1.8}\text{; }m=0.5\text{; }r_b=2\text{; for }a=0.5\text{, }1.0\text{ and }1.5$ as 
function of $\tilde{R}=R/m$.} \label{fig_11}
\end{figure}

\begin{figure} 
\centering
\includegraphics[scale=0.6]{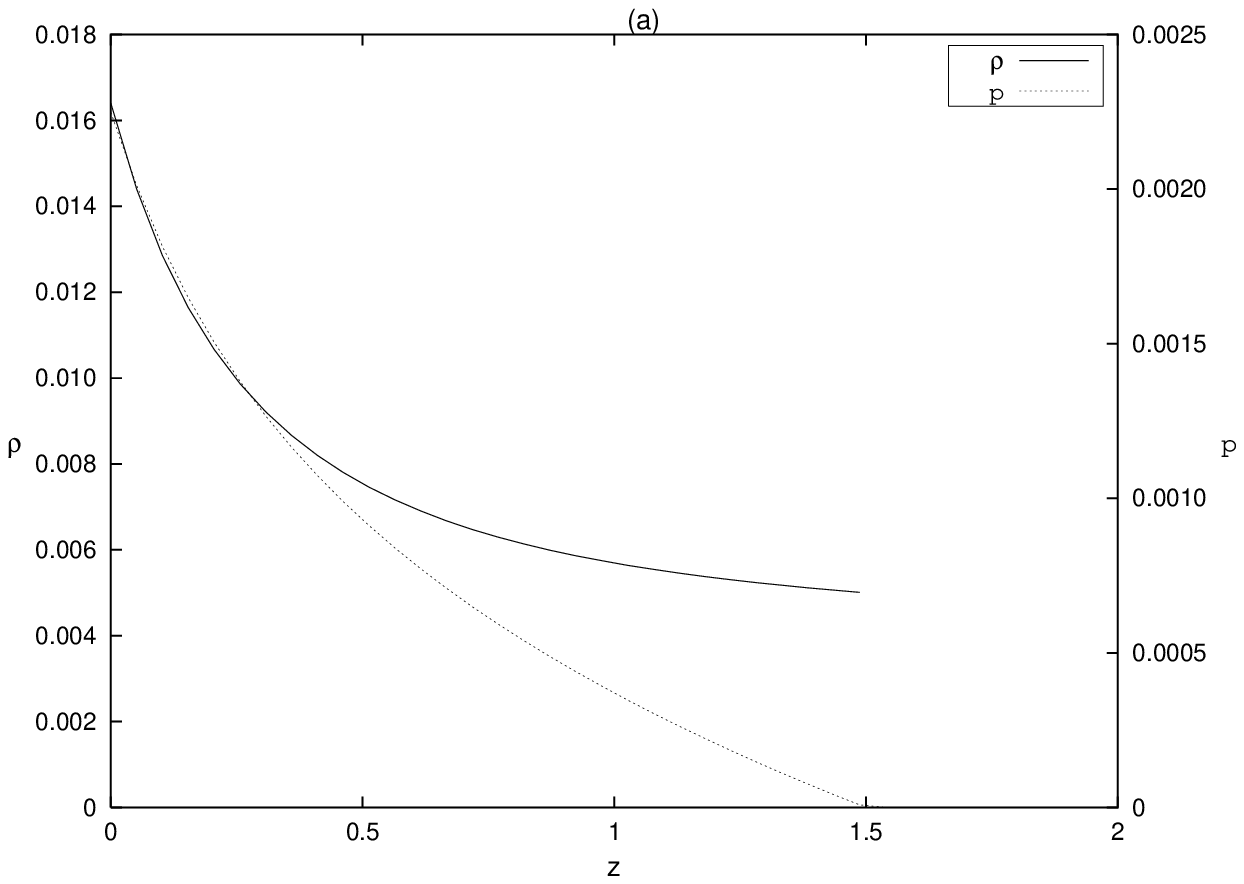}%
\hspace{0.2cm}
\includegraphics[scale=0.6]{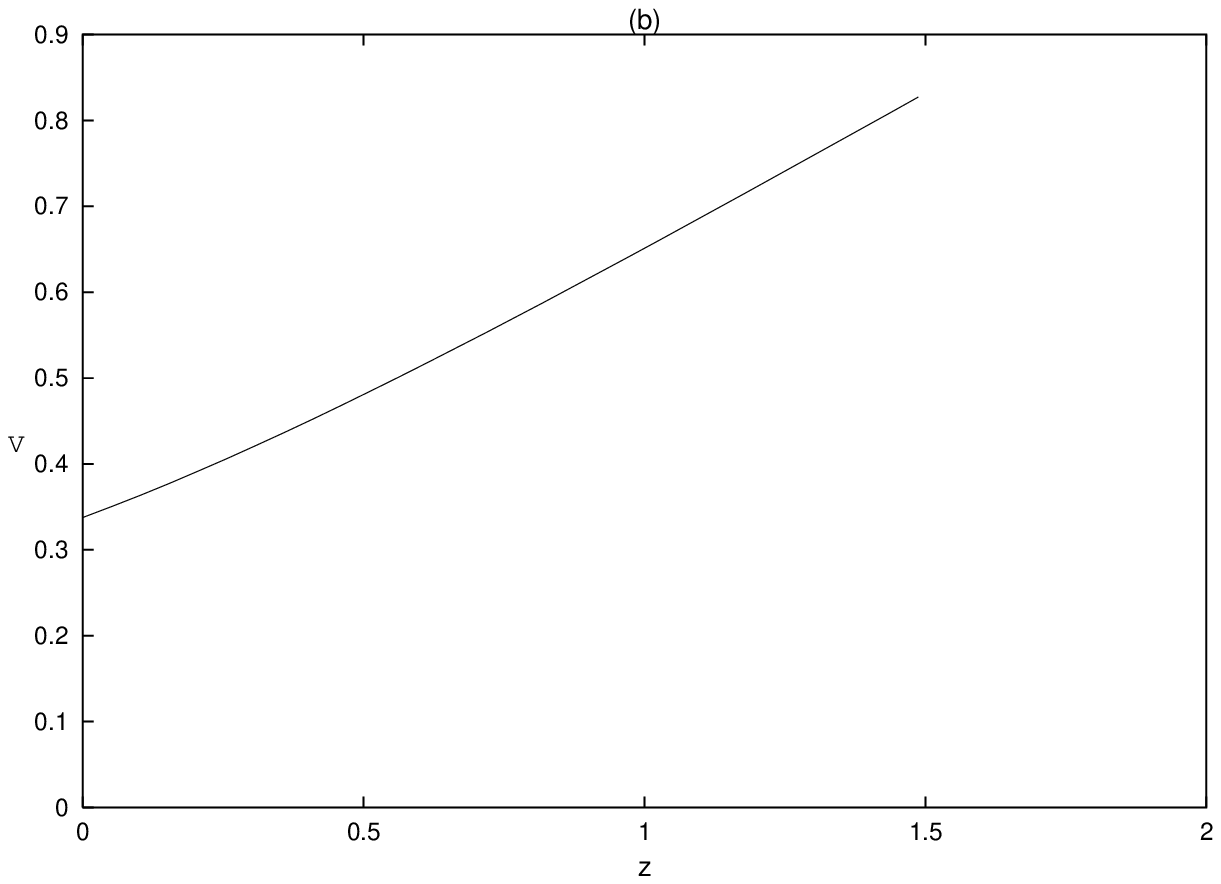}
\caption{(a) The density $\rho$ Eq.\ (\ref{eq_rho_nar2}) and pressure $p$ 
Eq.\ (\ref{eq_pa_nar2}), (b) the velocity of sound V Eq.\ (\ref{eq_va_nar2}) for the halo with $\mathbf{n=1.8}\text{; }m=0.5\text{; }
r_b=2\text{; for }a=0.5$ along the axis $z$.} \label{fig_12}
\end{figure}

\begin{figure} 
\centering
\includegraphics[scale=0.6]{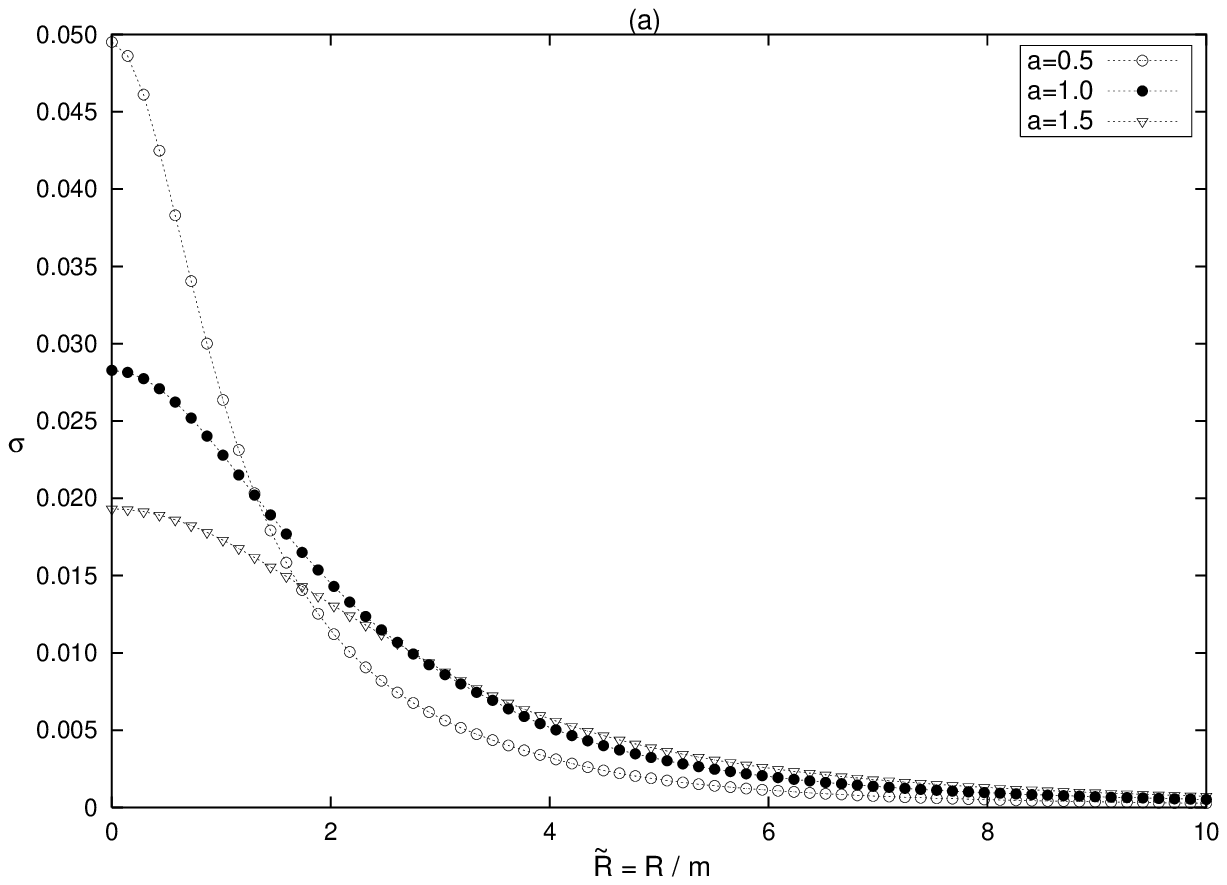}%
\includegraphics[scale=0.6]{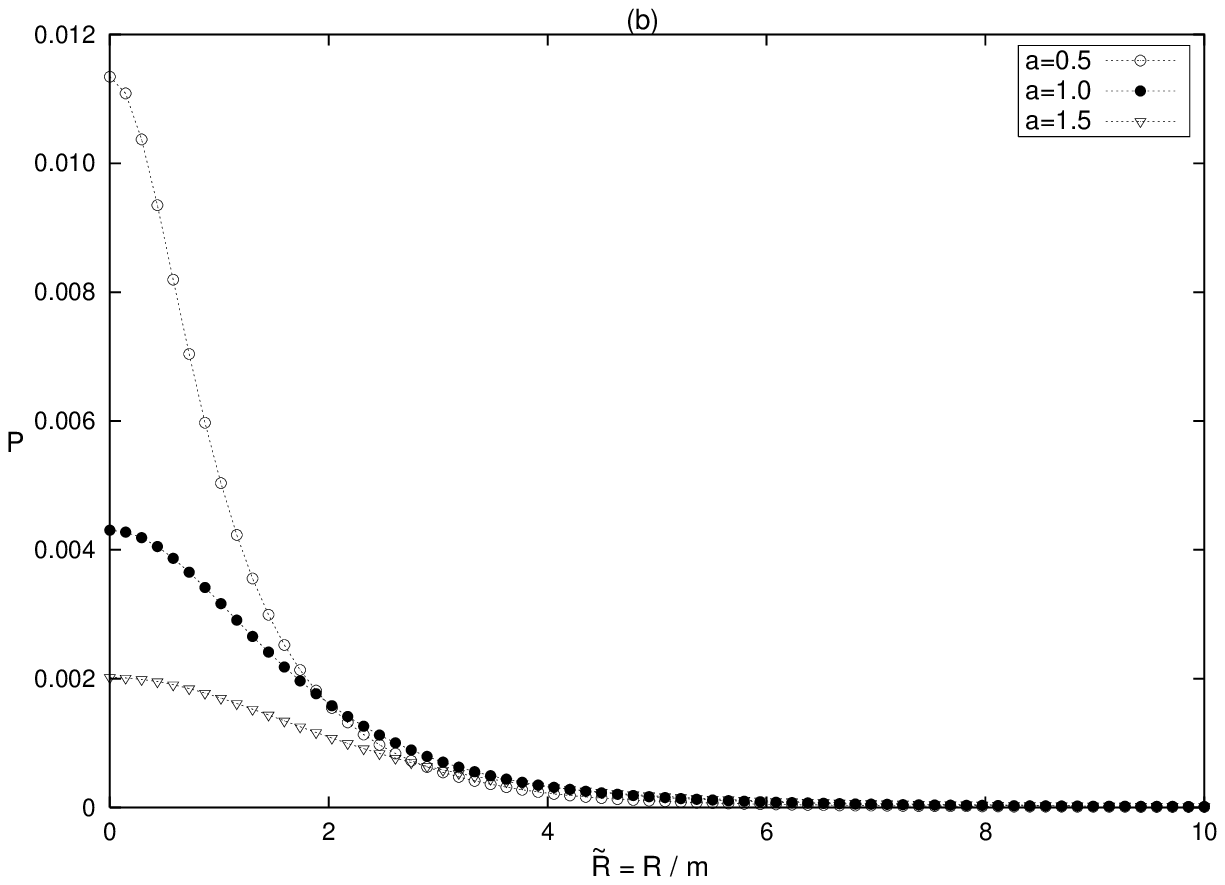} \\
\vspace{0.3cm}
\includegraphics[scale=0.6]{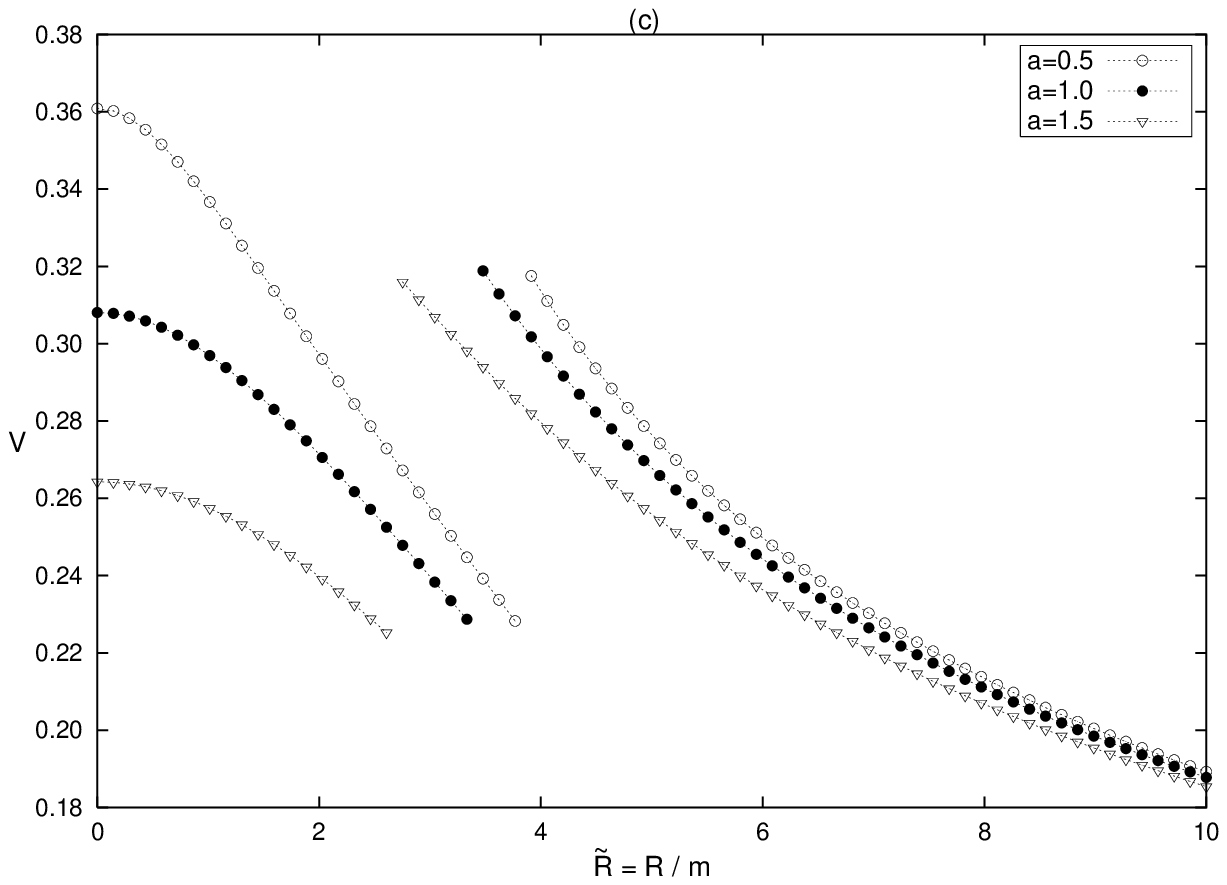}
\caption{(a) The surface energy density $\sigma$ Eq.\ (\ref{eq_en_nar2}), (b) the pressure $P$ Eq.\ (\ref{eq_Pb_nar2}), (c ) the velocity of sound $V$ Eq.\ (\ref{eq_Vb_nar2})  for the disk with 
$n=\sqrt{2}\text{; }r_b=2\text{; }m=0.5\text{; for }a=0.5\text{, }1.0\text{ and }1.5$ 
as function of $\tilde{R}=R/m$.} \label{fig_13}
\end{figure}

\begin{figure} 
\centering
\includegraphics[scale=0.6]{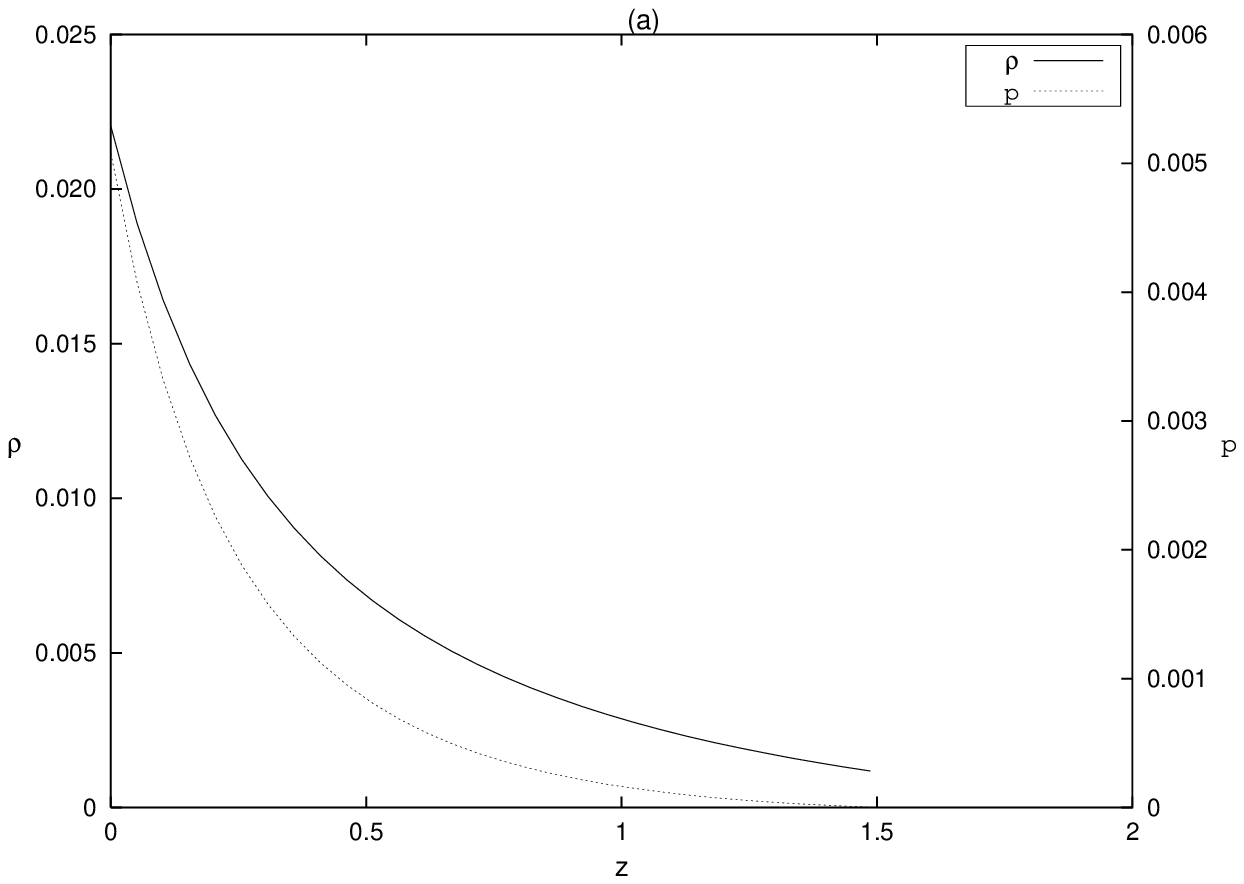}%
\hspace{0.2cm}
\includegraphics[scale=0.6]{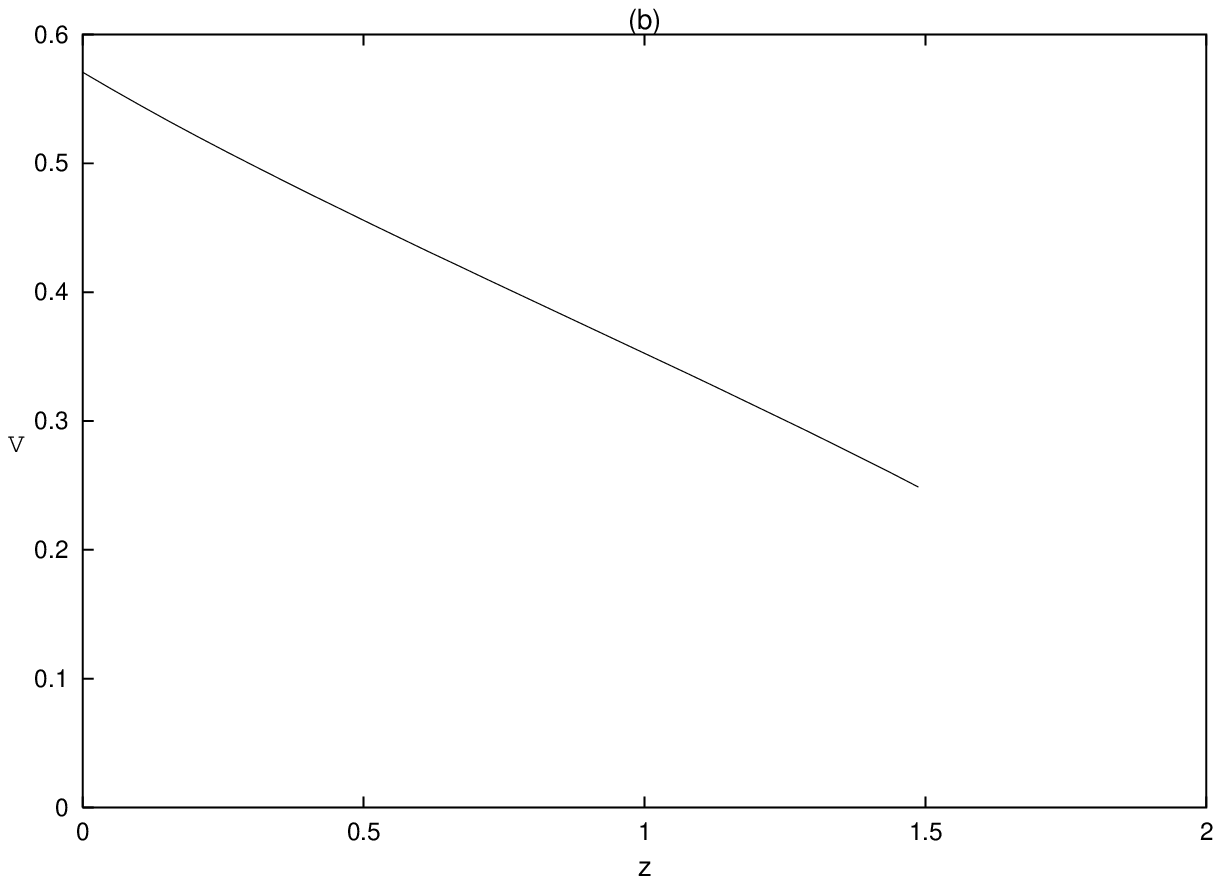}
\caption{(a) The density $\rho$ Eq.\ (\ref{eq_rho_nar2}) and pressure $p$ 
Eq.\ (\ref{eq_pb_nar2}), (b) the velocity of sound V Eq.\ (\ref{eq_vb_nar2})  for the halo with $n=\sqrt{2}\text{; }r_b=2
\text{; for }a=0.5$ along the axis $z$.} \label{fig_14}
\end{figure}

Using Eq.\ (\ref{eq_nar_sol2a})--(\ref{eq_nar_sol2c}) in Eq.\ (\ref{eq_qtt})--(\ref{eq_disk_surf}), we get the expressions for the energy density, pressure and sound velocity of the disk:
\begin{align}
\sigma &=\frac{a}{4 \pi} \left[ A_1(2+n)\mathcal{R}^{-1/2+n/4}+ A_2(2-n)\mathcal{R}^{-1/2-n/4} \right] \text{,} \label{eq_en_nar2}\\
P_{2a} &=-\frac{a}{8 \pi \left( B_{1a}\mathcal{R}^{1/2+x/4}+B_{2a}\mathcal{R}^{1/2-x/4} \right)} \left[ B_{1a}A_1(2+2n-x)\mathcal{R}^{(x+n)/4} \right. \notag \\
& \left. +B_{1a}A_2(2-2n-x)\mathcal{R}^{(x-n)/4}+B_{2a}A_1(2+2n+x)\mathcal{R}^{(-x+n)/4} \right. \notag \\
& \left. +B_{2a}A_2(2-2n+x)\mathcal{R}^{-(x+n)/4} \right] \mbox{,} \label{eq_Pa_nar2}\\
P_{2b} &=-\frac{a}{4 \pi [2B_{1b}+B_{2b}\ln (\mathcal{R})]}\left\{ 2(1+\sqrt{2})B_{1b}A_1\mathcal{R}^{-1/2+
\sqrt{2}/4}\right. \notag \\
& \left. +2(1-\sqrt{2})B_{1b}A_2\mathcal{R}^{-1/2-\sqrt{2}/4}
+\left[ (1+\sqrt{2}) \ln (\mathcal{R})-2 \right] B_{2b}A_1\mathcal{R}^{-1/2+\sqrt{2}/4}\right. \notag \\
& \left. +\left[ (1-\sqrt{2}) \ln (\mathcal{R})-2 \right] B_{2b}A_2\mathcal{R}^{-1/2-\sqrt{2}/4} \right\} \mbox{,} \label{eq_Pb_nar2}\\
V_{2a}^2 &=\frac{1}{2(4-n^2)\left[A_2+A_1\mathcal{R}^{n/2}\right] \left[ B_{2a}+B_{1a}\mathcal{R}^{x/2}\right]^2}
\left\{ A_1\mathcal{R}^{n/2} \right. \notag \\
& \left. \left[ B_{1a}^2(n-2)(2n-x+2)\mathcal{R}^{x}+B_{2a}^2(n-2)(2n+x+2) \right. \right. \notag \\
& \left. \left. -4B_{1a}B_{2a}(-n^2+n+x^2+2)\mathcal{R}^{x/2} \right] +A_2 \left[ B_{1a}^2(n+2)(2n+x-2)\mathcal{R}^{x} \right. \right. \notag \\
& \left. \left. +B_{2a}^2(n+2)(2n-x-2)-4B_{1a}B_{2a}(-n^2-n+x^2+2)\mathcal{R}^{x/2} \right] \right\} \mbox{,} \label{eq_Va_nar2}\\
V_{2b}^2 &=- \frac{2B_{1b}+2B_{2b}+B_{2b}\ln (\mathcal{R})}{2[2B_{1b}+B_{2b}\ln (\mathcal{R})]^2[A_2+A_1\mathcal{R}^{1/\sqrt{2}}]}\left\{ \left[ \sqrt{2}\ln (\mathcal{R})-4 \right]\right. \notag \\
& \left. \times B_{2b}A_1\mathcal{R}^ {1/\sqrt{2}}-\left[ \sqrt{2}\ln (\mathcal{R})+4 \right]B_{2b}A_2 
+2\sqrt{2}B_{1b}\left[ A_1\mathcal{R}^{1/\sqrt{2}}-A_2 \right] \right\} \mbox{,} \label{eq_Vb_nar2}
\end{align}
where $\mathcal{R}=R^2+a^2$. The curves of $\sigma$, $P$ and $V$ as function of $\tilde{R}=R/m$ with parameters $n=1.8\text{; }m=0.5\text{; }r_b=2\text{ and }a=0.5\text{, }1.0\text{,}1.5$ are displayed in figure \ref{fig_11} (a) -- (c), respectively. Figure \ref{fig_12} (a) -- (b) shows the density $\rho$, pressure $p$ and velocity of sound V for the halo with parameters $n=1.8\text{; }m=0.5\text{; }r_b=2\text{, for }a=0.5$ along the axis $z$. The same physical quantities are shown in figures \ref{fig_13} and \ref{fig_14} with $n=\sqrt{2}$. 

\begin{figure} 
\centering
\includegraphics[scale=0.6]{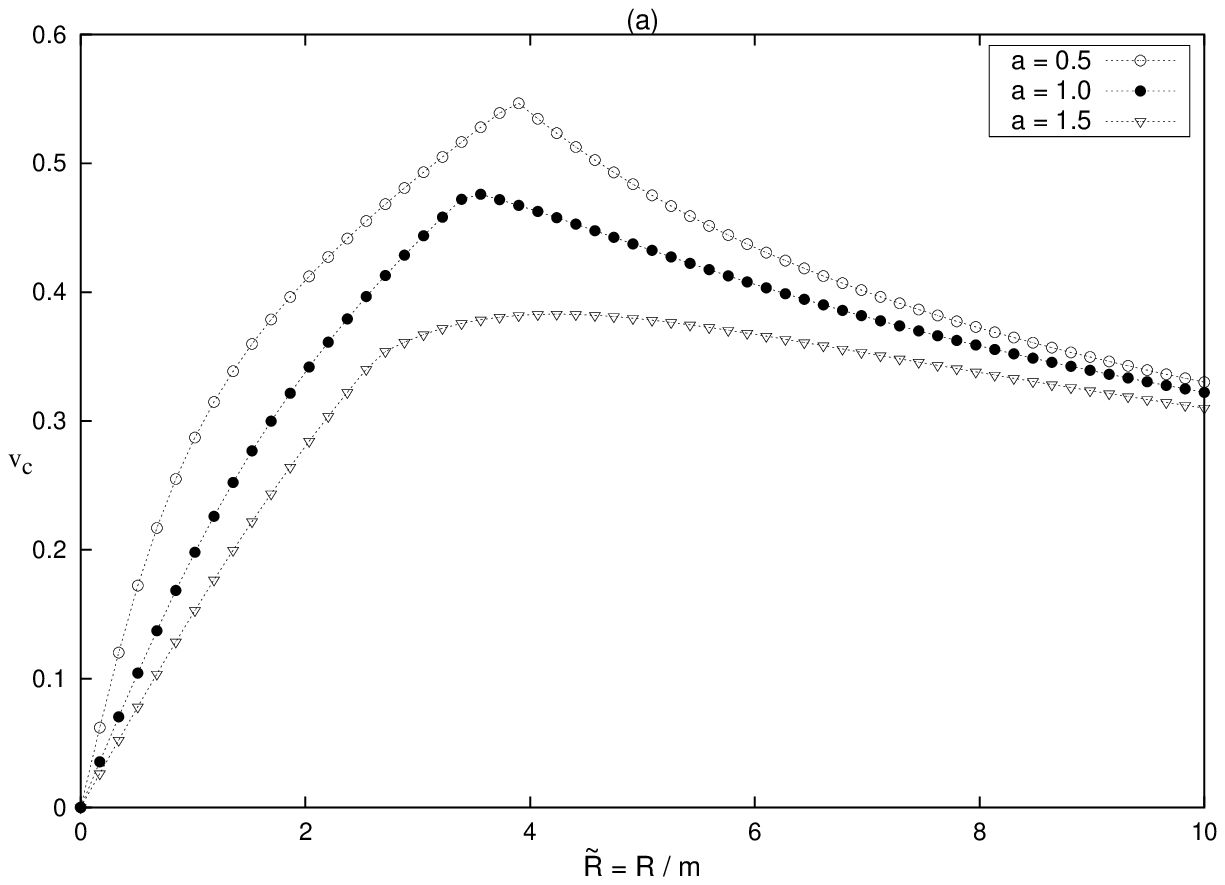}%
\includegraphics[scale=0.6]{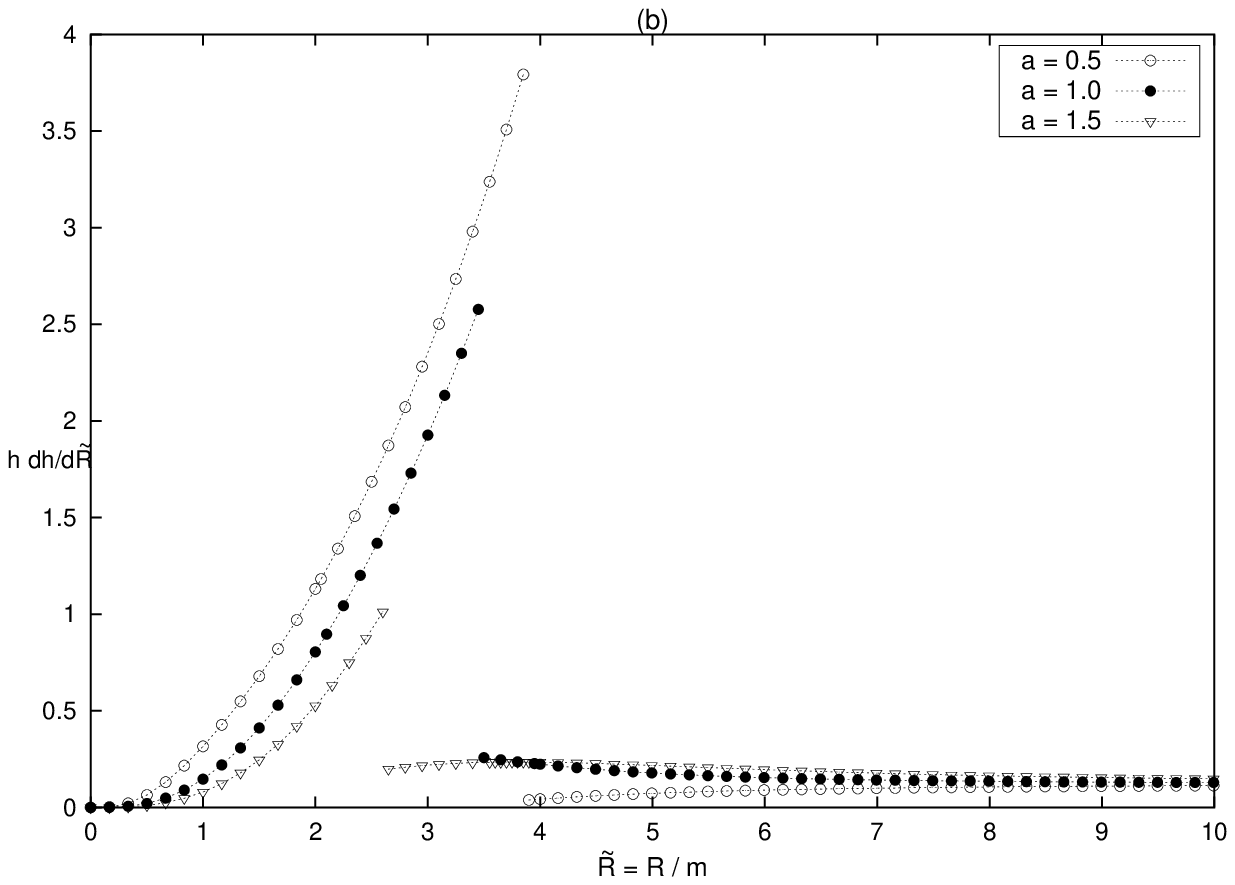}
\caption{(a) The tangential velocity $\mathrm{v}_{ca}$ Eq.\ (\ref{eq_v_fi_nara}), (b) the curves of $h\frac{dh}{d\tilde{R}}$ Eq.\ (\ref{eq_h_nara}) with $\mathbf{n=1.8}\text{; }m=0.5\text{; }r_b=2\text{; for }a=0.5 \text{, }1.0\text{, and }1.5$
as function of $\tilde{R}=R/m$. The disks have no unstable orbits
for these parameters.} \label{fig_15}
\end{figure}

\begin{figure} 
\centering
\includegraphics[scale=0.6]{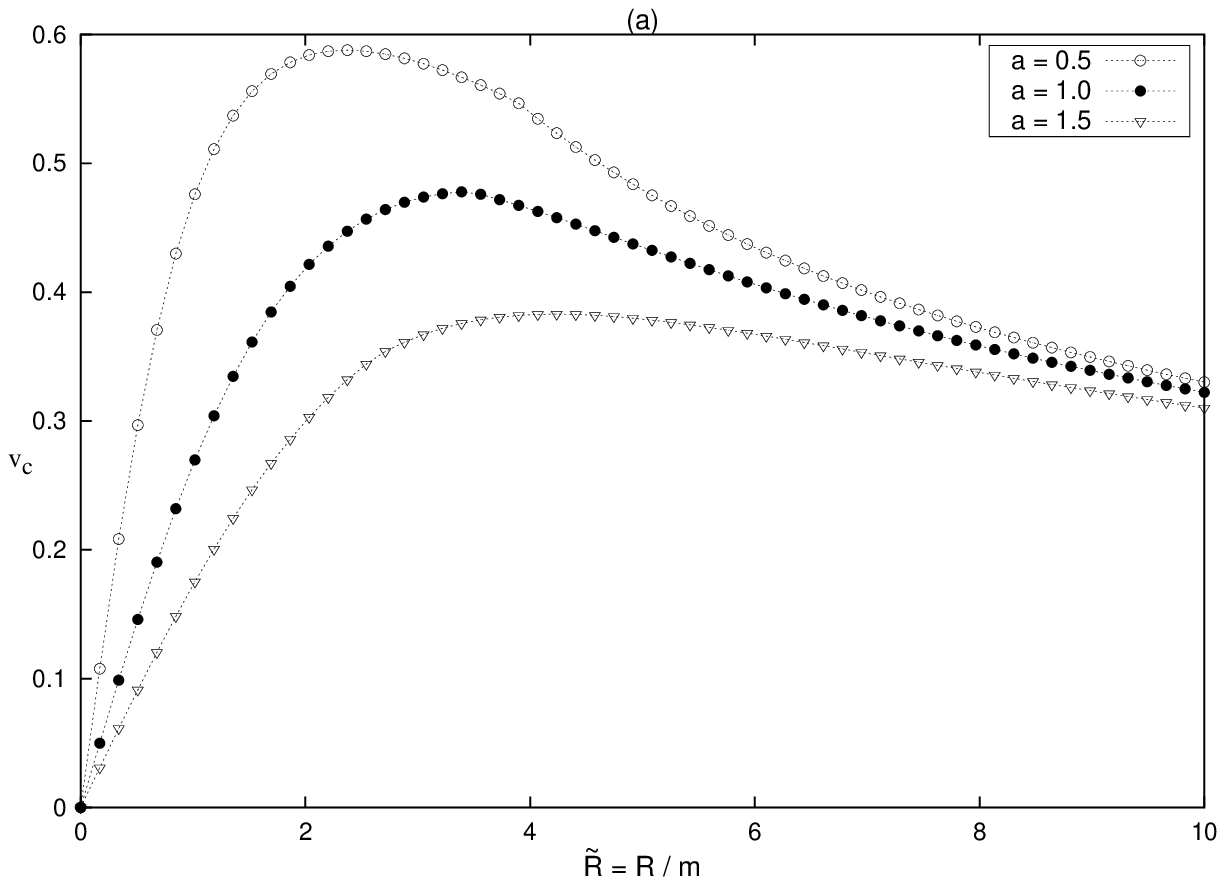}%
\includegraphics[scale=0.6]{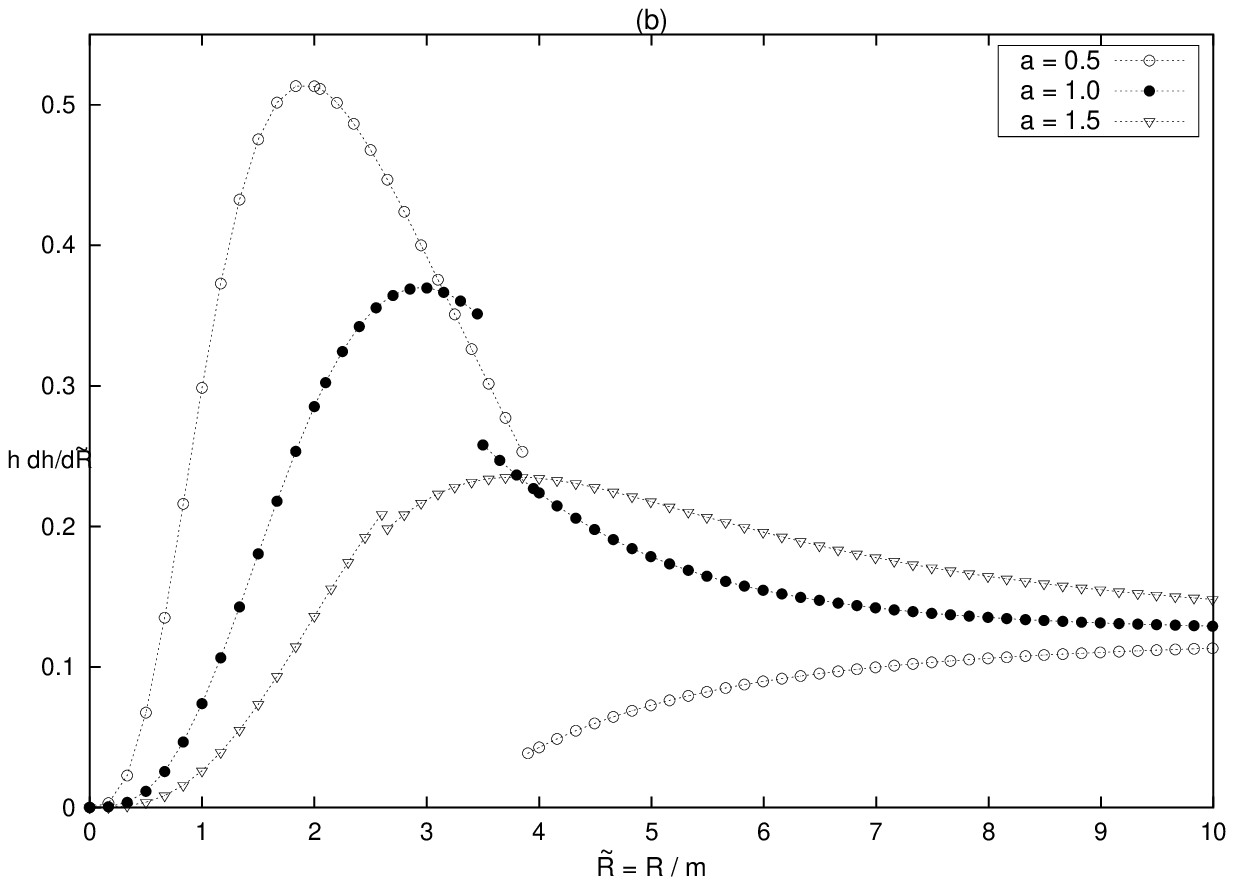}
\caption{(a) The tangential velocity $\mathrm{v}_{cb}$ Eq.\ (\ref{eq_v_fi_narb}), (b) The curves of $h\frac{dh}{d\tilde{R}}$ Eq.\ (\ref{eq_h_narb}) with $n=\sqrt{2}\text{; }m=0.5\text{; }r_b=2\text{; for }a=0.5 \text{, }1.0\text{, and }1.5$
as function of $\tilde{R}=R/m$. The disks have no unstable orbits
for these parameters.} \label{fig_16}
\end{figure}

The tangential velocity $\mathrm{v}_{c}$ is given by
\begin{align}
\mathrm{v}_{c2a}^2 &=R^2\frac{A_2[B_{1a}(n+x)\mathcal{R}^{x/2}+B_{2a}(n-x)]-A_1\mathcal{R}^{n/2}[B_{1a}
(n-x)\mathcal{R}^{x/2}+B_{2a}(n+x)]}{[B_{1a}\mathcal{R}^{x/2}+B_{2a}][A_1\mathcal{R}^{n/2}(2a^2-nR^2)
+A_2(2a^2+nR^2)]} \mbox{,} \label{eq_v_fi_nara} \\
\mathrm{v}_{c2b}^2 &=R^2\frac{A_2[2\sqrt{2}B_{1b}+B_{2b}(4+\sqrt{2}\ln (\mathcal{R}))]-A_1\mathcal{R}^{\sqrt{2}/2}
[2\sqrt{2}B_{1b}+B_{2b}(-4+\sqrt{2}\ln (\mathcal{R}))]}{[2B_{1b}+B_{2b}\ln (\mathcal{R})][A_1\mathcal{R}^{\sqrt{2}/2}
(2a^2-\sqrt{2}R^2)+A_2(2a^2+\sqrt{2}R^2)]}\mbox{,} \label{eq_v_fi_narb}
\end{align}
and the specific angular momentum $h$
\begin{align}
h_{2a} &=\frac{R^2\mathcal{R}^{-1/2+n/4}}{(A_1\mathcal{R}^{n/2}+A_2)^{3/2}
\sqrt{B_{1a}(2a^2-xR^2)\mathcal{R}^{x/2}+B_{2a}(2a^2+xR^2)}} \times \notag \\
 &\left\{ A_2[B_{1a}(n+x)\mathcal{R}^{x/2}+B_{2a}(n-x)]
- A_1\mathcal{R}^{n/2}[B_{1a}(n-x)\mathcal{R}^{x/2}+B_{2a}(n+x)]
\right\}^{1/2} \mbox{,} \label{eq_h_nara}\\
h_{2b} &=\frac{R^2\mathcal{R}^{-1/2+\sqrt{2}/4}}{[A_1\mathcal{R}^{\sqrt{2}/2}+A_2]^{3/2}
\sqrt{4a^2B_{1b}+2B_{2b}[-2R^2+a^2\ln(\mathcal{R})]}} \times \notag \\
&\left\{ A_2[2\sqrt{2}B_{1b}+B_{2b}(4+\sqrt{2}
\ln (\mathcal{R}))]
 -A_1\mathcal{R}^{\sqrt{2}/2}[2\sqrt{2}B_{1b}+B_{2b}(-4+\sqrt{2}\ln (\mathcal{R}))]
\right\}^{1/2} \mbox{.} \label{eq_h_narb}
\end{align}
In figure \ref{fig_15} (a)--(b), the curves of tangential velocities Eq.\ (\ref{eq_v_fi_nara}) and $h\frac{dh}{d\tilde{R}}$ Eq.\ (\ref{eq_h_nara}), respectively, are displayed as functions of $\tilde{R}=R/m$ with $n=1.8\text{; }m=0.5\text{; }r_b=2\text{; }a=0.5\text{, }1.0\text{,}1.5$. Figure \ref{fig_16} shows the same quantities with $n=\sqrt{2}$. Unlike solution 2, no unstable circular 
orbits are present for the disks constructed with these parameters. 

\section{Disks with composite halos from spherical solutions}

We study two examples of  disks with halos constructed from spheres of 
fluids with two layers 
as the  ones  depicted  in Fig.\ \ref{fig_schem3}.

\subsection{ Internal Schwarzschild solution and Buchdahl solution}

Let us consider a fluid sphere is formed by two layers:
 The internal layer,  $0 \leq r <r_1$, will be taken as the 
internal Schwarzschild solution (solution 2a with n=2),
\begin{equation} \label{eq_layer1}
e^{\nu}=\frac{\left( B_1r^2+B_2 \right)^2}{\left( A_1r^2+A_2\right)^{2}} \text{,} \quad e^{\lambda}=\frac{1}{\left( A_1r^2+A_2\right)^{2}} \mbox{.}
\end{equation}
The external layer,  $r>r_1$ is taken as the Buchdahl solution, 
\begin{equation} \label{eq_layer2}
e^{\nu}=\left( 
\frac{1-\frac{C}{\sqrt{1+kr^2}}}{1+\frac{C}{\sqrt{1+kr^2}}}\right)^2 \text{,} 
\quad e^{\lambda}=\left( 1+\frac{C}{\sqrt{1+kr^2}}\right)^4 \mbox{.}
\end{equation}
Note that the external layer has no boundary, i.e., this layer has infinite radius.

According to the continuity conditions at $r=r_1$, the constants are related through:
\begin{align}
A_1 &=\frac{Ck}{\left( C+\sqrt{1+kr_1^2}\right)^3} \text{, } \quad
A_2=\frac{1+\frac{C}{(1+kr_1^2)^{3/2}}}{\left( 1+\frac{C}{\sqrt{1+kr_1^2}}\right)^3} \mbox{,} \label{eq_cond_l1}\\
B_1 &=\frac{Ck}{(1+kr_1^2)\left( 1+\frac{C}{\sqrt{1+kr_1^2}}\right)^3}D,\\
D &= \frac{\left[
C(1-3kr_1^2)-\frac{C^2}{\sqrt{1+kr_1^2}}(1-kr_1^2)+2(1+kr_1^2)^{3/2}\right]}
{\left[ (1+kr_1^2)^2+2C\sqrt{1+kr_1^2}+C^2(1-kr_1^2)\right]} \mbox{,}\\
B_2 &=\frac{1-\frac{C}{\sqrt{1+kr_1^2}}}{\left( 1+\frac{C}{\sqrt{1+kr_1^2}}\right)^3}
-B_1r_1^2 \mbox{.} \label{eq_cond_l2}
\end{align}
With these relations, one verifies that, using Eq.\ (\ref{eq_en_buch}) and Eq.\ (\ref{eq_en_nar2}), Eq.\ (\ref{eq_P_buch}) and Eq.\ (\ref{eq_Pa_nar2}), both the energy density and the pressure are continuous at the radius $R=\sqrt{r_1^2-a^2}$ of the disk.

\begin{figure} 
\centering
\includegraphics[scale=0.6]{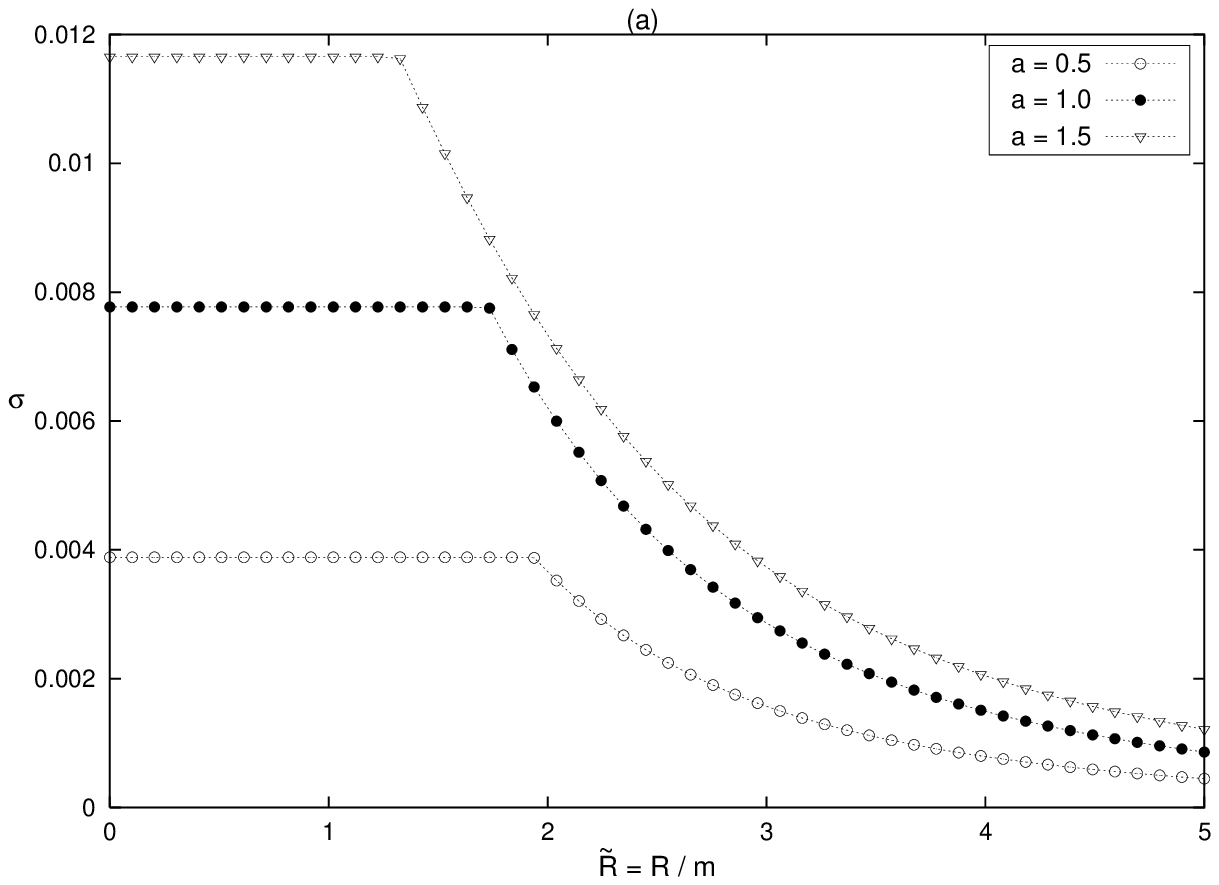}%
\includegraphics[scale=0.6]{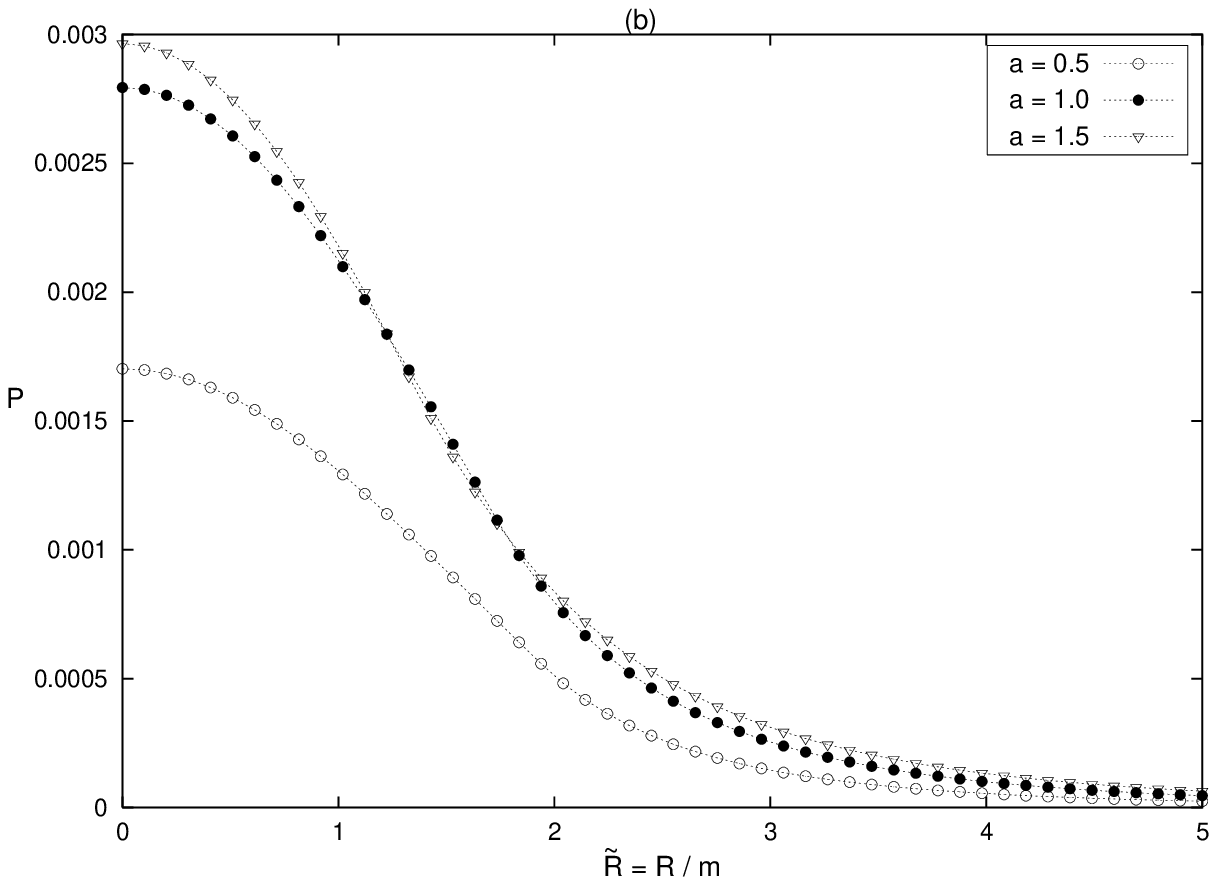} \\
\vspace{0.3cm}
\includegraphics[scale=0.6]{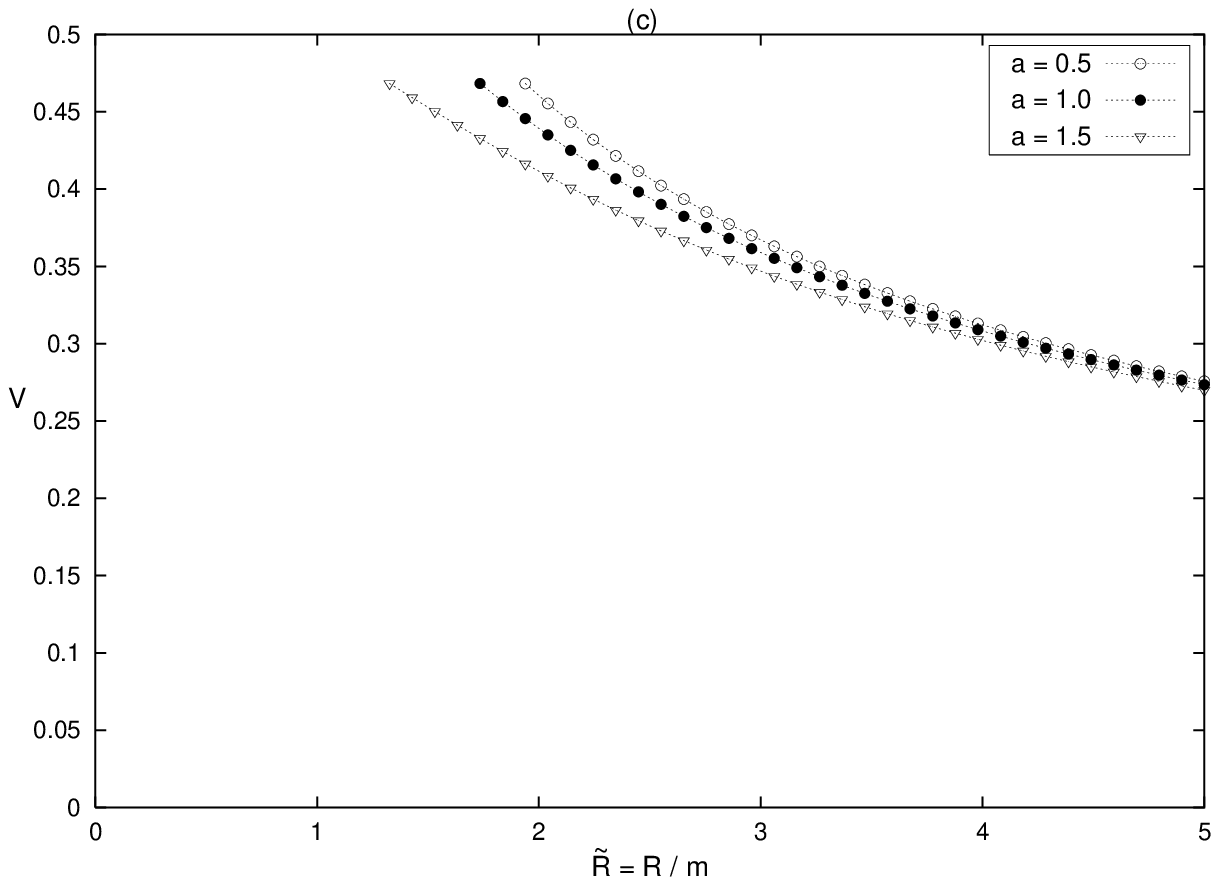}
\caption{(a) The surface energy density $\sigma$, (b) the pressure $P$, (c ) the velocity of sound $V$ for the disk generated from spherical fluid 
layers Eq.\ (\ref{eq_layer1})--(\ref{eq_layer2}) with $m=1\text{; }k=1\text{; }r_1=2\text{; for }
a=0.5\text{, }1.0\text{ and }1.5$ as function of $\tilde{R}=R/m$.} \label{fig_17}
\end{figure}

\begin{figure} 
\centering
\includegraphics[scale=0.6]{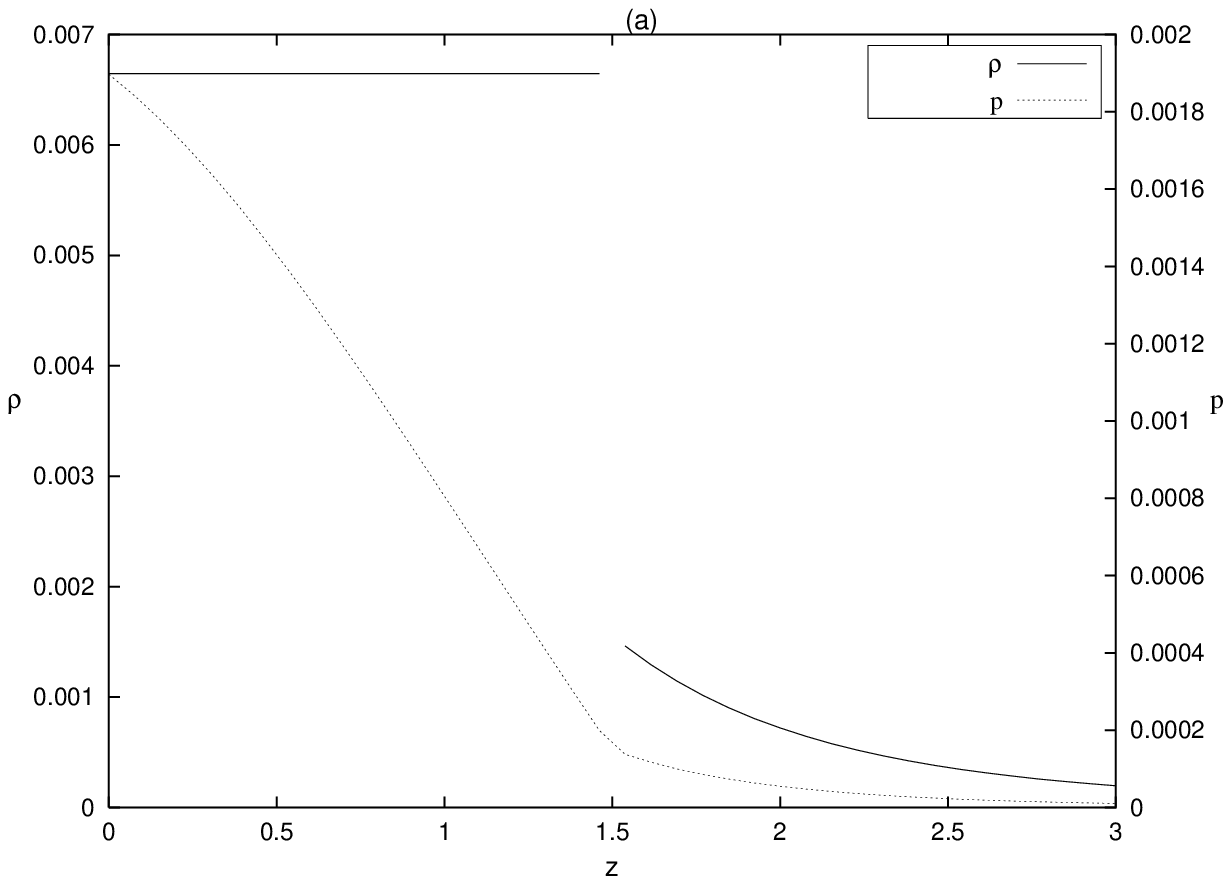}%
\hspace{0.2cm}
\includegraphics[scale=0.6]{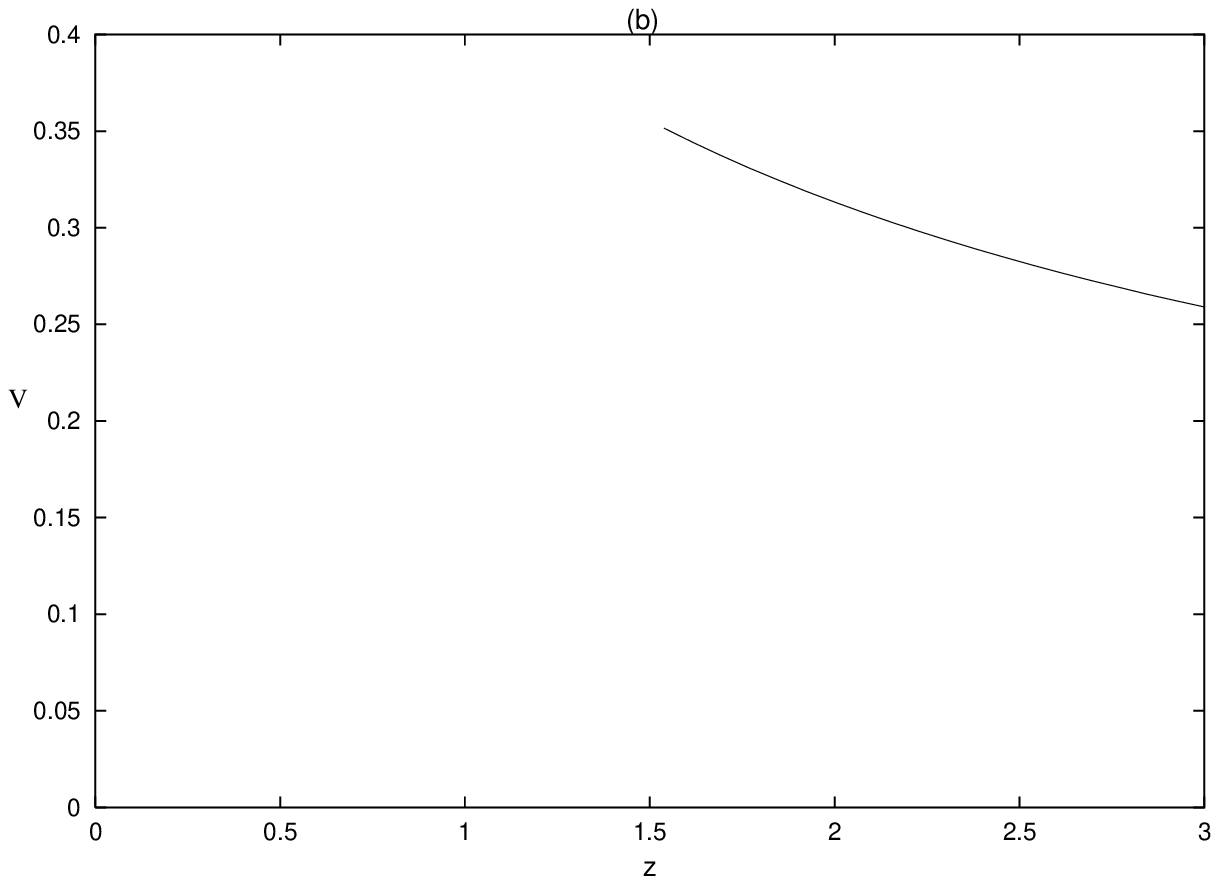}
\caption{(a) The density $\rho$ and pressure $p$, (b) the velocity of sound V for the halo formed by fluid layers Eq.\ (\ref{eq_layer1})--(\ref{eq_layer2}) with $m=1\text{; }k=1\text{; }r_1=2\text{ and }
a=0.5$ along the axis $z$.} \label{fig_18}
\end{figure}

Figure \ref{fig_17} (a)--(c) shows, respectively, $\sigma$, $P$ and $V$ for the disk obtained from fluid layers Eq.\ (\ref{eq_layer1})--(\ref{eq_layer2}) with parameters $m=1\text{; }k=1\text{; }r_1=2\text{; for }a=0.5\text{, }1.0\text{ and }1.5$ as function of $\tilde{R}=R/m$. The density $\rho$, pressure $p$ and velocity of sound V for the halo along the $z$ axis with the same parameters for $a=0.5$ is shown in figure \ref{fig_18}.

\begin{figure}
\centering
\includegraphics[scale=0.6]{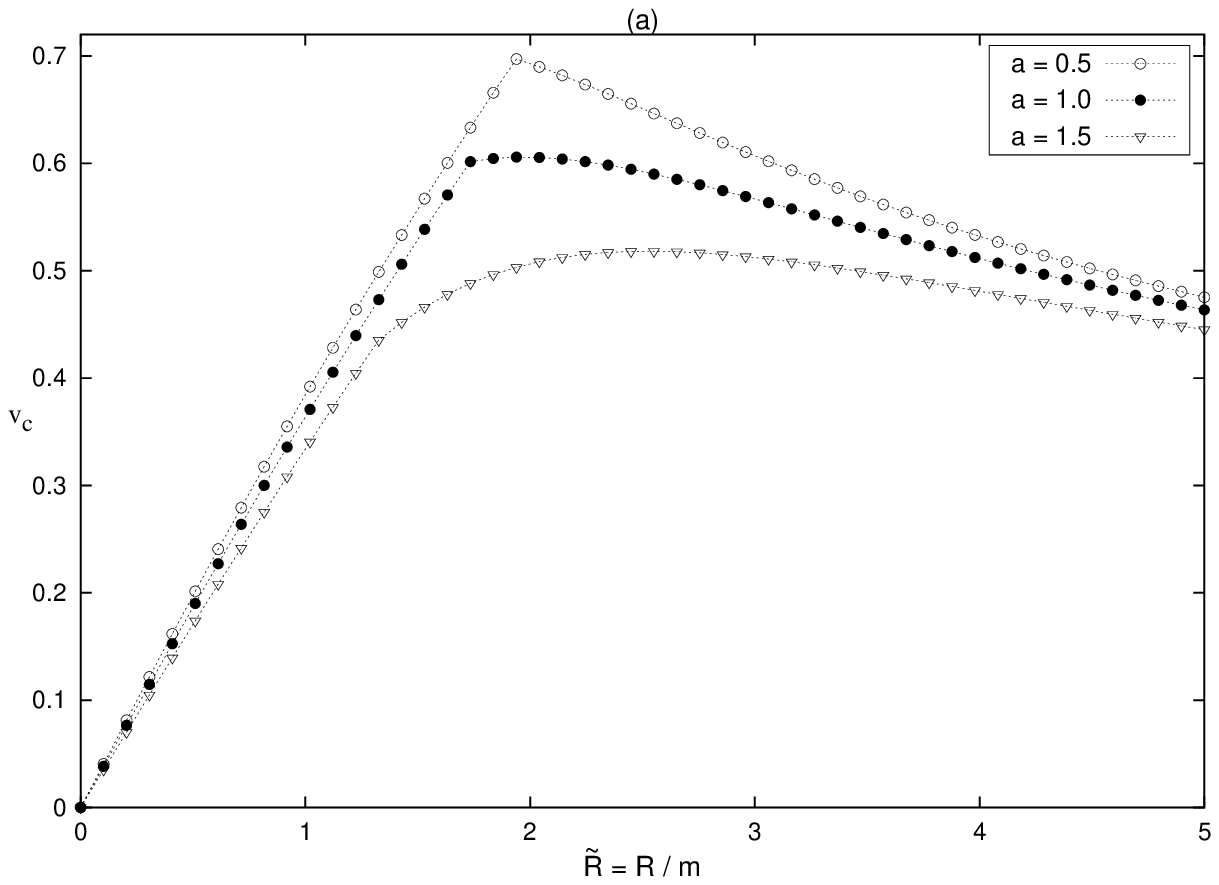}%
\includegraphics[scale=0.6]{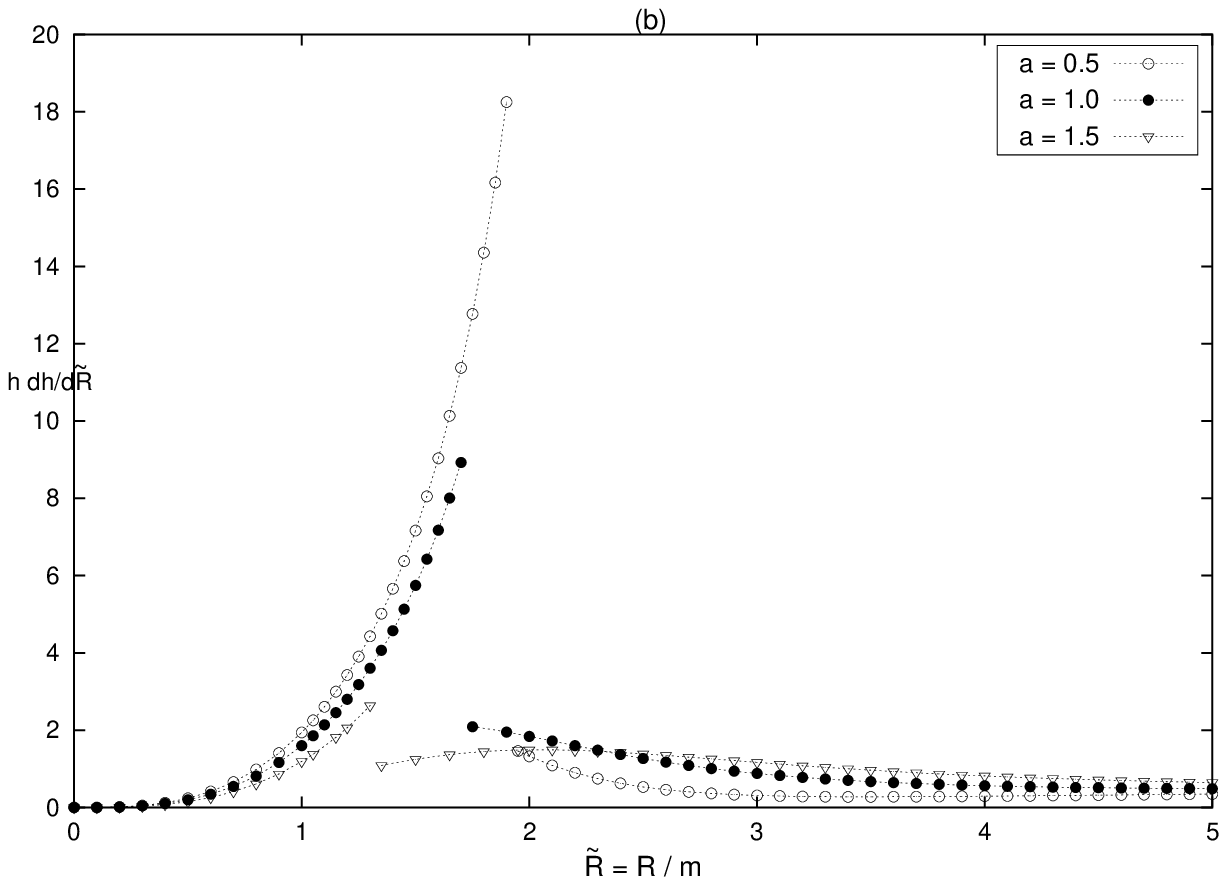}
\caption{(a) The tangential velocity $\mathrm{v}_c$, (b) the curves of $h\frac{dh}{d\tilde{R}}$ for the disk generated from fluid layers
Eq.\ (\ref{eq_layer1})--(\ref{eq_layer2}) with $m=1\text{; }k=1\text{; }r_1=2\text{; for }
a=0.5\text{, }1.0\text{ and }1.5$ as function of $\tilde{R}=R/m$. The disks obtained with 
these parameters have no unstable orbits.} \label{fig_19}
\end{figure}

In figure \ref{fig_19} (a)--(b), the curves of tangential velocities  
and of $h\frac{dh}{d\tilde{R}}$, respectively, are displayed as functions of $\tilde{R}=R/m$.

\subsection{NPV Solution 2b  with $n=\sqrt{2}$ and NPV solution 1b
 with $k=-2+\sqrt{2}$}
 
Now we consider a sphere composed with two finite layers: The internal layer, $0\leq r <r_1$, is taken as the NPV solution 2b with  $n=\sqrt{2}$,
\begin{equation} \label{eq_layer3}
e^{\nu}=\frac{\left( B_1+B_2 \ln (r)\right)^2}{\left( A_1r^{\sqrt{2}/2}+A_2r^{-\sqrt{2}/2}\right)^{2}}
\text{,} \quad e^{\lambda}=\frac{1}{\left( A_1r^{1+\sqrt{2}/2}+A_2r^{1-\sqrt{2}/2}\right)^{2}} \mbox{.}
\end{equation}
The external layer, $r_1<r<r_2$, is taken as the NPV solution 1b
 with $k=-2+\sqrt{2}$,
\begin{equation} \label{eq_layer4}
e^{\nu}= r^{\sqrt{2}}\left( A_3+B_3 \ln (r) \right)^2 \text{,} \quad
e^{\lambda}=Cr^{-2+\sqrt{2}} \mbox{.}
\end{equation}
The spacetime outside the sphere, $r>r_2$, will be taken as the 
Schwarzschild's vacuum solution is isotropic coordinates.
\begin{equation} \label{eq_sch_l2} 
e^{\nu}= \frac{\left(1-\frac{m}{2r}\right)^2}{\left(1+\frac{m}{2r}\right)^{2}} \text{,} \quad
e^{\lambda}=\left( 1+\frac{m}{2r} \right)^4 \mbox{.}
\end{equation}
In this case the pressure should be zero at $r=r_2$. The continuity conditions at $r=r_1$ and $r=r_2$ give the relations:
\begin{align}
\frac{m}{r_2} &=\frac{\sqrt{2}(2-\sqrt{2})}{1+\sqrt{2}}\text{, }C =\frac{64r_2^{2-\sqrt{2}}}{(1+\sqrt{2})^4} \text{, } B_3=\frac{1}{4r_2^{1/\sqrt{2}}}
\text{, }A_3=-\frac{2\sqrt{2}+\ln (r_2)}{r_2^{1/\sqrt{2}}}\mbox{,} \label{eq_cond_l3}\\
A_1 &=0 \text{, }A_2=\frac{1}{\sqrt{C}}\text{, } B_2=\frac{r_1^{\sqrt{2}}}{2\sqrt{2C}}\left[
\left(A_3+B_3 \ln (r_1)\right)(1-r_1^{-\sqrt{2}})+2\sqrt{2}B_3 \right] \mbox{,}\\
B_1 &=\frac{r_1^{\sqrt{2}}}{2\sqrt{2C}}\left[ \left(A_3+B_3 \ln (r_1)\right)\left( 2\sqrt{2}r_1^{-\sqrt{2}}
-\ln (r_1)+\ln (r_1)r_1^{-\sqrt{2}}\right)-2\sqrt{2}B_3 \ln (r_1) \right] \mbox{.} \label{eq_cond_l4}
\end{align}
Using Eq.\ (\ref{eq_en_nar1}) and Eq.\ (\ref{eq_en_nar2}), the energy density of the disk at $R=\sqrt{r_1^2-a^2}$ is continuous, but not the pressure. The difference between Eq. (\ref{eq_Pb_nar2}) and Eq. (\ref{eq_P2_nar1}) is
\begin{equation}
\Delta P=\frac{a(r_1^{\sqrt{2}/2}-r_1^{-\sqrt{2}/2})}{16 \pi \sqrt{C}r_1
(A_3+B_3 \ln (r_1))}\left[ \sqrt{2}(A_3+B_3 \ln (r_1)) +4B_3 \right] \mbox{.}
\end{equation}
The pressure is continuous if $r_1^{\sqrt{2}/2}-r_1^{-\sqrt{2}/2}=0 \rightarrow r_1=1$.

\begin{figure} 
\centering
\includegraphics[scale=0.6]{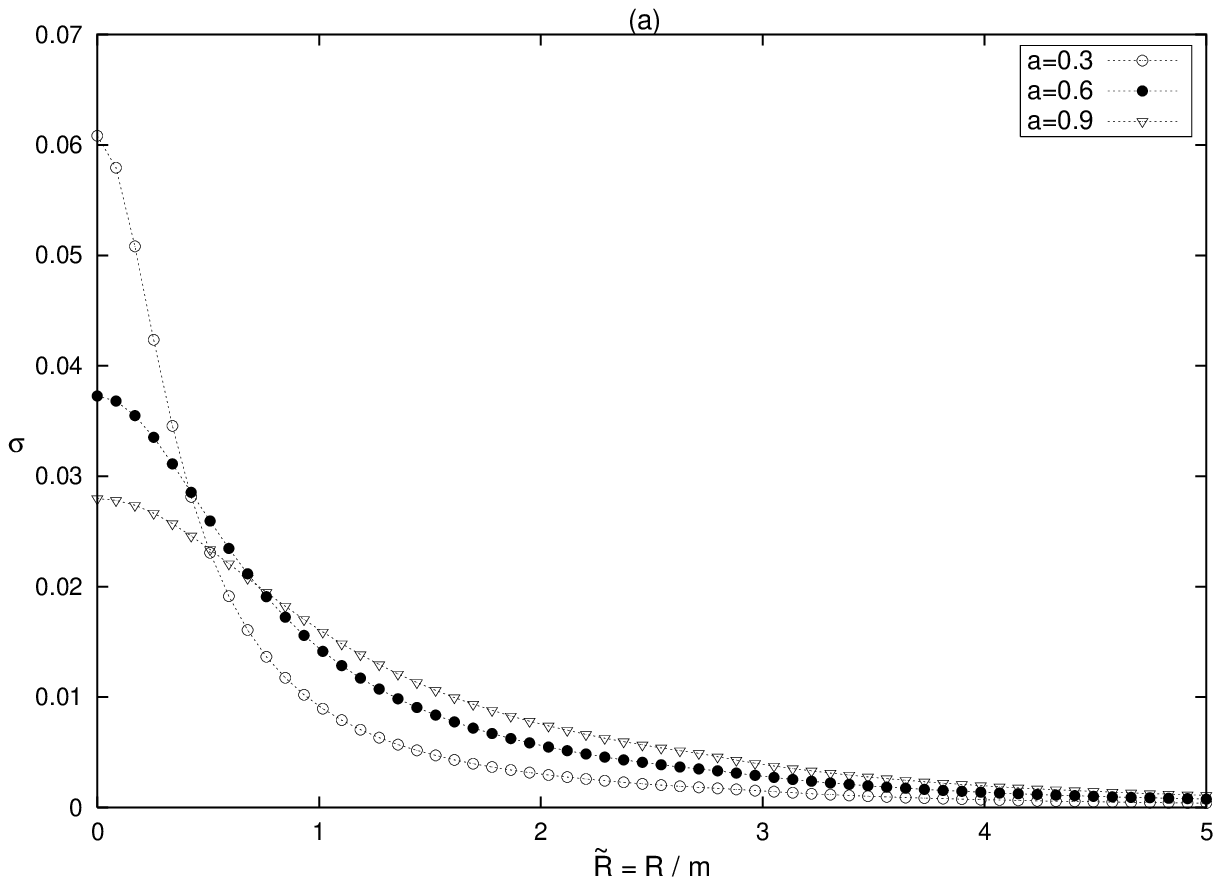}%
\includegraphics[scale=0.6]{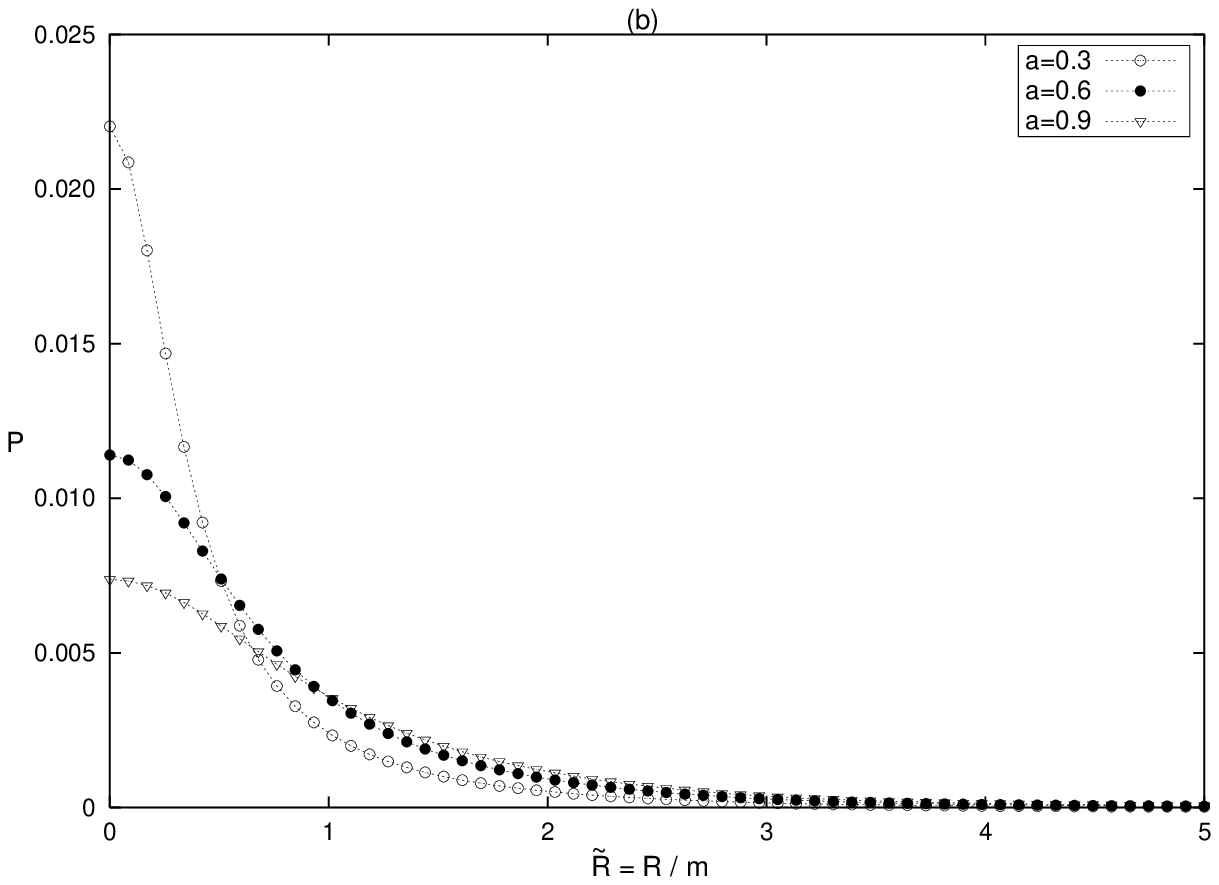} \\
\vspace{0.3cm}
\includegraphics[scale=0.6]{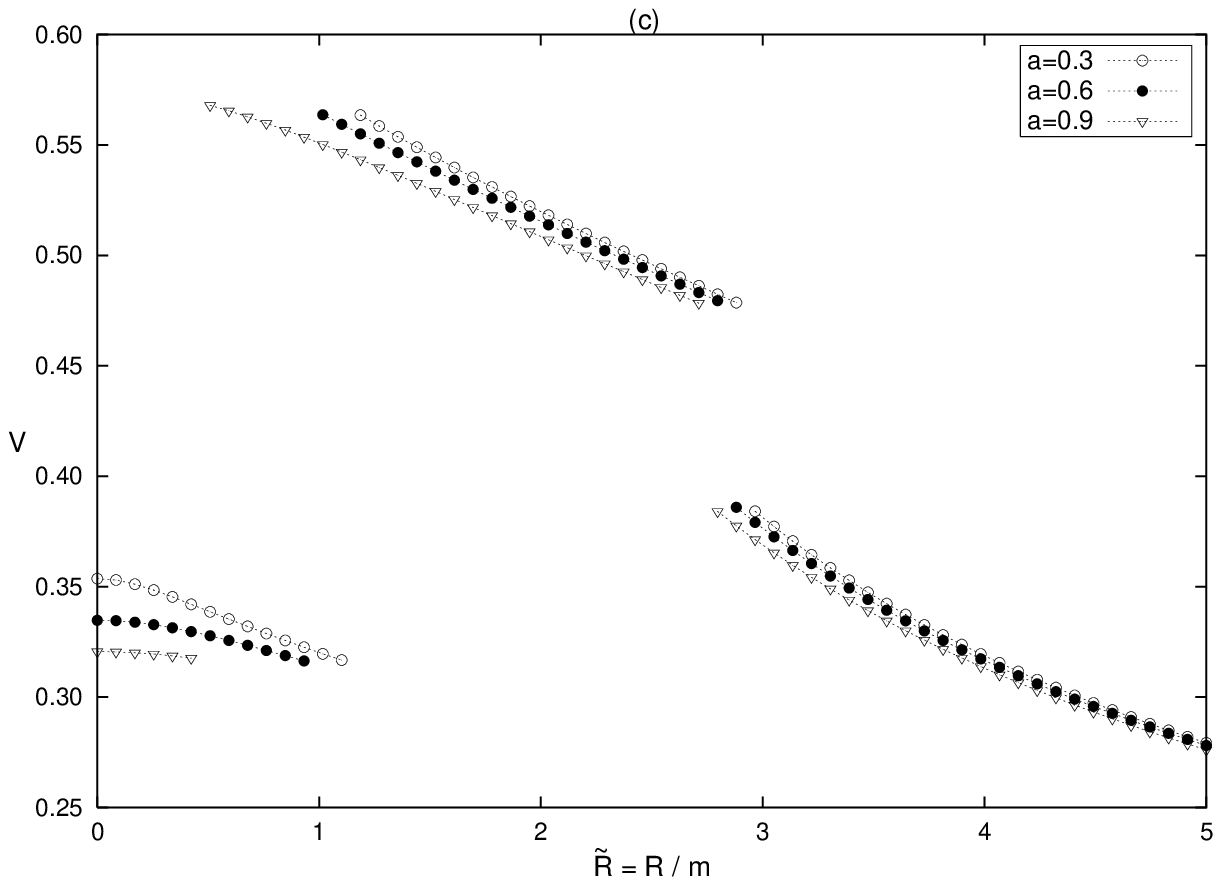}
\caption{(a) The surface energy density $\sigma$, (b) the pressure $P$, (c ) the velocity of sound $V$ for the disk generated from spherical 
fluid layers Eq.\ (\ref{eq_layer3})--(\ref{eq_layer4}) with $r_1=1\text{; }r_2=2\text{; for }
a=0.3\text{, }0.6\text{ and }0.9$ as function of $\tilde{R}=R/m$.} \label{fig_20}
\end{figure}

\begin{figure} 
\centering
\includegraphics[scale=0.6]{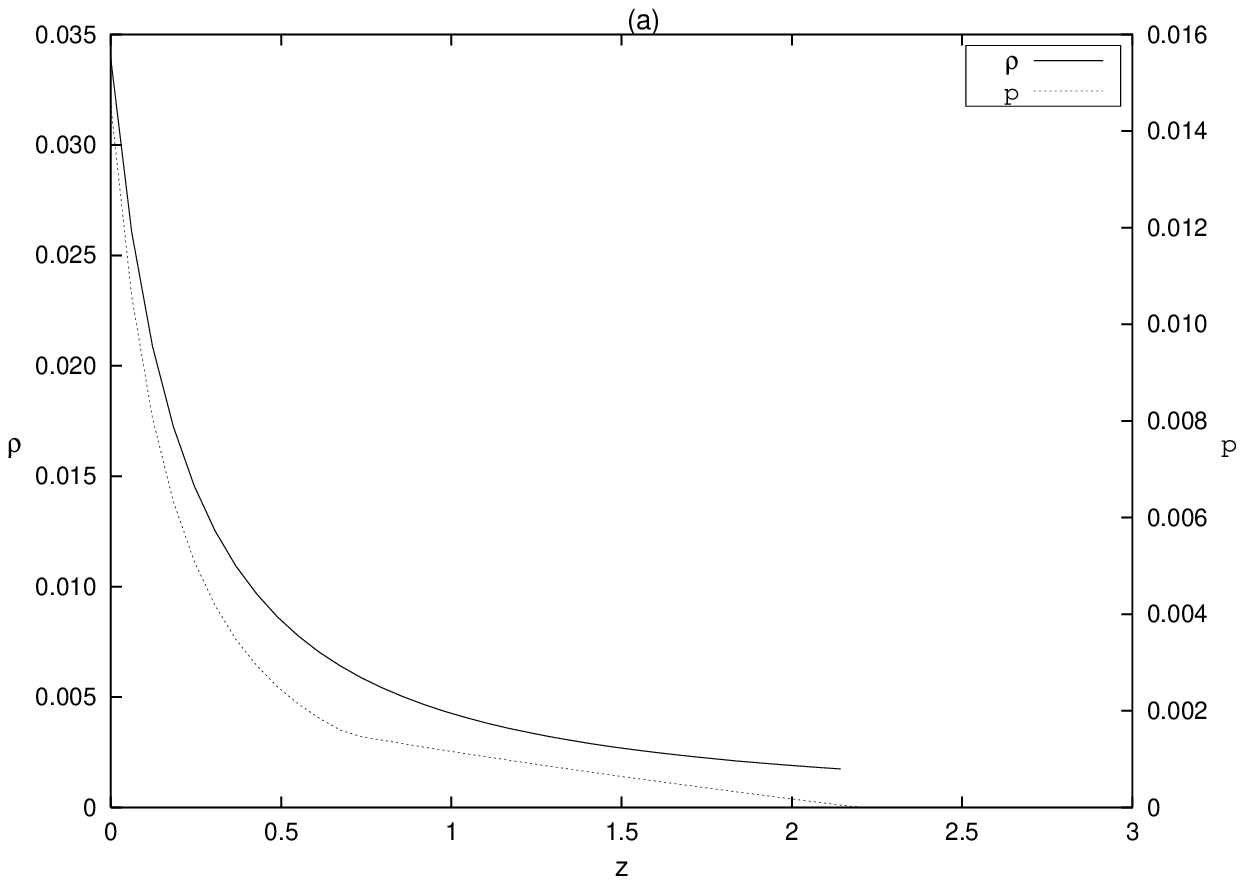}%
\hspace{0.2cm}
\includegraphics[scale=0.6]{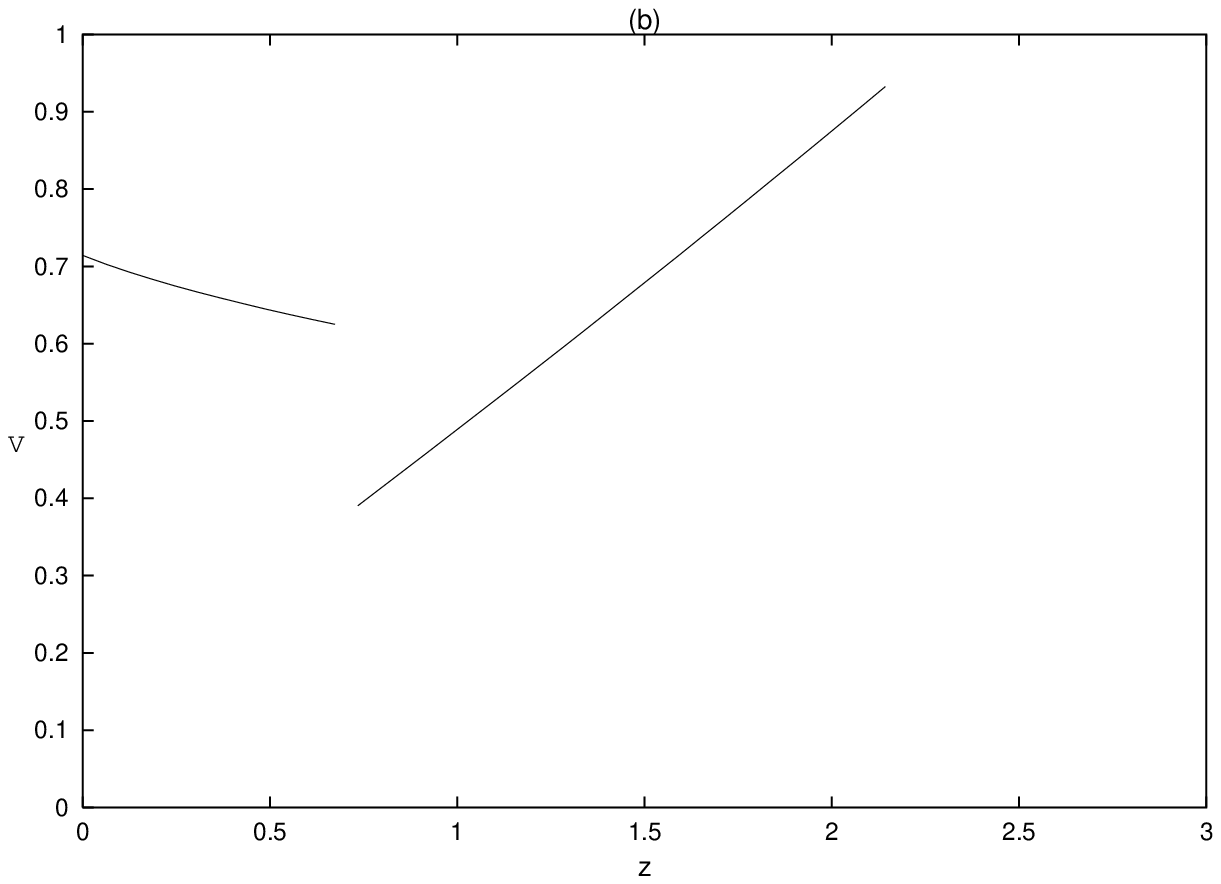}
\caption{(a) The density $\rho$ and pressure $p$, (b) the velocity of sound V for the halo formed by fluid layers 
Eq.\ (\ref{eq_layer3})--(\ref{eq_layer4}) with $r_1=1\text{; }r_2=2\text{ and }a=0.3$ 
along the axis $z$.} \label{fig_21}
\end{figure}

Figure \ref{fig_20} (a)--(c) shows, respectively, $\sigma$, $P$ and $V$ for the disk obtained from fluid layers (\ref{eq_layer3})--(\ref{eq_layer4}) with parameters $ r_1=1\text{; }r_2=2 \text{; for }a=0.3\text{, }0.6\text{ and }0.9$ as function of $\tilde{R}=R/m$. The density $\rho$, pressure $p$ and velocity of sound V for the halo along the $z$ axis with the same parameters for $a=0.3$ is shown in figure \ref{fig_21}.

\begin{figure}
\centering
\includegraphics[scale=0.6]{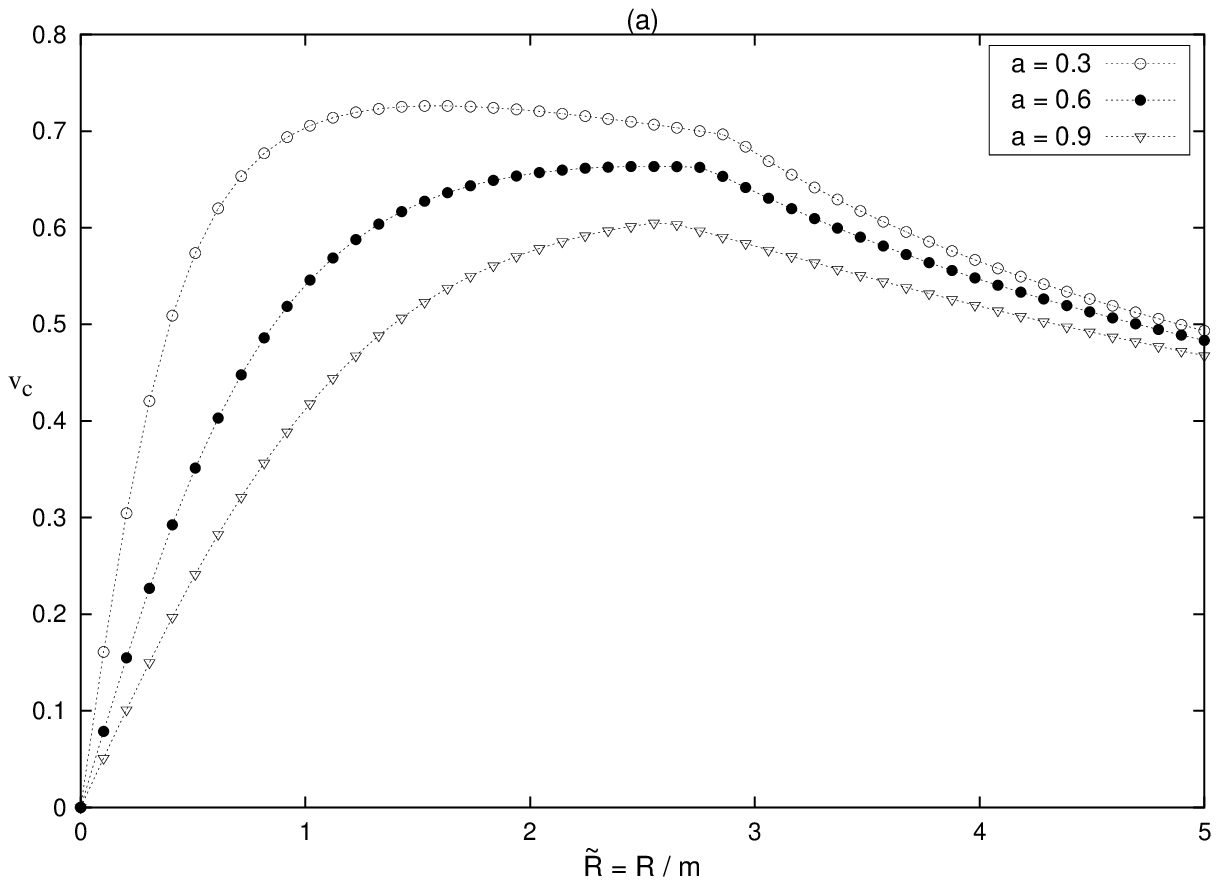}%
\includegraphics[scale=0.6]{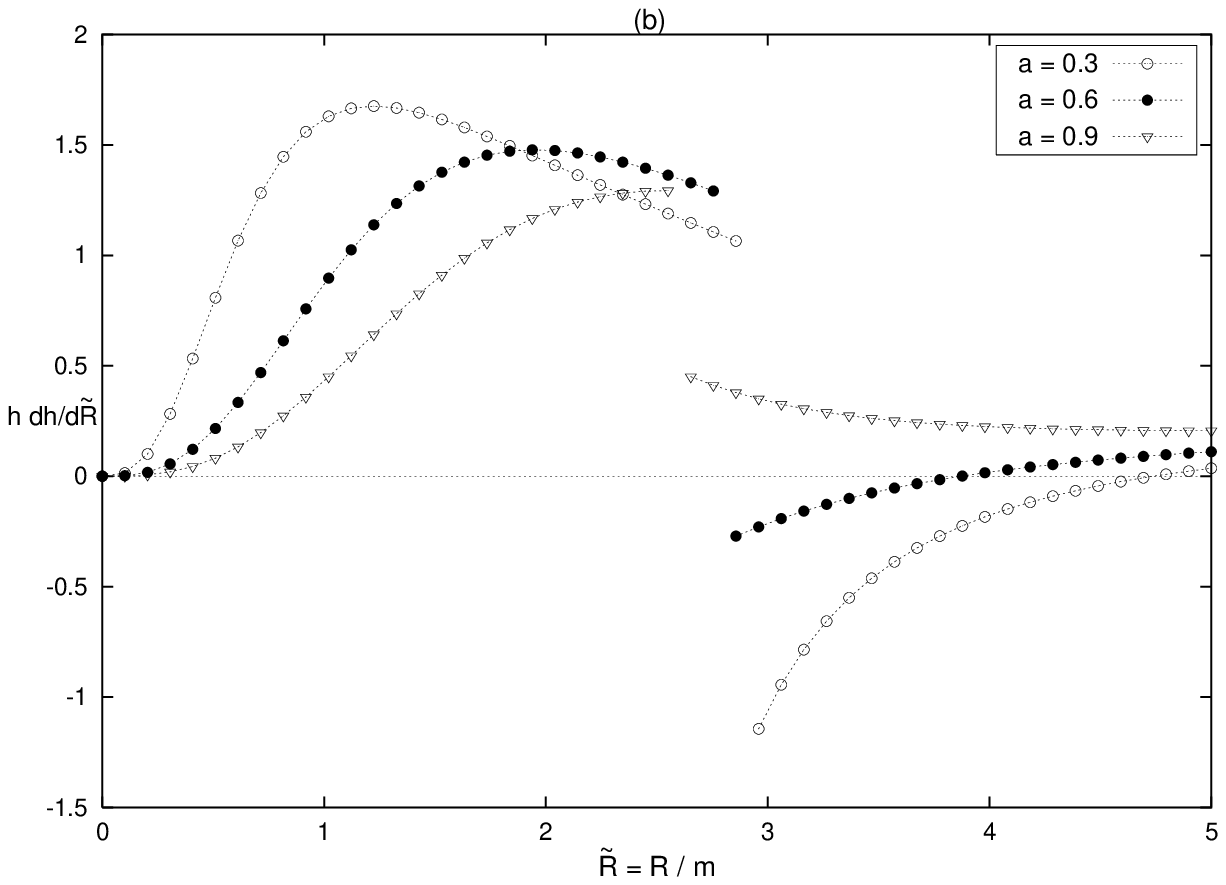}
\caption{(a) The tangential velocity $\mathrm{v}_c$, (b) the curves of $h\frac{dh}{d\tilde{R}}$  for the disk generated from spherical fluid layers
Eq.\ (\ref{eq_layer3})--(\ref{eq_layer4}) with $r_1=1\text{; }r_2=2\text{; for }a=0.3\text{, }
0.6\text{ and }0.9$ as function of $\tilde{R}=R/m$. Regions of unstable circular orbits appear for the 
disks obtained with parameters $a=0.3$ and $0.6$.} \label{fig_22}
\end{figure}

In figure \ref{fig_22} (a)--(b), the curves of tangential velocities  
and of $h\frac{dh}{d\tilde{R}}$, respectively, are displayed as functions of $\tilde{R}=R/m$. In this 
case, regions of unstable orbits exist for parameters $a=0.3$ and $a=0.6$.  

\section{Discussion}

The ``displace, cut and reflect'' method applied to solutions of Einstein field 
equations in isotropic coordinates can generate disks with positive energy density and 
equal radial and azimuthal pressures (perfect fluid). With solutions of static spheres 
of perfect fluid it is possible to construct disks of perfect fluid surrounded also by 
perfect fluid matter. As far we know these are the first disk models of this kind in the literature.

All disks constructed as examples have some common features: surface energy density 
and pressures decrease monotonically and rapidly with radius. As the ``cut'' parameter $a$ decreases, 
the disks become more relativistic, with surface energy density and pressure more concentrated 
near the center. Also regions of unstable circular orbits are more likely to 
appear for high relativistic disks. Parameters can be chosen so 
that the sound velocity in the fluid and the tangential velocity of test particles 
in circular motion are less then the velocity of light. This tangential velocity first increases with
radius and reaches a maximum. Then, for large radii, it decreases as ${1/\sqrt{R}}$, in case of disks generated from Schwarzschild 
and Buchdahl's solutions. The sound velocity is also a decreasing function of radius, except in solution
NPV 2a with $\sqrt{2}<n \leq 2$, where it reaches its maximum value
 at the boundary.
In principle other solutions of static spheres of perfect fluid could be used to
generate other disk + halo configurations, but it is not guaranteed that the disks will have the
characteristics of normal fluid matter.

We believe that the presented  disks can be used to describe   more realistic 
model of galaxies  than most of  the
 already studied disks since the  counter rotation hypothesis is not needed
to have a stable configuration. 

We want to finish our discussion by
 presenting a table that summarizes  our results about disks in an unify manner.

\subsection*{Disks properties}

\begin{center}
\begin{tabular}{|c|c|c|c|c|c|c|}
\hline
\textbf{Solution}&\textbf{metric}&\textbf{matching}&\textbf{energy}&\textbf{pressure}&\textbf{sound}&\textbf{angular}\\
 &\textbf{coefficients}&\textbf{conditions}&\textbf{density}& &\textbf{velocity}&\textbf{momentum} \\ \hline
External Schwarzschild&(\ref{eq_metrica_sch})&--&(\ref{eq_en_sch})&(\ref{eq_P_sch})&(\ref{eq_sound_sch})&(\ref{eq_h_sch})\\ \hline
Buchdahl&(\ref{eq_sol_buch})&--&(\ref{eq_en_buch})&(\ref{eq_P_buch})&(\ref{eq_V_buch})&(\ref{eq_h_buch})\\ \hline
NPV 1a&(\ref{eq_nar_sol1a})--(\ref{eq_nar_sol1b})&(\ref{eq_cond_sol1a})--(\ref{eq_cond_sol1b})&(\ref{eq_en_nar1})&(\ref{eq_P1_nar1})&(\ref{eq_V1_nar1})&(\ref{eq_h_nar1})\\ \hline
NPV 1b&(\ref{eq_nar_sol1a}), (\ref{eq_nar_sol1c})&(\ref{eq_cond_sol1a}), (\ref{eq_cond_sol1c})&(\ref{eq_en_nar1})&(\ref{eq_P2_nar1})&(\ref{eq_V2_nar1})&(\ref{eq_h_nar2})\\ \hline
NPV 2a&(\ref{eq_nar_sol2a})--(\ref{eq_nar_sol2b})&(\ref{eq_cond_sol2a})--(\ref{eq_cond_sol2d})&(\ref{eq_en_nar2})&(\ref{eq_Pa_nar2})&(\ref{eq_Va_nar2})&(\ref{eq_h_nara})\\ \hline
NPV 2b&(\ref{eq_nar_sol2a}), (\ref{eq_nar_sol2c})&(\ref{eq_cond_sol2a})--(\ref{eq_cond_sol2b}), (\ref{eq_cond_sol2e})--(\ref{eq_cond_sol2f})&(\ref{eq_en_nar2})&(\ref{eq_Pb_nar2})&(\ref{eq_Vb_nar2})&(\ref{eq_h_narb})\\ \hline
Internal Schwarzschild&&&&&& \\
+Buchdahl&(\ref{eq_layer1})--(\ref{eq_layer2})&(\ref{eq_cond_l1})--(\ref{eq_cond_l2})&(\ref{eq_en_nar2}), (\ref{eq_en_buch})&(\ref{eq_Pa_nar2}), (\ref{eq_P_buch})&(\ref{eq_Va_nar2}), (\ref{eq_V_buch})&(\ref{eq_h_nara}), (\ref{eq_h_buch})\\ \hline
NPV 2b+NPV 1b&(\ref{eq_layer3}), (\ref{eq_layer4}),&(\ref{eq_cond_l3})--(\ref{eq_cond_l4})&(\ref{eq_en_nar2}), (\ref{eq_en_nar1}),&(\ref{eq_Pb_nar2}), (\ref{eq_P2_nar1}),&(\ref{eq_Vb_nar2}), (\ref{eq_V2_nar1}),&(\ref{eq_h_narb}), (\ref{eq_h_nar2}),\\
+external Schwarzschild&(\ref{eq_sch_l2})&&(\ref{eq_en_sch})&(\ref{eq_P_sch})&(\ref{eq_sound_sch})&(\ref{eq_h_sch})\\ \hline
\end{tabular}
\end{center}

In the table we list the seed metric coefficients, matching conditions at the boundaries  and relevant physical quantities of
all disks studied in this work. The numbers refer to  the  equations presented along the paper and  NPV  stands for Narlikar, Patwardhan and Vaidya as before.

\acknowledgments{We want to thank FAPESP, CAPES and CNPQ for financial support.}


\begin{thebibliography}{99}
\bibitem{Bonnor} W. A. Bonnor and A. Sackfield, \textit{Commun. Math. Phys.} \textbf{8}, 
338 (1968).
\bibitem{Morgan1} T. Morgan and L. Morgan, \textit{Phys. Rev.} \textbf{183}, 1097 
(1969).
\bibitem{Morgan2} L. Morgan and T. Morgan, \textit{Phys. Rev. D} \textbf{2}, 2756 (1970).
\bibitem{Gonzalez} G. Gonz\'{a}lez and P. S. Letelier, \textit{Class. Quantum Grav.} 
\textbf{16}, 479 (1999).
\bibitem{Bicak} T. Ledvinka, M. Zofka, and J. Bi\v{c}\'{a}k, in 
\textit{Proceedings of the 8th Marcel Grossman Meeting in General Relativity},
edited by T. Piran (World Scientific, Singapore, 1999), pp. 339-341.
\bibitem{Letelier1} P. S. Letelier, \textit{Phys. Rev. D} \textbf{60}, 104042 (1999).
\bibitem{Katz} J. Katz, J. Bi\v{c}\'{a}k, and D. Lynden-Bell, \textit{Class. Quantum 
Grav.} \textbf{16}, 4023 (1999).
\bibitem{LP} D. Lynden-Bell and S. Pineault, Mon. Not. R. Astron. Soc.
185, 679 (1978)
\bibitem{LEM} J. P. S. Lemos, Class. Quan. Grav. 6, 1219 (1989)
\bibitem{LL1} J. P. S. Lemos and P. S. Letelier, Class. Quan. Grav. 10,
L75 (1993)
\bibitem{LL2}
J. P. S. Lemos and P. S. Letelier, Phys. Rev D49, 5135 (1994)
\bibitem{LL3} J. P. S. Lemos and P. S. Letelier, Int. J. Mod. Phys. D5,
53 (1996)
\bibitem{KLE} C. Klein, Class. Quan. Grav. 14, 2267 (1997)
\bibitem{BLK} J. Bi\u{c}\'{a}k, D. Lynden-Bell and J. Katz, Phys. Rev.
D47, 4334 (1993) 
\bibitem{BLP} J. Bi\u{c}\'{a}k, D. Lynden-Bell and C. Pichon, Mon. Not.
R. Astron. Soc. 265, 126 (1993)
\bibitem{Semerak} O. Semer\'{a}k, \textit{Towards gravitating disks around stationary black holes}, (April, 2002), available at http://xxx.lanl.gov/abs/gr-qc/0204025.
\bibitem{Kuzmin} G. G. Kuzmin, \textit{Astron. Zh.}, \textbf{33}, 27 (1956). 
\bibitem{Taub} A. H. Taub, J. Math. Phys. 21, 1423 (1980) 
\bibitem{LLstrings} J.P.S. Lemos and P.S. Letelier, Phys Lett. A 153, 288 (1991).
\bibitem{lord} Lord Rayleigh,  Proc. R. Soc. Lond.
 Ser. A 93, 148 (1916).
\bibitem{Landau} L. D. Landau, E. M. Lifshitz, \textit{Fluid Mechanics}, 2nd Ed. (Pergamon Press, Oxford, 1987), \S 27.
\bibitem{Kuchowicz} B. Kuchowicz, \textit{Acta Phys. Pol.}, 
\textbf{B} 3, 209 (1972).
\bibitem{Buchdahl} H. A. Buchdahl, \textit{Astrophys. J.}, \textbf{140}, 1512 (1964).

\bibitem{Narlikar} V. V. Narlikar, G. K. Patwardhan, P. C. Vaidya, \textit{Proc. Natl. Inst. Sci. India}, \textbf{9}, 229 (1943).
\end{thebibliography}
\end{document}